\documentclass[twocolumn]{aastex7}
\usepackage{hyperref}
\usepackage{graphicx}
\usepackage{color}
\usepackage{tipa}
\usepackage{CJK}
\usepackage{gensymb}
\usepackage{amsmath,amstext}
\usepackage{amssymb}
\usepackage{xspace}
\usepackage{siunitx}
\DeclareSIUnit{\angstrom}{\text{\AA}}
\newcommand{\dotsec}{\rlap{.}^\text{s}}
\newcommand{\kkoname}
{k'ni\textipa{P}atn k'l$\left._\mathrm{\smile}\right.$stk'masqt}
\newcommand{\halpha}{\ensuremath{\mathrm{H\alpha}}}
\newcommand{\frb}{FRB\,20250316A\xspace}
\def\torun{Toru\'n\xspace}
\def\nancay{Nan\c{c}ay\xspace}

\newcommand{\kkonamecaps}{K'ni\textipa{P}atn k'l$\left._\mathrm{\smile}\right.$stk'masqt}

\newcommand{\hbeta}{\ensuremath{\mathrm{H\beta}}}
\newcommand{\dmunits}{\ensuremath{\mathrm{pc~cm^{-3}}}}
\newcommand{\dmhism}{\ensuremath{\mathrm{DM_{host, ISM}}}}
\newcommand{\emha}{\ensuremath{\mathrm{EM_{\halpha}}}}

\defcitealias{CHIME+2021}{CHIME/FRB Collaboration et~al.}
\defcitealias{outriggers_2025}{CHIME/FRB Collaboration et~al.}
\defcitealias{chimeoverview}{CHIME Collaboration et~al.}
\defcitealias{catIIsubmitted}{CHIME/FRB Collaboration et~al.}
\begin{document}
\begin{CJK*}{UTF8}{gbsn}
\title{\frb: A Brilliant and Nearby One-Off Fast Radio Burst Localized to 13~parsec Precision}
\shortauthors{CHIME/FRB Collaboration et al.}
  \collaboration{1}{The CHIME/FRB collaboration:}

\correspondingauthor{Amanda M. Cook}
\email{amanda.cook@mail.mcgill.ca}
\author[0000-0001-5002-0868]{Thomas C. Abbott}
\email{thomas.abbott@mail.mcgill.ca}
  \affiliation{Department of Physics, McGill University, 3600 rue University, Montr\'eal, QC H3A 2T8, Canada}
  \affiliation{Trottier Space Institute, McGill University, 3550 rue University, Montr\'eal, QC H3A 2A7, Canada}
  \author[0009-0005-5370-7653]{Daniel Amouyal}
\email{daniel.amouyal@mail.mcgill.ca}
  \affiliation{Department of Physics, McGill University, 3600 rue University, Montr\'eal, QC H3A 2T8, Canada}
  \affiliation{Trottier Space Institute, McGill University, 3550 rue University, Montr\'eal, QC H3A 2A7, Canada}
\author[0000-0002-3980-815X]{Shion E.~Andrew}
  \email{shiona@mit.edu}
  \affiliation{MIT Kavli Institute for Astrophysics and Space Research, Massachusetts Institute of Technology, 77 Massachusetts Ave, Cambridge, MA 02139, USA}
  \affiliation{Department of Physics, Massachusetts Institute of Technology, 77 Massachusetts Ave, Cambridge, MA 02139, USA}
\author[0000-0003-3772-2798]{Kevin Bandura}
  \email{kevin.bandura@mail.wvu.edu}
  \affiliation{Lane Department of Computer Science and Electrical Engineering, 1220 Evansdale Drive, PO Box 6109, Morgantown, WV 26506, USA}
  \affiliation{Center for Gravitational Waves and Cosmology, West Virginia University, Chestnut Ridge Research Building, Morgantown, WV 26505, USA}
  \author[0000-0002-3615-3514]{Mohit Bhardwaj}
  \email{mohitb@andrew.cmu.edu}
  \affiliation{McWilliams Center for Cosmology \& Astrophysics, Department of Physics, Carnegie Mellon University, Pittsburgh, PA 15213, USA}
  \author[0000-0002-9218-1624]{Kalyani Bhopi}
  \email{kalyani.bhopi@mail.wvu.edu}
  \affiliation{Lane Department of Computer Science and Electrical Engineering, 1220 Evansdale Drive, PO Box 6109, Morgantown, WV 26506, USA}
  \affiliation{Center for Gravitational Waves and Cosmology, West Virginia University, Chestnut Ridge Research Building, Morgantown, WV 26505, USA}
\author[0000-0002-5342-163X]{Yash Bhusare}
  \email{ybhusare@ncra.tifr.res.in}
  \affiliation{National Centre for Radio Astrophysics, Post Bag 3, Ganeshkhind, Pune, 411007, India}
  \affiliation{Department of Astronomy and Astrophysics, Tata Institute of Fundamental Research, Mumbai, 400005, India}
  
\author[0000-0002-1800-8233]{Charanjot Brar}
  \email{charanjot.brar@nrc-cnrc.gc.ca}
  \affiliation{National Research Council of Canada, Herzberg Astronomy and Astrophysics, 5071 West Saanich Road, Victoria, BC V9E 2E7, Canada}
\author[0009-0001-0983-623X]{Alice Cai}
  \email{acai@u.northwestern.edu}
    \affiliation{Department of Physics and Astronomy, Northwestern University, Evanston, IL 60208, USA}
  \affiliation{Center for Interdisciplinary Exploration and Research in Astronomy, Northwestern University, 1800 Sherman Avenue, Evanston, IL 60201, USA}
  \author[0000-0003-2047-5276]{Tomas Cassanelli}
  \email{tcassanelli@ing.uchile.cl}
  \affiliation{Department of Electrical Engineering, Universidad de Chile, Av. Tupper 2007, Santiago 8370451, Chile}
\author[0000-0002-2878-1502]{Shami Chatterjee}
  \email{shami@astro.cornell.edu}
  \affiliation{Cornell Center for Astrophysics and Planetary Science, Cornell University, Ithaca, NY 14853, USA}
  \author[0000-0001-6509-8430]{Jean-Fran\c{c}ois Cliche}
\email{jfcliche@jfcliche.com}
\affiliation{Department of Physics, McGill University, 3600 rue University, Montr\'eal, QC H3A 2T8, Canada}
  \affiliation{Trottier Space Institute, McGill University, 3550 rue University, Montr\'eal, QC H3A 2A7, Canada}
\author[0000-0001-6422-8125]{Amanda M.~Cook}
  \email{amanda.cook@mail.mcgill.ca}
  \affiliation{Department of Physics, McGill University, 3600 rue University, Montr\'eal, QC H3A 2T8, Canada}
  \affiliation{Trottier Space Institute, McGill University, 3550 rue University, Montr\'eal, QC H3A 2A7, Canada}
  \affiliation{Anton Pannekoek Institute for Astronomy, University of Amsterdam, Science Park 904, 1098 XH Amsterdam, The Netherlands}
\author[0000-0002-8376-1563]{Alice P.~Curtin}
  \email{alice.curtin@mail.mcgill.ca}
  \affiliation{Department of Physics, McGill University, 3600 rue University, Montr\'eal, QC H3A 2T8, Canada}
  \affiliation{Trottier Space Institute, McGill University, 3550 rue University, Montr\'eal, QC H3A 2A7, Canada}
  \author[0009-0002-1944-6398]{Evan Davies-Velie}
  \email{evan.davies-velie@mail.mcgill.ca}
  \affiliation{Department of Physics, McGill University, 3600 rue University, Montr\'eal, QC H3A 2T8, Canada}
  \affiliation{Trottier Space Institute, McGill University, 3550 rue University, Montr\'eal, QC H3A 2A7, Canada}
\author[0000-0001-7166-6422]{Matt Dobbs}
  \email{matt.dobbs@mcgill.ca}
  \affiliation{Department of Physics, McGill University, 3600 rue University, Montr\'eal, QC H3A 2T8, Canada}
  \affiliation{Trottier Space Institute, McGill University, 3550 rue University, Montr\'eal, QC H3A 2A7, Canada}
\author{Fengqiu Adam Dong}
  \email{adong@nrao.edu}
  \affiliation{National Radio Astronomy Observatory, 520 Edgemont Rd, Charlottesville, VA 22903, USA}
\author[0000-0002-9363-8606]{Yuxin (董雨欣) Dong}
  \email{yuxin.dong@northwestern.edu}
    \affiliation{Department of Physics and Astronomy, Northwestern University, Evanston, IL 60208, USA}
  \affiliation{Center for Interdisciplinary Exploration and Research in Astronomy, Northwestern University, 1800 Sherman Avenue, Evanston, IL 60201, USA}
  \author[0000-0003-3734-8177]{Gwendolyn  Eadie}
  \email{gwen.eadie@utoronto.ca}
  \affiliation{David A.\ Dunlap Department of Astronomy and Astrophysics, 50 St. George Street, University of Toronto, ON M5S 3H4, Canada}
    \affiliation{Department of Statistical Sciences, University of Toronto, Toronto, ON M5S 3G3, Canada}
    \affiliation{Data Sciences Institute, University of Toronto, 17th Floor, Ontario Power Building, 700 University Ave, Toronto, ON M5G 1Z5, Canada}

\author[0000-0003-0307-9984]{Tarraneh Eftekhari}
  \email{teftekhari@northwestern.edu}
  \affiliation{Center for Interdisciplinary Exploration and Research in Astronomy, Northwestern University, 1800 Sherman Avenue, Evanston, IL 60201, USA}
\author[0000-0002-7374-935X]{Wen-fai Fong}
  \email{wfong@northwestern.edu}
    \affiliation{Department of Physics and Astronomy, Northwestern University, Evanston, IL 60208, USA}
  \affiliation{Center for Interdisciplinary Exploration and Research in Astronomy, Northwestern University, 1800 Sherman Avenue, Evanston, IL 60201, USA}
\author[0000-0001-8384-5049]{Emmanuel Fonseca}
  \email{emmanuel.fonseca@mail.wvu.edu}
  \affiliation{Department of Physics and Astronomy, West Virginia University, PO Box 6315, Morgantown, WV 26506, USA }
  \affiliation{Center for Gravitational Waves and Cosmology, West Virginia University, Chestnut Ridge Research Building, Morgantown, WV 26505, USA}
\author[0000-0002-3382-9558]{B.~M.~Gaensler}
  \email{gaensler@ucsc.edu}
  \affiliation{Department of Astronomy and Astrophysics, University of California, Santa Cruz, 1156 High Street, Santa Cruz, CA 95060, USA}
  \affiliation{Dunlap Institute for Astronomy and Astrophysics, 50 St. George Street, University of Toronto, ON M5S 3H4, Canada}
  \affiliation{David A.\ Dunlap Department of Astronomy and Astrophysics, 50 St. George Street, University of Toronto, ON M5S 3H4, Canada}
\author[0000-0001-6128-3735]{Nina Gusinskaia}
  \email{gusinskaia@astron.nl}
  \affiliation{ASTRON, Netherlands Institute for Radio Astronomy, Oude Hoogeveensedijk 4, 7991 PD Dwingeloo, The Netherlands}
  \affiliation{Anton Pannekoek Institute for Astronomy, University of Amsterdam, Science Park 904, 1098 XH Amsterdam, The Netherlands}
\author[0000-0003-2317-1446]{Jason W. T. Hessels}
  \email{jason.hessels@mcgill.ca}
  \affiliation{Department of Physics, McGill University, 3600 rue University, Montr\'eal, QC H3A 2T8, Canada}
  \affiliation{Trottier Space Institute, McGill University, 3550 rue University, Montr\'eal, QC H3A 2A7, Canada}
  \affiliation{Anton Pannekoek Institute for Astronomy, University of Amsterdam, Science Park 904, 1098 XH Amsterdam, The Netherlands}
  \affiliation{ASTRON, Netherlands Institute for Radio Astronomy, Oude Hoogeveensedijk 4, 7991 PD Dwingeloo, The Netherlands}
\author[0000-0002-5794-2360]{Dant\'e M. Hewitt}
\email{d.m.hewitt@uva.nl}
  \affiliation{Anton Pannekoek Institute for Astronomy, University of Amsterdam, Science Park 904, 1098 XH Amsterdam, The Netherlands}
\author[0000-0002-8043-0048]{Jeff Huang}
  \email{jeff.huang@mail.mcgill.ca}
  \affiliation{Department of Physics, McGill University, 3600 rue University, Montr\'eal, QC H3A 2T8, Canada}
  \affiliation{Trottier Space Institute, McGill University, 3550 rue University, Montr\'eal, QC H3A 2A7, Canada}
  \author[0009-0009-0938-1595]{Naman Jain}
    \email{naman.jain@mail.mcgill.ca}
    \affiliation{Department of Physics, McGill University, 3600 rue University, Montr\'eal, QC H3A 2T8, Canada}
    \affiliation{Trottier Space Institute, McGill University, 3550 rue University, Montr\'eal, QC H3A 2A7, Canada}
\author[0000-0003-3457-4670]{Ronniy.~C.~Joseph}
  \email{ronniy.joseph@mcgill.ca}
  \affiliation{Department of Physics, McGill University, 3600 rue University, Montr\'eal, QC H3A 2T8, Canada}
  \affiliation{Trottier Space Institute, McGill University, 3550 rue University, Montr\'eal, QC H3A 2A7, Canada}
\author[0009-0007-5296-4046]{Lordrick Kahinga}
  \email{lkahinga@ucsc.edu}
  \affiliation{Department of Astronomy and Astrophysics, University of California, Santa Cruz, 1156 High Street, Santa Cruz, CA 95060, USA}
\author[0000-0001-9345-0307]{Victoria M.~Kaspi}
  \email{victoria.kaspi@mcgill.ca}
  \affiliation{Department of Physics, McGill University, 3600 rue University, Montr\'eal, QC H3A 2T8, Canada}
  \affiliation{Trottier Space Institute, McGill University, 3550 rue University, Montr\'eal, QC H3A 2A7, Canada}
\author[0009-0004-4176-0062]{Afrasiyab (Afrokk) Khan}
\email{afrasiyab.khan@mcgill.ca}
  \affiliation{Department of Physics, McGill University, 3600 rue University, Montr\'eal, QC H3A 2T8, Canada}
  \affiliation{Trottier Space Institute, McGill University, 3550 rue University, Montr\'eal, QC H3A 2A7, Canada}
\author[0009-0008-6166-1095]{Bikash Kharel}
  \email{bk0055@mix.wvu.edu}
  \affiliation{Department of Physics and Astronomy, West Virginia University, PO Box 6315, Morgantown, WV 26506, USA }
  \affiliation{Center for Gravitational Waves and Cosmology, West Virginia University, Chestnut Ridge Research Building, Morgantown, WV 26505, USA}
\author[0000-0003-2116-3573]{Adam E.~Lanman}
  \email{alanman@mit.edu}
  \affiliation{MIT Kavli Institute for Astrophysics and Space Research, Massachusetts Institute of Technology, 77 Massachusetts Ave, Cambridge, MA 02139, USA}
  \affiliation{Department of Physics, Massachusetts Institute of Technology, 77 Massachusetts Ave, Cambridge, MA 02139, USA}
\author[0000-0001-5523-6051]{Magnus L'Argent}
  \email{magnus.largent@mail.mcgill.ca}
  \affiliation{Department of Physics, McGill University, 3600 rue University, Montr\'eal, QC H3A 2T8, Canada}
  \affiliation{Trottier Space Institute, McGill University, 3550 rue University, Montr\'eal, QC H3A 2A7, Canada}
\author[0000-0002-5857-4264]{Mattias Lazda}
  \email{mattias.lazda@mail.utoronto.ca}
  \affiliation{Dunlap Institute for Astronomy and Astrophysics, 50 St. George Street, University of Toronto, ON M5S 3H4, Canada}
  \affiliation{David A.\ Dunlap Department of Astronomy and Astrophysics, 50 St. George Street, University of Toronto, ON M5S 3H4, Canada}
\author[0000-0002-4209-7408]{Calvin Leung}
  \email{calvin_leung@berkeley.edu}
  \affiliation{Miller Institute for Basic Research in Science, Stanley Hall, Room 206B, Berkeley, CA 94720, USA}
 \affiliation{Department of Astronomy, University of California, Berkeley, CA 94720-3411, USA}
\author[0000-0002-7164-9507]{Robert Main}
  \email{robert.main@mcgill.ca}
  \affiliation{Department of Physics, McGill University, 3600 rue University, Montr\'eal, QC H3A 2T8, Canada}
  \affiliation{Trottier Space Institute, McGill University, 3550 rue University, Montr\'eal, QC H3A 2A7, Canada}
\author[0000-0003-4584-8841]{Lluis Mas-Ribas}
  \email{lmr@ucsc.edu}
  \affiliation{Department of Astronomy and Astrophysics, University of California, Santa Cruz, 1156 High Street, Santa Cruz, CA 95060, USA}
\author[0000-0002-4279-6946]{Kiyoshi W.~Masui}
  \email{kmasui@mit.edu}
  \affiliation{MIT Kavli Institute for Astrophysics and Space Research, Massachusetts Institute of Technology, 77 Massachusetts Ave, Cambridge, MA 02139, USA}
  \affiliation{Department of Physics, Massachusetts Institute of Technology, 77 Massachusetts Ave, Cambridge, MA 02139, USA}
\author[0000-0003-2111-3437]{Kyle McGregor}
  \email{kyle.mcgregor@mail.mcgill.ca}
  \affiliation{Trottier Space Institute, McGill University, 3550 rue University, Montr\'eal, QC H3A 2A7, Canada}
  \affiliation{Department of Physics, McGill University, 3600 rue University, Montr\'eal, QC H3A 2T8, Canada}
  \author[0000-0001-7348-6900]{Ryan Mckinven}
  \email{ryan.mckinven@mcgill.ca}
    \affiliation{Department of Physics, McGill University, 3600 rue University, Montr\'eal, QC H3A 2T8, Canada}
    \affiliation{Trottier Space Institute, McGill University, 3550 rue University, Montr\'eal, QC H3A 2A7, Canada}
\author[0000-0002-0772-9326]{Juan Mena-Parra}
  \email{juan.menaparra@utoronto.ca}
  \affiliation{Dunlap Institute for Astronomy and Astrophysics, 50 St. George Street, University of Toronto, ON M5S 3H4, Canada}
  \affiliation{David A.\ Dunlap Department of Astronomy and Astrophysics, 50 St. George Street, University of Toronto, ON M5S 3H4, Canada}
\author[0000-0002-2551-7554]{Daniele Michilli}
  \email{danielemichilli@gmail.com}
    \affiliation{Laboratoire d'Astrophysique de Marseille, Aix-Marseille Univ., CNRS, CNES, Marseille, France}
\author[0000-0001-7491-046X]{Nicole Mulyk}
\email{nicole.mulyk@mail.mcgill.ca}
  \affiliation{Department of Physics, McGill University, 3600 rue University, Montr\'eal, QC H3A 2T8, Canada}
  \affiliation{Trottier Space Institute, McGill University, 3550 rue University, Montr\'eal, QC H3A 2A7, Canada}
\author[0000-0002-0940-6563]{Mason Ng}
  \email{mason.ng@mcgill.ca}
  \affiliation{Department of Physics, McGill University, 3600 rue University, Montr\'eal, QC H3A 2T8, Canada}
  \affiliation{Trottier Space Institute, McGill University, 3550 rue University, Montr\'eal, QC H3A 2A7, Canada}
  %\affiliation{FRQNT Postdoctoral Fellow}
\author[0000-0003-0510-0740]{Kenzie Nimmo}
  \email{knimmo@mit.edu}
  \affiliation{MIT Kavli Institute for Astrophysics and Space Research, Massachusetts Institute of Technology, 77 Massachusetts Ave, Cambridge, MA 02139, USA}
\author[0000-0002-8897-1973]{Ayush Pandhi}
  \email{ayush.pandhi@mail.utoronto.ca}
  \affiliation{David A.\ Dunlap Department of Astronomy and Astrophysics, 50 St. George Street, University of Toronto, ON M5S 3H4, Canada}
  \affiliation{Dunlap Institute for Astronomy and Astrophysics, 50 St. George Street, University of Toronto, ON M5S 3H4, Canada}
\author[0009-0008-7264-1778]{Swarali Shivraj Patil}
  \email{sp00049@mix.wvu.edu}
  \affiliation{Department of Physics and Astronomy, West Virginia University, PO Box 6315, Morgantown, WV 26506, USA }
  \affiliation{Center for Gravitational Waves and Cosmology, West Virginia University, Chestnut Ridge Research Building, Morgantown, WV 26505, USA}
\author[0000-0002-8912-0732]{Aaron B.~Pearlman}
  \email{aaron.b.pearlman@physics.mcgill.ca}
  \affiliation{Department of Physics, McGill University, 3600 rue University, Montr\'eal, QC H3A 2T8, Canada}
  \affiliation{Trottier Space Institute, McGill University, 3550 rue University, Montr\'eal, QC H3A 2A7, Canada}
  \altaffiliation{Banting Fellow, McGill Space Institute Fellow, and FRQNT Postdoctoral Fellow}
\author[0000-0003-2155-9578]{Ue-Li Pen}
  \email{pen@cita.utoronto.ca}
  \affiliation{Institute of Astronomy and Astrophysics, Academia Sinica, Astronomy-Mathematics Building, No. 1, Sec. 4, Roosevelt Road, Taipei 10617, Taiwan}
  \affiliation{Canadian Institute for Theoretical Astrophysics, 60 St.~George Street, Toronto, ON M5S 3H8, Canada}
  \affiliation{Canadian Institute for Advanced Research, 180 Dundas St West, Toronto, ON M5G 1Z8, Canada}
  \affiliation{Dunlap Institute for Astronomy and Astrophysics, 50 St. George Street, University of Toronto, ON M5S 3H4, Canada}
  \affiliation{Perimeter Institute of Theoretical Physics, 31 Caroline Street North, Waterloo, ON N2L 2Y5, Canada}
\author[0000-0002-4795-697X]{Ziggy Pleunis}
  \email{z.pleunis@uva.nl}
  \affiliation{Anton Pannekoek Institute for Astronomy, University of Amsterdam, Science Park 904, 1098 XH Amsterdam, The Netherlands}
  \affiliation{ASTRON, Netherlands Institute for Radio Astronomy, Oude Hoogeveensedijk 4, 7991 PD Dwingeloo, The Netherlands}
\author[0000-0002-7738-6875]{J.~Xavier Prochaska}
  \email{jxp@ucsc.edu}
  \affiliation{Department of Astronomy and Astrophysics, University of California, Santa Cruz, 1156 High Street, Santa Cruz, CA 95060, USA}
  \affiliation{Kavli Institute for the Physics and Mathematics of the Universe (Kavli IPMU), 5-1-5 Kashiwanoha, Kashiwa, 277-8583, Japan}
  \affiliation{Division of Science, National Astronomical Observatory of Japan, 2-21-1 Osawa, Mitaka, Tokyo 181-8588, Japan}
  \author[0000-0001-7694-6650]{Masoud Rafiei-Ravandi}
  \email{masoudrafieiravandi@gmail.com}
    \affiliation{Department of Physics, McGill University, 3600 rue University, Montr\'eal, QC H3A 2T8, Canada}
\author[0000-0001-5799-9714]{Scott M.~Ransom}
  \email{sransom@nrao.edu}
  \affiliation{National Radio Astronomy Observatory, 520 Edgemont Rd, Charlottesville, VA 22903, USA}
 \author[0009-0005-6633-3945]{Gurman Sachdeva}
  \email{gurman.sachdeva@mail.utoronto.ca}
  \affiliation{Dunlap Institute for Astronomy and Astrophysics, 50 St. George Street, University of Toronto, ON M5S 3H4, Canada}
  \affiliation{David A.\ Dunlap Department of Astronomy and Astrophysics, 50 St. George Street, University of Toronto, ON M5S 3H4, Canada}
\author[0000-0002-4623-5329]{Mawson W.~Sammons}
  \email{mawson.sammons@mcgill.ca}
  \affiliation{Department of Physics, McGill University, 3600 rue University, Montr\'eal, QC H3A 2T8, Canada}
  \affiliation{Trottier Space Institute, McGill University, 3550 rue University, Montr\'eal, QC H3A 2A7, Canada}
\author[0000-0003-3154-3676]{Ketan R.~Sand}
  \email{ketan.sand@mail.mcgill.ca}
  \affiliation{Department of Physics, McGill University, 3600 rue University, Montr\'eal, QC H3A 2T8, Canada}
  \affiliation{Trottier Space Institute, McGill University, 3550 rue University, Montr\'eal, QC H3A 2A7, Canada}
\author[0000-0002-7374-7119]{Paul Scholz}
  \email{pscholz@yorku.ca}
  \affiliation{Department of Physics and Astronomy, York University, 4700 Keele Street, Toronto, ON MJ3 1P3, Canada}
  \affiliation{Dunlap Institute for Astronomy and Astrophysics, 50 St. George Street, University of Toronto, ON M5S 3H4, Canada}
\author[0000-0002-4823-1946]{Vishwangi Shah}
  \email{vishwangi.shah@mail.mcgill.ca}
  \affiliation{Trottier Space Institute, McGill University, 3550 rue University, Montr\'eal, QC H3A 2A7, Canada}
  \affiliation{Department of Physics, McGill University, 3600 rue University, Montr\'eal, QC H3A 2T8, Canada}
\author[0000-0002-6823-2073]{Kaitlyn Shin}
  \email{kshin@mit.edu}
  \affiliation{MIT Kavli Institute for Astrophysics and Space Research, Massachusetts Institute of Technology, 77 Massachusetts Ave, Cambridge, MA 02139, USA}
  \affiliation{Department of Physics, Massachusetts Institute of Technology, 77 Massachusetts Ave, Cambridge, MA 02139, USA}
  \author[0000-0003-2631-6217]{Seth R. Siegel}
  \email{ssiegel@perimeterinstitute.ca}
    \affiliation{Perimeter Institute of Theoretical Physics, 31 Caroline Street North, Waterloo, ON N2L 2Y5, Canada}
    \affiliation{Department of Physics, McGill University, 3600 rue University, Montr\'eal, QC H3A 2T8, Canada}
    \affiliation{Trottier Space Institute, McGill University, 3550 rue University, Montr\'eal, QC H3A 2A7, Canada}
\author[0000-0003-3801-1496]{Sunil Simha}
  \email{sunil.simha@northwestern.edu}
  \affiliation{Center for Interdisciplinary Exploration and Research in Astronomy, Northwestern University, 1800 Sherman Avenue, Evanston, IL 60201, USA}
  \affiliation{Department of Astronomy and Astrophysics, University of Chicago, William Eckhardt Research Center, 5640 S Ellis Ave, Chicago, IL 60637, USA}
  \author[0000-0002-2088-3125]{Kendrick Smith}
  \email{kmsmith@perimeterinstitute.ca}	
  \affiliation{Perimeter Institute of Theoretical Physics, 31 Caroline Street North, Waterloo, ON N2L 2Y5, Canada}
\author[0000-0001-9784-8670]{Ingrid Stairs}
  \email{stairs@astro.ubc.ca}
  \affiliation{Department of Physics and Astronomy, University of British Columbia, 6224 Agricultural Road, Vancouver, BC V6T 1Z1 Canada}
  \author[0000-0002-9761-4353]{David C. Stenning}
\email{dstennin@sfu.ca}
\affiliation{Department of Statistics and Actuarial Science, Simon Fraser University, 8888 University Dr W, Burnaby, BC V5A 1S6, Canada}	
  \author[0000-0002-1491-3738]{Haochen Wang}
  \email{hcwang96@mit.edu}
  \affiliation{MIT Kavli Institute for Astrophysics and Space Research, Massachusetts Institute of Technology, 77 Massachusetts Ave, Cambridge, MA 02139, USA}
  \affiliation{Department of Physics, Massachusetts Institute of Technology, 77 Massachusetts Ave, Cambridge, MA 02139, USA}
  \collaboration{1000}{and}

    \author{Thomas Boles}
  \email{tomboles@coddenhamobservatories.org}
  \affiliation{Coddenham Obervatories, Peel House, High Street Coddenham, Suffolk, IP6 9QY, UK}
 \author[0000-0002-1775-9692]{Isma\"{e}l Cognard}
 \email{icognard@cnrs-orleans.fr}
 \affiliation{LPC2E, OSUC, Univ Orleans, CNRS, CNES, Observatoire de Paris, F-45071 Orleans, France and Observatoire Radioastronomique de Nan\c cay, Observatoire de Paris, Universit\'e PSL, Universit\'e d’Orl\'eans, CNRS, 18330 Nan\c cay, France}

  \author[0000-0001-7551-4493]{Tammo Jan Dijkema}
  \email{dijkema@astron.nl}
    \affiliation{ASTRON, Netherlands Institute for Radio Astronomy, Oude Hoogeveensedijk 4, 7991 PD Dwingeloo, The Netherlands}
  \author[0000-0003-3460-0103]{Alexei V. Filippenko}
  \email{afilippenko@berkeley.edu}
 \affiliation{Department of Astronomy, University of California, Berkeley, CA 94720-3411, USA}
\author[0000-0003-4056-4903]{Marcin P.~Gawro\'nski}
  \email{motylek@astro.umk.pl}
  \affiliation{Institute of Astronomy, Faculty of Physics, Astronomy and Informatics, Nicolaus Copernicus University, Grudziadzka 5, PL-87-100 Toru\'n Poland}
 \author[0000-0001-5806-446X]{Wolfgang Herrmann}
 \email{messbetrieb@astropeiler.de}
 \affiliation{Astropeiler Stockert e.V. Astropeiler 1-4, 53903 Bad M\"{u}nstereifel, Germany}
  \author[0000-0002-5740-7747]{Charles D.~Kilpatrick}
  \email{ckilpatrick@northwestern.edu}
  \affiliation{Center for Interdisciplinary Exploration and Research in Astronomy, Northwestern University, 1800 Sherman Avenue, Evanston, IL 60201, USA}
  
 \author[0000-0001-6664-8668]{Franz Kirsten}
 \email{franz.kirsten@chalmers.se}
 \affiliation{Department of Space, Earth and Environment, Chalmers University of Technology, Onsala Space Observatory, 439 92, Onsala, Sweden}
  \affiliation{ASTRON, Netherlands Institute for Radio Astronomy, Oude Hoogeveensedijk 4, 7991 PD Dwingeloo, The Netherlands}
  
  \author[0000-0001-5110-6241]{Shawn Knabel}
  \email{shawnknabel@astro.ucla.edu}
  \affiliation{Physics and Astronomy Department, University of California, Los Angeles, CA, 90095, USA}
  \author[0000-0001-9381-8466]{Omar S. Ould-Boukattine}
  \email{o.s.ouldboukattine@uva.nl}
 \affiliation{ASTRON, Netherlands Institute for Radio Astronomy, Oude Hoogeveensedijk 4, 7991 PD Dwingeloo, The Netherlands}
   \affiliation{Anton Pannekoek Institute for Astronomy, University of Amsterdam, Science Park 904, 1098 XH Amsterdam, The Netherlands}
  \author[0000-0002-2603-6031]{Hadrien Paugnat}
  \email{hpaugnat@astro.ucla.edu}
  \affiliation{Physics and Astronomy Department, University of California, Los Angeles, CA, 90095, USA}
\author[0000-0003-2422-6605]{Weronika Puchalska}
  \email{wpuchalska@doktorant.umk.pl}
  \affiliation{Institute of Astronomy, Faculty of Physics, Astronomy and Informatics, Nicolaus Copernicus University, Grudziadzka 5, PL-87-100 Toru\'n Poland}
\author[0000-0003-1889-0227]{William Sheu}
  \email{wsheu@astro.ucla.edu}
  \affiliation{Physics and Astronomy Department, University of California, Los Angeles, CA, 90095, USA}
  \author[0009-0005-8230-030X]{Aswin Suresh}
  \email{aswinsuresh2029@u.northwestern.edu}
   \affiliation{Department of Physics and Astronomy, Northwestern University, Evanston, IL 60208, USA}
  \affiliation{Center for Interdisciplinary Exploration and Research in Astronomy, Northwestern University, 1800 Sherman Avenue, Evanston, IL 60201, USA}
  \author[0000-0002-2810-8764]{Aaron Tohuvavohu}
  \email{tohuvavo@caltech.edu}
\affiliation{Cahill Center for Astronomy and Astrophysics, California Institute of Technology,
Pasadena, CA 91125, USA}
  \author[0000-0002-8460-0390]{Tommaso Treu}
  \email{tt@astro.ucla.edu}
  \affiliation{Physics and Astronomy Department, University of California, Los Angeles, CA, 90095, USA}
  \author[0000-0002-2636-6508]{WeiKang Zheng}
  \email{weikang@berkeley.edu}
 \affiliation{Department of Astronomy, University of California, Berkeley, CA 94720-3411, USA}

\begin{abstract}
 Precise localizations of a small number of repeating fast radio bursts (FRBs) using very long baseline interferometry (VLBI) have enabled multiwavelength follow-up observations revealing diverse local environments. However, the 2--3\% of FRB sources that are observed to repeat may not be representative of the full population. Here we use the VLBI capabilities of the full CHIME Outriggers array for the first time to localize a nearby (40 Mpc), bright (kJy), and apparently one-off FRB source, \frb, to its environment on 13-pc scales. We use optical and radio observations 
 to place deep constraints on associated transient emission and the properties of its local environment. 
 We place a $5\sigma$ upper limit of $L_{\mathrm{9.9~\mathrm{GHz}}} < 2.1\times10^{25}~\mathrm{erg~s^{-1}~Hz^{-1}}$ on spatially coincident radio emission, a factor of 100 lower than any known compact persistent radio source associated with an FRB. 
 Our KCWI observations allow us to characterize the gas density, metallicity, nature of gas ionization, dust extinction and star-formation rate through emission line fluxes. We leverage the exceptional brightness and proximity of this source to place deep constraints on the repetition of \frb, and find it is inconsistent with all well-studied repeaters given the non-detection of bursts at lower spectral energies. We explore the implications of a measured offset of 190$\pm20$ pc from the center of the nearest star-formation region, in the context of progenitor channels. \frb marks the beginning of an era of routine localizations for one-off FRBs on tens of mas-scales, enabling large-scale studies of their local environments.
\end{abstract}

\section{Introduction}
Fast radio bursts (FRBs) are roughly millisecond-duration flashes of radio waves with Jy-level peak fluxes, originating from cosmological distances \citep{lorimer}. The implied typical isotropic emitting energies of $10^{36}-10^{42}$ erg \citep[e.g.,][]{2013Sci...341...53T,2023ApJ...944..105S,2023Sci...382..294R} prompted many early interpretations to favor cataclysmic origins.
However, the discovery of the first repeating FRB challenged this picture, immediately ruling out cataclysmic theories for that source \citep{2016Natur.531..202S}. 
Since then, we have discovered roughly 4000 unique FRBs of which we have identified $\sim$100 as repeaters (\citetalias{CHIME+2021} \citeyear{CHIME+2021}; \citealt{CHIME+2023}; \citealt{catIIsubmitted}; Cook et al., in prep.). Even still, the answers to fundamental questions such as ``Do all FRBs arise from a single progenitor class?,'' ``Do all FRBs repeat?,'' and ``What are the sources of FRBs?,'' remain elusive.

There is a clear phenomenological difference in apparent one-off FRBs and repeaters: bursts from repeaters are statistically narrower in bandwidth and longer in duration than apparent one-offs\footnote{From this point onwards, we refer to these as simply `one-off FRBs' or `non-repeaters' for conciseness with the understanding that such a claim is inherently observationally limited and hence `apparent' ought to be implied.} \citep{2021ApJ...923....1P, 2024arXiv241102870C}. Compared to one-off FRBs, some of the most active repeaters appear anomalous given their  association with compact persistent radio sources (PRSs; \citealt{Marcote_2017,Niu_2022,Bhandari_2023,bruni2024nebularoriginpersistentradio,Bruni_2025}) and extreme magneto-ionic environments, as evidenced by high and variable Faraday rotation measure (RM) values \citep{2018Natur.553..182M,2023Sci...380..599A}. 
There also are hints of differences in host galaxy populations (repeater hosts may have generally lower stellar masses and optical luminosities; \citealt{Gordon+2023,2024ApJ...971L..51B,2024ApJ...977L...4H,Sharma+2024}) but the significance is currently marginal.  

On the other hand, high-cadence, long-exposure studies of repeaters have shown flattening power-law slopes in burst energy distributions towards higher energies (above $E_\nu \sim 10^{32}$ erg Hz$^{-1}$) which relieves tension between the two phenomena, suggesting that one-off FRB sources may simply be the rarest/highest-energy bursts from repeaters \citep{2024NatAs...8..337K,2024arXiv241017024O}. Detailed population modeling has shown that assuming all FRBs repeat with some reasonable distribution of rates, one can reproduce the fraction of repeaters to one-offs as well as their dispersion measures (DMs), and declinations \citep{2023PASA...40...57J, 2024ApJ...961...10M,2024MNRAS.52711158Y,2021A&A...647A..30G}.  

Environmental studies offer a complementary approach to determine whether one-off FRBs and repeaters originate from distinct progenitor populations. 
 For example, high-resolution studies of the repeating FRBs 20121102A and 20201124A find they are embedded in regions of active star formation \citep{2017ApJ...843L...8B, 2021A&A...656L..15P,2024ApJ...961...44D}, in line with models that attribute the emission to a young source, such as a magnetar.  
This is commensurate with studies of their host galaxy demographics, which find FRBs to primarily arise in star-forming galaxies, and trace star formation or a combination of stellar mass and star formation \citep{Bhandari+2022,Gordon+2023,2025arXiv250215566L,2025arXiv250408038H}. Moreover, a substantial fraction favor locations in or near spiral arms \citep{Mannings+2021,2024ApJ...971L..51B,Gordon2025}.
However, the growing number of FRBs localized to environments without active star-formation suggest that some may be formed through delayed channels. FRB 20180916B, the repeater that exhibits periodic active phases, is located slightly offset ($250 \pm 190$ pc) from the nearest knot of star formation \citep{Tendulkar+2021}. Repeating FRB 20240209A was localized to the outskirts of a quiescent galaxy with a mass-weighted stellar population age of $\sim$ 11 Gyr \citep{Eftekhari+2025, Shah+2025}. 
FRB 20200120E, the nearest known extragalactic FRB ($d_L = 3.6$ Mpc), was localized to a $\sim 9.2$ Gyr-old globular cluster in the halo of M81 \citep{Bhardwaj+2021,Kirsten+2022}.

Notably, FRB 20200120E and FRB 20240114A, the latter being located in a satellite galaxy of a larger galactic system \citep{2025arXiv250611915B}, demonstrate that the properties of the local environment of an FRB do not necessarily correlate with the properties of the most massive nearby galaxy. 

While these local environment studies have revealed great diversity, 
such access to tens of parsec scale localization precision has, hitherto, been limited to repeaters with high enough burst rates to permit detection in follow-up campaigns with very long baseline interferometry (VBLI) arrays \citep[e.g. ][]{2020Natur.577..190M,2022ApJ...927L...3N,2024MNRAS.529.1814H}. Only 2--3\% of FRBs have been observed to repeat, and even fewer are active enough to have been detected by telescopes other than their discovery machine, and those sources may not be representative of the full population \citep{CHIME+2023}.

Recently, construction was completed on three `Outrigger' stations for CHIME/FRB: CHIME-like reflectors oriented to observe the same sky as the core telescope, but located on baselines of 66, 956, and 3370\,km (\citetalias{outriggers_2025} \citeyear{outriggers_2025}, \citealt{lanman_kko}). The full Outrigger array enables sub-arcsecond precision localization for hundreds of FRBs a year, and thus host galaxy associations and, for the nearest sources, deep studies of their local environments.  

On 2025 March 16, shortly after the final Outrigger station began operating,
CHIME/FRB detected its highest-ever signal-to-noise ratio (S/N) extragalactic FRB, \frb\footnote{We colloquially refer to \frb as RBFLOAT, for the Radio Brightest FLash Of All Time. We use “flash” to refer to perceived intensity (S/N), noting that there have been $\sim10$ other FRBs detected in the CHIME/FRB side-lobes of even higher fluence than \frb \citep{CHIME+2020,2024ApJ...975...75L}.}. \frb was exceptional only in its proximity, representing an average FRB luminosity. 
Since the source is nearby and we captured burst data from all four telescope sites, it represents one of the best  opportunities to study the local environment of a one-off FRB.
The full array provided a 68-mas level localization, corresponding to a projected resolution of 13 pc, the finest physical scale achieved to date for an apparently one-off source.

In this Letter, we describe the radio properties of \frb, and
 examine the consistency of the host environment and energetics of this FRB with those of other FRBs, especially the known population of repeating FRBs, leveraging the spatial resolution. In \S\ref{sec:discovery}, we detail the discovery of \frb by CHIME/FRB and its $68 \times 57$ mas localization. We then detail our deep, multi-wavelength follow-up campaign in \S\ref{sec:followup}. We describe the properties of the radio burst in \S\ref{sec:radioburst}, including morphology, polarimetry, and observed two-screen scattering and scintillation. We also constrain the repetition of the source and compute the V/\texorpdfstring{V\textsubscript{max}}{V max} of the burst. In \S\ref{sec:environment} we infer the properties of the FRB's host galaxy (NGC 4141) and local environment. In \S\ref{sec:discussion} we discuss the multi-wavelength implications of \frb, and contrast the properties of the local environment to those of the host galaxy more broadly. 

\section{DISCOVERY OF FRB 20250316A}
\label{sec:discovery}
\subsection{CHIME/FRB}
\label{sec:CHIMEFRB}
The Canadian Hydrogen Intensity Mapping Experiment (CHIME) is a transit radio telescope operating over 400--800 MHz that surveys the northern sky (i.e., above declinations of $-11$ degrees) daily (\citetalias{chimeoverview} \citeyear{chimeoverview}). 
The FRB backend of CHIME/FRB has enabled the capture of $\sim$2--3 FRBs daily, increasing the total number of known FRB sources from a few dozen to 3649 since its first light in 2018 (\citetalias{catIIsubmitted} \citeyear{catIIsubmitted}). The real-time FRB search is described in detail by \cite{chimefrboverview}. 

Upon detection of \frb, CHIME/FRB sent triggers to its full Outrigger array and a VOEvent \citep{2025AJ....169...39A} was automatically published. \frb\ was extremely bright, which led to its detection above the real-time pipeline S/N$> 8$ in a total of 44 of CHIME/FRB's 1024 formed beams. The brightness of this burst caused our realtime pipeline to apply a mask over the signal in the two highest S/N detection beams -- an intentional radio frequency interference (RFI) mitigation technique. The masks, however, caused a non-linear reduction in S/N and we primarily detected the signal from its `spectral ghosts': the dispersed spectral leakage originating from the frequency channelization process. This means that the detection S/Ns in these beams are not representative of the true S/N, and the non-linearity caused the second brightest beam to be reported as the highest S/N. Using the raw voltage data we captured for the burst at its native 2.56~$\mu s$ resolution, having beamformed the data towards the position of \frb (\S\ref{sec:vlbi}), dedispersed the data to the FRB's structure optimizing DM (\S\ref{sec:radioburst}), and summed over all frequencies, the peak S/N is 865.  

CHIME/FRB had 269.77 hours of exposure towards the position of \frb between 2018 August 28 and 2025 May 30 above a fluence sensitivity threshold of 0.5 Jy ms. Details of how we estimate the exposure and sensitivity towards \frb are provided in Appendix~\ref{app:sens_exp}. 
 
\subsection{CHIME Outriggers \& Full-Array VLBI Localization}

\begin{deluxetable}{ll}\label{table:properties}
\caption{\frb Measured Radio Properties. \frb is a two component burst and we report burst width and scattering timescale per component. Parenthetical numbers indicate uncertainties in the least-significant digits.}\label{tab:radioprops}%
\tablehead{
\colhead{Property} & \colhead{Value}
}
\startdata
Right Ascension (ICRS)  & 12$^\text{h}$09$^\text{m}$44$\dotsec$319 \\
Declination (ICRS)    & $+58^\circ50'56.708''$  \\ 
Error Ellipse Semi-major Radius     & $a = 68$\,mas\\
Error Ellipse Semi-minor Radius & $b = 57$\,mas\\
Error Ellipse Position Angle & $\theta = -0.26^{\circ}$ East of North \\
\hline
Time of Arrival (UTC) & 2025-03-16 08:33:50.859038(3)\tablenotemark{a} \\
Dispersion Measure (DM)& $161.82(2)$pc cm$^{-3}$ \\ 
Peak Flux Density & 1.2(1) kJy\\ 
Fluence & $1.7(2) $ kJy\,ms\tablenotemark{b}\\ 
Burst Width (Component 1)& $0.0228(4)\,$ms\\ 
Burst Width (Component 2)& $0.226(3)\,$ms\\
Scattering Timescale 1& $0.0851(6)$ ms\tablenotemark{c} \\
Scattering Timescale 2& $0.405(1)\,$ms\tablenotemark{c} \\
Scattering Frequency Evolution 1& $-4.47(2)$ \\
Scattering Frequency Evolution 2& $-3.08(1)$ \\
Linear Polarization Fraction & $0.9547(2)$\\ 
Rotation Measure (RM)& $+16.79(85)$ rad m$^{-2}$\\ 
\enddata
\tablenotetext{a}{Topocentric at 400 MHz.}
\tablenotetext{b}{Fluence is band-averaged.}
\tablenotetext{c}{Referenced to 600 MHz.}
\end{deluxetable}

\label{sec:vlbi}
\frb marks the first FRB localization achieved using the full CHIME Outrigger Array (\citetalias{outriggers_2025} \citeyear{outriggers_2025}). 
The first Outrigger, \kkoname{} (KKO)\footnote{The name of the first Outrigger \kkoname{} was a generous gift from the Upper Similkameen Indian Band and means ``a listening device for outer space.''}, is located in British Columbia, 66\,km west of CHIME, and provides improved localizations of order an arcsecond in R.A. \citep{lanman_kko}. 
The second CHIME Outrigger is located at the National Radio Astronomy Observatory in Green Bank, West Virginia and together with the CHIME-KKO baseline can provide 50 milliarcsecond by 20 arcsecond localization capabilities (\citealt{curtinsubmitted}; Andrew et al. in prep.). 
The third and final CHIME Outrigger is located in Hat Creek, California, at the Hat Creek Radio Observatory and provides the $uv$ coverage necessary for a full 2D localization of $\sim 50 \times$ 150 mas. The full four-station (three Outriggers and the core CHIME telescope) array is still being commissioned, and its astrometric performance will be fully characterized in an upcoming work. Here we summarize our procedure for localizing this burst as well as a brief justification for the error bars reported. 

We used \texttt{pyfx} \citep{Leung_2024} to obtain cross-correlated visibilities between station beams formed at CHIME and each Outrigger site. VLBI fringes on \frb were clearly detected with cross-correlation S/N exceeding 100 on all CHIME-Outrigger baselines. As described by \cite{ATel17114}, the target visibilities were delay-referenced using the in-beam Radio Fundamental Catalog (RFC) calibrator RFC J1204+5202 located approximately 6 degrees from the FRB \citep{Andrew+2025,Petrov_2025}. 

Two main tests were performed to derive an error budget for this FRB localization. Similar to the test localizations presented by \cite{lanman_kko}, the first was archival localizations of over 200 RFC calibrators and VLBI-localized pulsars with the full Outriggers array to derive a representative error budget on each CHIME-Outrigger baseline as a function of cross-correlation S/N, bandwidth, and calibrator-target separation. For test localizations with similarly high S/N, bandwidth, and calibrator-target separations as \frb, our RMS error budget on the CHIME-HCO and CHIME-GBO baselines were 57 mas and 68 mas, respectively. 
The second test was localizing all detected in-beam calibrators by phase referencing delays to one another so that angular separations on the sky were maximized in our tests. We note that the delay offsets from our bootstrapping method were consistent with those obtained from our archival localizations under similar target-calibrator separations and at lower cross-correlation S/N than observed in this dataset. These calibrations provided the localization ellipse parameters given in Table \ref{tab:radioprops}: Right Ascension (ICRS) = $12^\text{h}09^\text{m}44\dotsec319$, Declination (ICRS) = $+58^\circ50'56.708''$, $a_{\text{err}}$ = 68 milliarcsec, $b_{\text{err}}$ = 57 milliarcsec, $\theta$ = $-$0.26 degrees East of North. 

This localization confirms the findings of \cite{Ng+2025} and \cite{ATel17086} that \frb is robustly associated with the nearby galaxy NGC 4141 in the vicinity of a star forming clump (see~\S\ref{ss:PATH} and \S\ref{sec:clump}). We display the 1-, 2-, and 3-$\sigma$ localization ellipses on top of our MMT optical image (\S\ref{sec:MMT}) in Figure~\ref{fig:mmt_outrigger}. NGC 4141 has a Tully-Fisher distance of 37--44 Mpc \citep{2014MNRAS.444..527S,2016AJ....152...50T}, and redshift of $z\sim0.0067$ \citep{2025ATel17091....1C}. 

\section{Multi-wavelength Follow Up}
\label{sec:followup}
\subsection{Radio}
\subsubsection{Radio Burst Monitoring}
\label{sec:hyperflash}
After the initial detection, \frb was observed at high cadence as part of the HyperFlash project (PI: O.~S.~Ould-Boukattine), which monitors known FRB sources using a network of 25$-$32 meter European radio telescopes with fluence sensitivities better than $\sim 15$\,Jy ms. Observations are carried out on a best-effort basis and, whenever possible, the telescopes observe simultaneously at complementary wavelengths. During this campaign, the position of \frb\, was observed using telescopes at Westerbork (the Netherlands), Dwingeloo (the Netherlands), Stockert (Germany), \torun (Poland), and Onsala (Sweden). A full description of the observing strategy is given by \cite{2024NatAs...8..337K}. 
We also observed with the \nancay Radio Telescope (NRT; France) as part of the Extragalactic Coherent Light from Astrophysical Transients (\'ECLAT, PI: D.~M.~Hewitt) monitoring campaign. These observations used the \nancay Ultimate Pulsar Processing Instrument \citep{2011AIPC.1357..349D} and recorded data at a central observing frequency of 1.484\,GHz, with 512\,MHz of bandwidth. 

\frb was observed between 2025 March 16 (14 hours after the detection) and 2025 May 15 for a total non-overlapping exposure at L-band of 244.06 hours, and 30.94 hours at P band. The first $\sim 120$\,hours of this campaign were summarized in \cite{2025ATel17124....1O}. There were no detections of bursts in this campaign. All telescopes would have been sensitive to bursts above 15 Jy ms, but, e.g., NRT observed for 13.69 hours and detected no bursts above 0.59 Jy ms. 

\subsubsection{Radio Continuum Imaging}
\label{sec:cradio}
We performed continuum radio follow-up observations using the European VLBI Network (EVN), High Sensitivity Array (HSA), Karl G. Jansky Very-Large-Array (VLA) and the upgraded Giant Metre-Wave Radio Telescope (uGMRT) to search for compact, persistent radio emission and characterize the radio environment surrounding FRB\,20250316A. Full details regarding the individual observations and calibration are provided in Appendix \ref{app:crcal}.
\paragraph{EVN \& HSA} EVN observations were carried out on 2025 March 25 from 17:00 UTC to 05:00 (+1) UTC (ID: RL008; PI: M. Lazda). Observations were performed at a central frequency of  $4.86~\mathrm{GHz}$ and data were recorded at a sampling rate of 4096 Mbps (512 MHz of bandwidth, full polarizations, 16 subbands, 32 MHz per subband). The integrated time resolution of the observations was set to 2 seconds per sample. The data were correlated at the Joint Institute for VLBI European Research Infrastructure Consortium (JIVE) using the Super FX Correlator (SFXC; \citealt{2015ExA....39..259K}). \par
HSA observations were carried out on 2025 April 01 from 02:30 UTC to 10:45 UTC with participation of all standard 10 VLBA dishes and the phased VLA in D-configuration (ID: Bl327A; PI: M. Lazda). Observations were performed at a central frequency of $4.87~\mathrm{GHz}$ and data were recorded at a sampling rate of 4096 Mbps (512 MHz of bandwidth, full polarization, four subbands, 128 MHz per subband). The integrated time resolution of the observations was set to 2 seconds per sample. The data were correlated at Socorro using the DiFX correlator \citep{Deller_2007}. Further details on phase calibration for both datasets can be found in Appendix~\ref{app:crcal}. \par

We identified no significant radio emission with either the HSA or EVN within an arcsecond of the quoted localization region in Table \ref{tab:radioprops}, placing a 5-$\sigma$ upper limit on any compact, persistent radio emission of $\leq22~\mu\mathrm{Jy}$ and $\leq46~\mu\mathrm{Jy}$ with the HSA and EVN, respectively. Residual dirty images obtained with both VLBI arrays are provided in Appendix Figure \ref{fig:vlbi_dirtyimages}.  
\paragraph{VLA}
We obtained multi-frequency ($3-27$~GHz) VLA observations of the field of FRB\,20250316A in S-, C-, X-, and K-band (central frequencies of $3.2,6.1,9.9$ and $21.8~\mathrm{GHz}$, respectively) on 2025 April 4 and 2025 May 10 UTC (ID: 25A-434; PI: M. Lazda) while the VLA was in the D-Configuration. We obtained a third epoch at X-band while the VLA was in C-configuration on 2025 May 30. For observations at C-, X-, and K-band, we utilized the 3-bit samplers with the full 4 GHz of bandwidth at C- and X-band and 8 GHz of bandwidth at K-band (excluding the excision of edge channels and RFI). For S-band, we used the standard 8-bit samplers with $\sim 2$ GHz of bandwidth. 

We processed and imaged the data using the standard VLA pipeline (version 6.6.1-17) as part of the Common Astronomy Software Applications (CASA; \citealt{McMullin2007}) software package. We performed bandpass and flux density calibration using NVSS J133108+303032 and complex gain calibration with NVSS J115324+493109, NVSS J121710+583526, and ICRF J121906.4+482956. We do not detect radio emission at the FRB location in the X- and K-band images, with 5-$\sigma$ rms values of $11$ and $34.5$ $\mu$Jy, respectively. At S- and C-band, the lower angular resolution of the array in D-configuration is ill suited for probing compact radio emission associated with the FRB position: we instead detect extended radio emission coincident with the host galaxy, which we use to characterize the radio star formation rate in \S\ref{ss:sfr_radio}. A subset of the radio images is provided in Figure  \ref{fig:2x2grid}.

\paragraph{uGMRT} We conducted uGMRT radio observations in Band~4 ($550-750$~MHz) and Band~5 ($1260-1460$~MHz) to search for any continuum radio emission associated with \frb (ID: ddtC-430; PI: Y. Bhusare). A total of eight hours were observed in Band~5 and three hours in Band~4. NVSS J133108+303032 was used as the flux density calibrator for both bands, while NVSS J131337+545824 and NVSS J120624+641337 were used as the complex gain calibrators for Band~5 and Band~4, respectively. The data were processed and imaged using the CAPTURE pipeline~\citep{Kale_2020}. We do not detect any continuum source at the target position above a 5-$\sigma$ significance threshold of 100 $\mu$Jy in Band~4 and 75 $\mu$Jy in Band~5. However, we detect some emission associated with the host galaxy in both bands. \par

\vspace{2pt}
\noindent A summary of all continuum radio observations is provided in Table \ref{tab:radiofluxes}. 

\subsection{Optical}
\label{sec:opt}

\begin{figure*}[t]
    \centering
    \includegraphics[width=\textwidth]{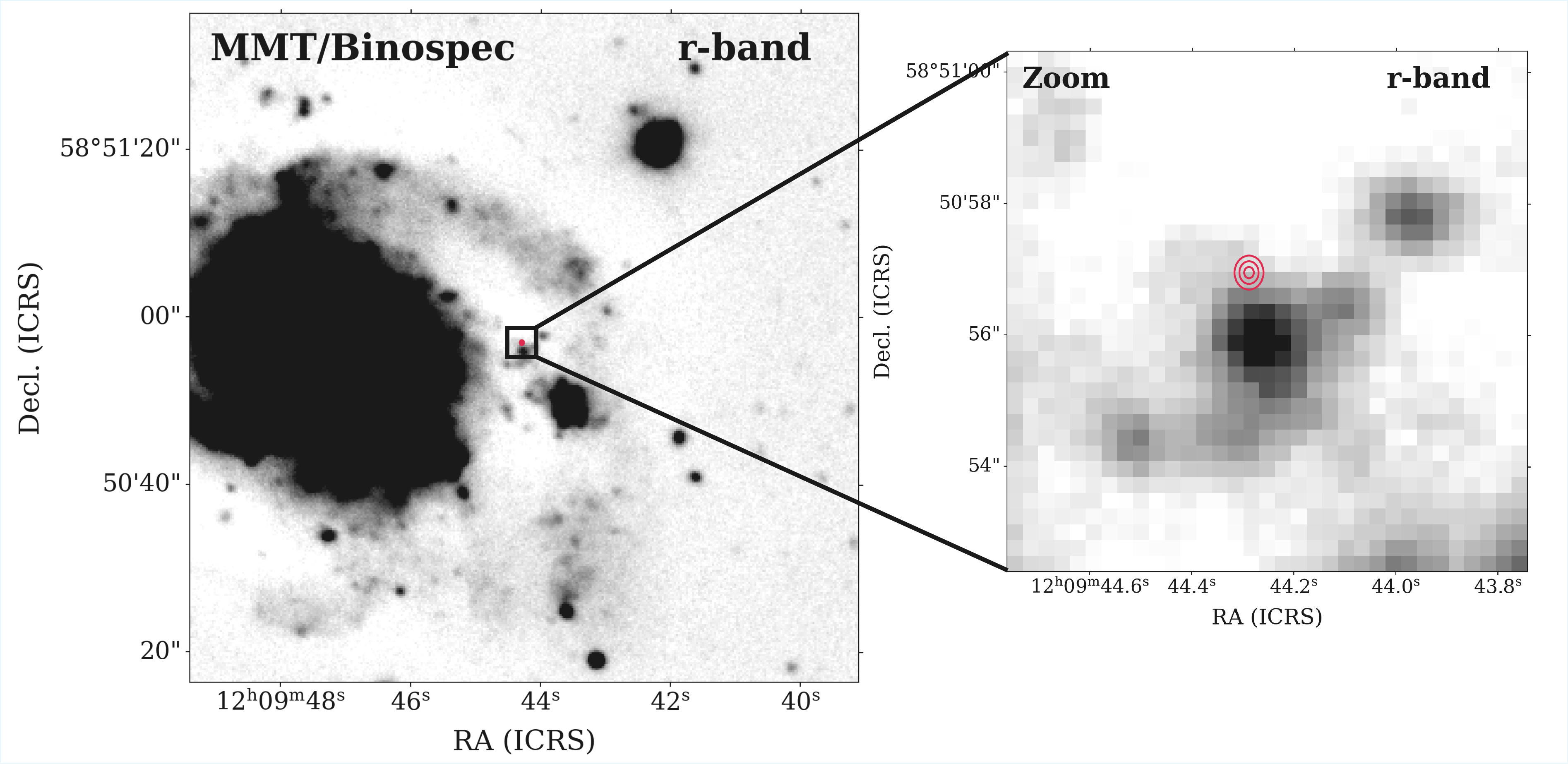}
    \caption{\textit{Left:} Linearly scaled MMT r-band image showing part of the host galaxy, NGC\,4141, of \frb~with the CHIME Outriggers VLBI localization overlaid in red; the ellipse incorporates both positional and absolute astrometric tie uncertainties. The 3-$\sigma$ limiting magnitude of the image is r $\gtrsim$ 25.6~mag (AB, corrected for Galactic extinction). The image is oriented North up and East left. \textit{Right:} Image zoomed in at the Outrigger position, 8\arcsec $\times$ 8\arcsec, and additionally showing the 1-, 2- and 3-$\sigma$ localization ellipses of FRB\,20250316A.}
    \label{fig:mmt_outrigger}
\end{figure*}

\subsubsection{MMT}
\label{sec:MMT}
We obtained $r$-band imaging of the field of \frb\, on 2025 March 23 (PI: J.~Rastinejad; ID: UAO-G169-25A) with the Binospec optical spectrograph mounted on the 6.5-m MMT telescope atop Mt. Hopkins, Arizona for a total exposure time of 0.98 hr at $\sim$8.23~days after the FRB detection (first reported by \citealt{ATel17112}). We reduced the images using the \texttt{POTPyRI} software package\footnote{\url{https://github.com/CIERA-Transients/POTPyRI}}. \texttt{POTPyRI} applies flat-field and bias calibration, performs absolute and relative alignment, and co-adds the science frames into a single stack (Figure~\ref{fig:mmt_outrigger}). We performed astrometric alignment relative to sources in common with the Gaia Data Release (DR) 3 catalog~\citep{GaiaDR3}; the resulting solution has astrometric uncertainties of $\Delta$RA=$0.053\arcsec$ and $\Delta$Dec=$0.046\arcsec$, which are added in quadrature to the localization ellipse parameters. With the photometric zeropoint calibrated to PanSTARRS \citep{Flewelling+2020}, there is no clear source at the Outrigger position to a 3-$\sigma$ limit of $r \gtrsim 25.6$~mag (here and after, all quoted magnitudes are in the AB system and are corrected for Galactic extinction in the direction of the FRB). We note that the FRB position is slightly to the North of an extended clump of emission (Figure~\ref{fig:mmt_outrigger}), with a total magnitude of $r \approx 21.8$~mag ($M_r \approx -11.2$~mag at the adopted distance of NGC 4141, $d_L \sim 40$~Mpc).

To assess any variability at or near the FRB position, we acquired a second set of MMT $r$-band observations on UT 2025 March 24 (PI: Y.~Dong; ID: UAO-G159-25A), 1.26~days after our initial observation for a total exposure of 0.98~hr. Performing digital image subtraction with \texttt{HOTPANTS} \citep{HOTPANTS}, we do not detect any residual emission to a 3-$\sigma$ limit of $r \gtrsim$ 25.0~mag, or $M_r \approx -8.0$~mag. We note that we can only comment on variability between these two epochs, and are not sensitive to longer-timescale variability.

\subsubsection{Gemini} \label{sec:gemini}

\paragraph{Imaging} We imaged the field of \frb\, with the Gemini Multi-Object Spectrograph (GMOS) mounted on the 8-m Gemini-North telescope on UT 2025 March 24 (PI: T. Eftekhari; ID: GN-2025A-LP-110) for a total exposure of 0.5 hr in the $g$-band. As in the MMT imaging, we used \texttt{POTPyRI} to apply bias and flat-field corrections, and to co-add the images. We carried out astrometric calibration using sources in common with the Gaia DR3 catalog, achieving a 1-$\sigma$ astrometric tie uncertainty of $\sim0.08\arcsec$. We do not detect any optical emission coincident with the FRB localization, to a 3-$\sigma$ point-source limiting magnitude of $g \gtrsim 23.7$~mag, as reported by \cite*{ATel17116}.

\paragraph{Spectroscopy}
We obtained $8 \times 900$~sec of long-slit optical spectroscopy of the localization region of \frb\ with GMOS on 2025 March 25 (PI: T. Eftekhari; ID: GN-2025A-LP-110). Observations were conducted using a 1$\arcsec$ slit width, the B480 grating, and the GG455 blocking filter at central wavelengths of 640 and 650 nm. The resulting spectra covered a wavelength range of \SIrange[range-phrase={\textendash}]{4480}{8560}{}\,\AA. We oriented the slit at a position angle of 10.6$^\circ$ East of North to cover both the initial FRB Outrigger position (as reported by \citealt{ATel17114}) and the nearby clump of optical emission directly to the South of the FRB Outrigger position. We reduced and co-added the spectra using the \texttt{PypeIt} software package\footnote{\label{pypeit}\href{https://github.com/pypeit/PypeIt}{https://github.com/pypeit/PypeIt}\\ (version 1.17; \texttt{kcwi\_dec\_2024} branch)} \citep{pypeit}. All frames were subject to basic image processing: bias subtraction, astrometric alignment, wavelength calibration, and flat-fielding to normalize pixel-to-pixel variation. We applied absolute flux calibration using spectrophotometric standard spectra. To avoid the masking of bright emission lines, the automatic cosmic-ray detection and masking was turned off. 

We do not detect any spectral features at the FRB position, but observe strong nebular emission (H$\beta$, $H\alpha$, and [OIII]) from the location of the nearby extended r-band emission detected in our MMT imaging with a common redshift of $z = 0.0065$. This is consistent with that emission originating from star formation within NGC 4141.

\subsubsection{KCWI}
\label{sec:kcwi}
On 2025 March 22, six days after the FRB event, we observed the location of the FRB with the Keck Cosmic Web Imager (KCWI) integral field spectrograph (IFS; PI T. Treu; ProgID U036). We used the ``Small'' slicer, which covers a field of view of $8^{\prime\prime}\times 20^{\prime\prime}$ with the highest spatial resolution offered by the instrument ($0.34^{\prime\prime}$ per pixel), and the BL and RL gratings on the blue and red arms ($R\approx 3600$ and $R>2000$, respectively). We obtained a total of 2060~s on the blue arm ($2\times1030$ s) and 1800~s on the red arm ($6\times300$ s) on-source without dithering. We also obtained off-source exposures of the same durations on a blank patch of sky offset by $30\arcsec$ from our initial pointing.
To calibrate our exposures, we obtained the standard bias frames, ThAr comparison lamps, internal flatfields, dome flatfields, sky flatfields, alignment frames, and standard-star exposures \citep[CALSpec standard BD+54 1216; ][]{Bohlin+2025}.

The data were also reduced using the PypeIt reduction package. As mentioned in \S\ref{sec:gemini} for our Gemini spectroscopic data, our KCWI frames were first subject to basic image-processing steps. However, as recommended by PypeIt documentation, in anticipation of co-adding the individual exposures, spectral illumination correction was turned off. Then, one of the sky flats (dome flats) for the blue side (red side) was subsequently reduced as a science frame to provide relative scale correction in the co-addition step. The resulting 2D science spectra were co-added with the \texttt{pypeit\_coadd\_datacube} script to produce spectral data cubes. As the field of view was entirely within the disk of the host galaxy on the sky, each science frame was paired with one of the blank sky exposures to better subtract the sky emission. During co-addition, the standard-star data were also employed for flux calibration.

After reduction and fluxing, we corrected the astrometric solution of the red and blue cubes. Since no obvious point source was identified on the cube, we attempted to align the centroids of the three line-emitting clumps in Figure \ref{fig:kcwi_mosaic} with the corresponding clumps in our deep MMT $r$-band image (shown in the right panel of Figure \ref{fig:mmt_outrigger}). To this end, narrow-band maps of \halpha\ and the [O~II] $\lambda\lambda3726$, 3729 doublet were generated and the sources were characterized through \texttt{Photutils} image segmentation. Simultaneously, sources were identified in the MMT image. Offsets between the source centroids in the KCWI and MMT images were computed and the KCWI world coordinate system (WCS) solution was translated by the median offset to the correct the cube WCS. No rotation correction was applied as both the data cubes and the MMT images have North and East directions aligned with the image axes from their respective reduction packages. Also, no higher-order WCS corrections were applied on account of the sparsity of sources to accurately derive these terms. The uncertainty in this astrometric correction was estimated with the standard deviation of the individual offsets to the three clumps: $0.1^{\prime\prime}$ for both the blue- and red-arm cubes. Our analyses with the KCWI data (i.e., as described in \S\ref{sec:clump}) incorporate this astrometric uncertainty by adding it in quadrature with the FRB localization uncertainty and the MMT-Gaia astrometric tie-in uncertainties.

Preliminary examination of the cubes revealed strong emission features, including the [OI~I] 3726, 3729 \AA\ doublet and \halpha, from clumps in the host-galaxy interstellar medium (ISM). The top panels of Figure~\ref{fig:kcwi_mosaic} show the clumps in close proximity to the FRB location. We compared these relatively compact features to our MMT imaging to calibrate the cube astrometry since the KCWI data do not cover any known star or point source in the field of view. Similar to the MMT imaging, we do not observe any obvious point source at the location of the FRB itself. Further analysis of the KCWI data, including 1D extraction in and around the FRB localization, is presented in \S\ref{sec:clump}.

\subsection{X-ray}
\label{sec:xray}
Upon receiving \frb's VOEvent from CHIME/FRB, the \textit{Neil Gehrels Swift Observatory} repointed to the position reported in the VOEvent\footnote{This was achieved through the `continuous commanding' capability that can achieve 10 second latency response time on-orbit to unscheduled ToO requested received on the ground \citep{2024ApJ...975L..19T}.}. However, \frb\ did not end up in the field of view of Swift/XRT as the initial VOEvent position was incorrect\footnote{The VOEvent position was reported based on an incorrect assignment of the beam with the highest S/N detection (see \S\ref{sec:CHIMEFRB}).}. The true position was outside of the reported 1-$\sigma$ uncertainty region, and hence was not in the FOV of Swift/XRT. We repointed Swift/XRT via ToO request once we had a refined baseband position for the burst, and began observing on 2025 March 16 at 21:10 UTC, 12.6 hours after the ToA of the burst (PI: A. Tohuvavohu) and observed the position for 6.2\,ks. An additional Swift ToO observation (PI: E. Troja) of 10.3 ks took place about a week after the event. Preliminary analyses of the Swift observations (target ID 19620) have been reported elsewhere \citep{2025ATel17101....1Y,2025ATel17109....1T}, but the observations have been re-analyzed with the updated position in this work. These led to 0.5--10.0~keV upper limits on the unabsorbed flux of $7 \times 10^{-14}$ and $1\times 10^{-13}$ erg\,cm$^{-2}$\,s$^{-1}$ for the two observations, respectively, assuming an isotropic emission with $\Gamma = 2 $ power-law spectrum, absorbed by a $N_H = 1.4\times10^{20}$ cm$^{-2}$ neutral hydrogen column. This neutral hydrogen column corresponds to the Galactic expectation \citep{2016A&A...594A.116H} and assumes negligible X-ray absorption local to the source. 

Follow-up observations by the Einstein Probe (EP) showed evidence of X-ray variability at the position consistent with \frb\ between 2025 March 18 and 21 \citep{ATel17100}. The subsequent follow-up Chandra/High Resolution Camera observation (PI: H. Sun), with its sub-arcsecond spatial resolution capability, combined with the updated sub-arcsecond localization from CHIME/FRB Outriggers \citep{ATel17114}, ruled out a physical association between the EP-detected variable X-ray source and \frb\ \citep{SunChandra}. Follow-up ToO observations (PI: A.~B.~Pearlman) with the Neutron star Interior Composition Explorer (NICER) between 2025 March 18 and 21 (ObsIDs 861701010[1-4]; $\sim6.7$~ks) yielded 3-$\sigma$ persistent X-ray absorbed 0.5--10.0~keV flux upper limits between \text{(0.07--3)}$\times$\,10$^{\text{--12}}$\,${\rm\,erg\,cm^{-2}\,s^{-1}}$ with similar spectral parameters assumed above~\citep{Pearlman+2025_ATel17117}.

\section{Radio Burst Analysis}
\label{sec:radioburst}
\subsection{Morphology}
\begin{figure}
    \centering
    \includegraphics[width=0.95\linewidth]{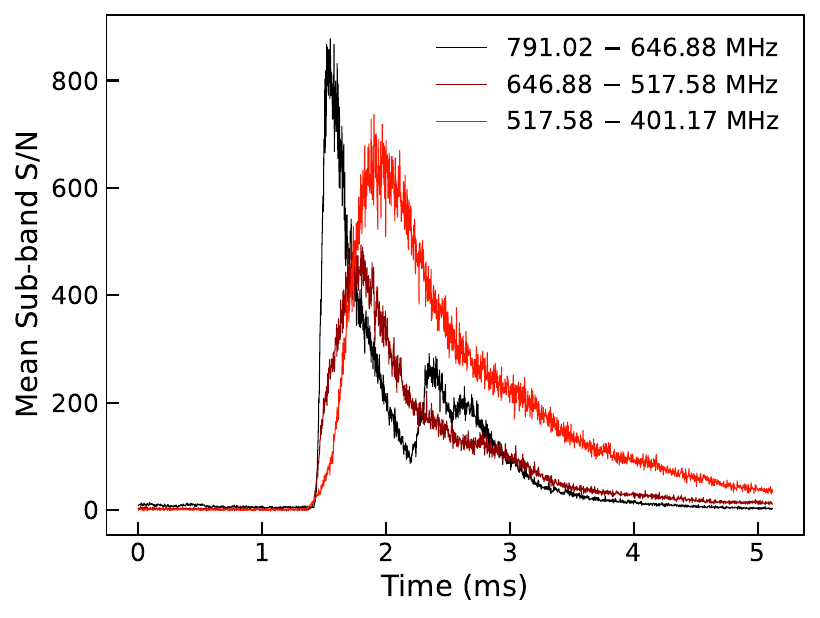}
    \caption{S/N of \frb averaged over three frequency sub-bands and at the $2.56\,\mu$s time resolution native to the raw voltage data.}
    \label{fig:risetime}
\end{figure}

\frb exhibits a scattered two component structure at the top of the observed band, transitioning to a single scattered component at lower frequencies. Using \texttt{DM\_phase} \citep{2019ascl.soft10004S} we find an incoherent structure optimizing DM of $161.82\pm0.02$ \dmunits. As shown in  
Figure \ref{fig:risetime}, the burst peak displays substantial drift in time as a function of frequency, consistent with a frequency dependent rise time expected for multiple scattering screens as seen by \cite{Kirsten+2022}.
A pulse broadening function associated with multiple screens is not a model currently available within the burst modeling software typically used for CHIME/FRB data, \texttt{fitburst} \citep{2024ApJS..271...49F}. To capture this behavior we define a new burst model from two Gaussian components convolved with two thin screen scattering kernels and fit it to the profile using the Bayesian Inference Library \citep[\texttt{Bilby}; ][]{bilby_paper}. The intrinsic components are centered at $t_1$ and $t_2$ and characterized by 1-$\sigma$ widths $w_1$ and $w_2$. The scattering kernels are each the canonical one--sided exponential decay with $1/e$ widths $\tau_1$ and $\tau_2$ referenced to $600\,$MHz that scale with frequency following $\alpha_1$ and $\alpha_2$. Further fit details are discussed in Appendix \ref{app:morphfit}, and best-fit parameters are listed in Table~\ref{tab:radioprops}.
\begin{figure}
    \centering
    \includegraphics[width=0.8\linewidth]{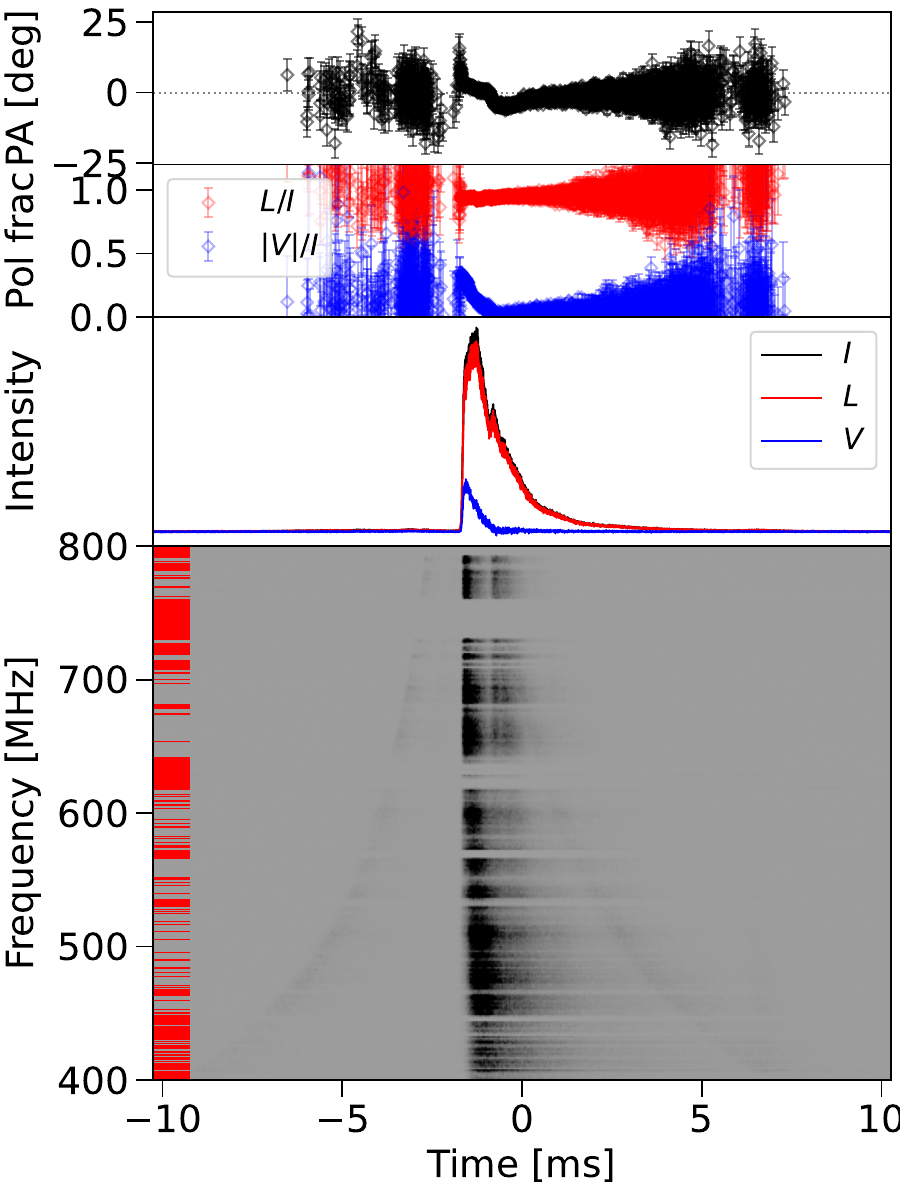}
    \caption{Bottom: total intensity waterfall de-dispersed to the structure optimized DM of $161.82 \pm 0.02~\mathrm{pc}~\mathrm{cm}^{-3}$ and temporal profiles of Stokes $I$ (black line), $L = \sqrt{Q^2 + U^2}$ (red line), $V$ (blue line), linear polarization fraction ($L/I$; red circles), circular polarization fraction ($|V|/I$; blue circles), and PA (black circles). Top: all points plotted on the $L/I$, $|V|/I$, and PA profiles exceed a linearly polarized S/N of 5. Channels masked out due to RFI are highlighted by red streaks on the left-hand side of the total intensity waterfall. The preceding and subsequent parabolic sweeps are artifacts due to `spectral ghosts' related to the instrumental signal chain.}
    \label{fig:pol_summary}
\end{figure}

\subsection{Polarimetry}
\begin{figure}
    \centering
    \includegraphics[width=0.8\linewidth]{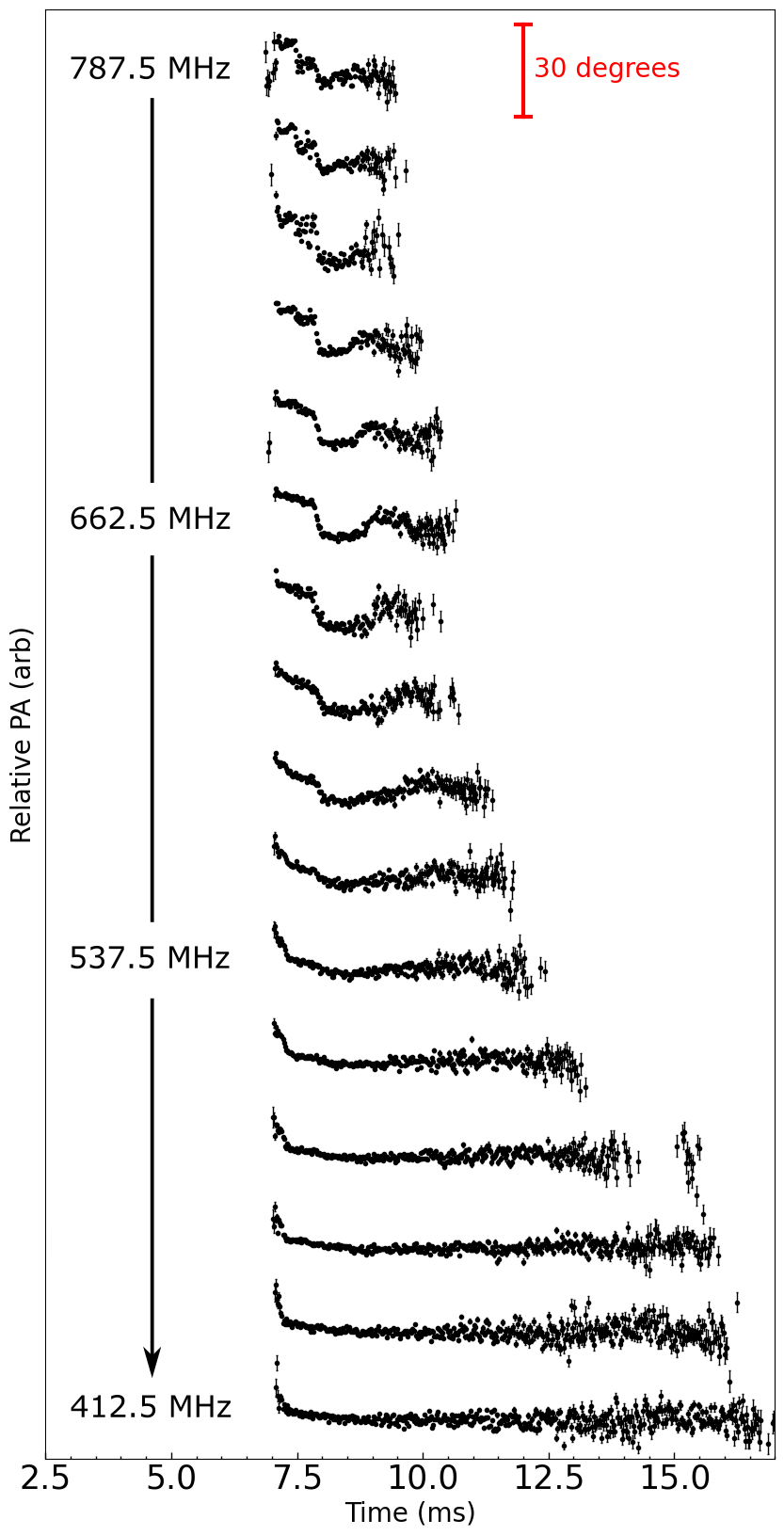}
    \caption{PA profile of FRB~20250316A as a function of frequency over $400-800$~MHz. The y-axis scale is arbitrary and a red marker is used to indicate PA variation on a $30\degree$ scale. The central frequencies of some frequency subbands are indicated on the left as a guide.}
    \label{fig:pa_freq}
\end{figure}

We applied the CHIME/FRB polarization pipeline \citep[for details, see][]{2021ApJ...920..138M} to the beamformed baseband data for FRB~20250316A to study its polarization properties. The total intensity waterfall plot, linear and circular polarization fractions, and polarization angle (PA) profile are shown in Figure~\ref{fig:pol_summary}. We measure an average linear polarization fraction $L/I = \sqrt{Q^2 + U^2}/I = 0.9547 \pm 0.0002$ across the burst. Note here that the small uncertainty is derived only from the Stokes $Q$ and $U$ spectra and it does not account for leakage into Stokes $V$, which may add additional uncertainty on the $L/I$ at the $\sim$~percent level. No circular polarization is detected above the expected instrumental polarization level. The linear polarization fraction remains constant across $400-800$~MHz and we find no evidence for beam depolarization in the signal. The frequency-averaged PA profile displays a complex morphology with a maximum variation of $\sim 20-30$~degrees across the burst profile. Given the high S/N of this burst, we split the burst into 16 frequency bands and measure the PA profile as a function of decreasing frequency, as shown in in Figure~\ref{fig:pa_freq}. We see that the PA profile has larger variations at the top of the CHIME/FRB band and flattens near the bottom as it becomes increasingly scatter broadened. Even at the lower frequencies measured by CHIME/FRB, the leading edge of the burst retains much of the PA variability before it becomes flat in the scattering tail. This PA profile flattening with frequency is a known effect in pulsars \citep[e.g.,][]{2003A&A...410..253L, 2023RAA....23j4002W}, however, this is the first reported instance for FRBs. Physically, this flattening is thought to arise from the averaging of PAs in a given time bin within the scattering tail, where signal from different pulse phases (and hence with different PAs) have become mixed.

To measure the Faraday rotation measure (RM), we employ the RM-synthesis algorithm \citep{burn_1966, brentjens_deBruyn_2005_RM_synth}, as implemented by \cite{2021ApJ...920..138M}, and search for peaks in the Faraday dispersion function exceeding a linearly polarized S/N of six. We observe an RM of $\mathrm{RM}_\mathrm{obs} = 16.7853~\mathrm{rad}~\mathrm{m}^{-2}$ with a statistical error of $0.0003~\mathrm{rad}~\mathrm{m}^{-2}$. In this case, the systematic error in RM measurements with CHIME/FRB \citep[$0.85~\mathrm{rad}~\mathrm{m}^{-2}$, as measured using multi-epoch, polarimetric monitoring of Crab giant pulses;][]{2023ApJ...950...12M} dominates the statistical error, so we quote an observed RM of $\mathrm{RM}_\mathrm{obs} = 16.79 \pm 0.85~\mathrm{rad}~\mathrm{m}^{-2}$. The Galactic RM contribution towards FRB~20250316A is $\mathrm{RM}_\mathrm{gal} = 17 \pm 4~\mathrm{rad}~\mathrm{m}^{-2}$ \citep{2024A&A...690A.314H}. Assuming negligible RM contributions from the intergalactic medium and Earth's ionosphere, and converting to the rest frame of the source, we constrain the host galaxy contribution to the RM to be $\mathrm{RM}_\mathrm{host} = 0 \pm 4~\mathrm{rad}~\mathrm{m}^{-2}$. However, the Galactic RM map by \cite{2024A&A...690A.314H} does not constrain RM variations on scales smaller than $\sim 1$ square degree, which can change the $\mathrm{RM}_\mathrm{gal}$ estimate by up to $\sim 40~\mathrm{rad}~\mathrm{m}^{-2}$ even away from the Galactic plane \citep{2025ApJ...982..146P}. This effectively places larger (unknown) uncertainties on our $\mathrm{RM}_\mathrm{host}$ estimate and we cannot conclusively determine whether FRB~20250316A originates from a weak magneto-ionic environment (i.e., one with $|\mathrm{RM}_\mathrm{host}| \sim 0~\mathrm{rad}~\mathrm{m}^{-2}$). Using our estimates of $\mathrm{DM}_\mathrm{host} = 60 \textcolor{red}~\mathrm{pc}~\mathrm{cm}^{-3}$ (assuming a conservative uncertainty of $20~\mathrm{pc}~\mathrm{cm}^{-3}$; see \S\ref{sec:ism}) and $\mathrm{RM}_\mathrm{host} = 0 \pm 4~\mathrm{rad}~\mathrm{m}^{-2}$, we calculate $\left<B_{\parallel,\mathrm{host}}\right> = 4 \pm 100~\mathrm{nG}$, noting again that our uncertainties may be underestimated.

Compared to the average non-repeating FRB population \citep{2024ApJ...968...50P}, FRB~20250316A stands out in a number of ways: (i) it is more highly linearly polarized than average, (ii) it undergoes a larger PA swing than most non-repeating FRBs, and (iii) it has the lowest RM contribution from its host galaxy (consistent with $0~\mathrm{rad}~\mathrm{m}^{-2}$) of any polarized non-repeater currently observed by CHIME/FRB.

\subsection{Scintillation}
\label{sec:scint}

\begin{figure}
    \centering
    \includegraphics[width=\linewidth]{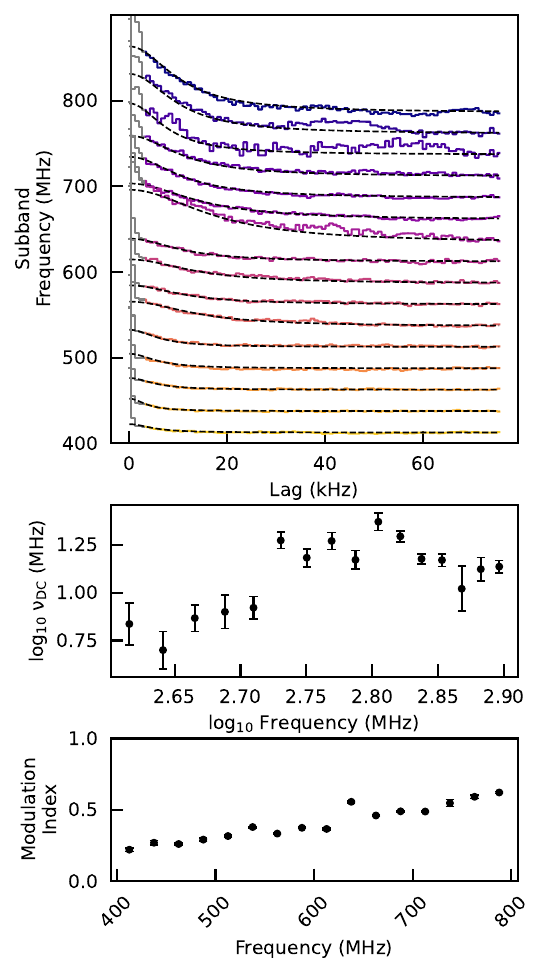}
    \caption{\textit{Top:} Autocorrelations of equal bandwidth subband spectra and corresponding Lorentzian fits to observed scintillation, gray regions denote data associated with distinct spectral structures on a separate scale which are excluded from the scintillation fit discussed in this work. \textit{Middle:} Decorrelation bandwidth ($\nu_{\text{DC}}$) as a function of frequency. \textit{Bottom:} Modulation index associated with the scintillation decorrelation bandwidth seen in the top panel.}
    \label{fig:scintResults}
\end{figure}

We perform a cursory scintillation analysis using the methodology outlined by \cite{Nimmo+2025}. We find convincing evidence for at least one scale of scintillation, with a decorrelation bandwidth of approximately $\nu_{\text{DC}}=14\pm4$ kHz at 600 MHz, as shown in Figure \ref{fig:scintResults}. We observed the scintillation to have a low modulation index consistent with interference effects from the interplay between two scattering surfaces \citep{Nimmo+2025, t2025scintillometryfastradiobursts}, in broad agreement with the two scattering screens inferred from the scattering tail and scattering-induced rise time in the burst morphology. We highlight that the observed $\nu_{\text{DC}}$ appear to flatten toward higher frequencies, contrary to the expectation of \cite{t2025scintillometryfastradiobursts} for evolution to steepen at higher frequencies. The extent to which this flattening deviates from theory, the impact of instrumental artifacts on the spectrum, a treatment of finer-scale spectral structures, and tests of consistency between spectral and temporal scattering effects will be considered more exhaustively in a future work.

As in previous studies \citep{sammons_two_screen_2023, Nimmo+2025} the scintillation and scattering times for separate screens may be used to constrain their geometries. \cite{t2025scintillometryfastradiobursts} show that as stronger scattering increases the size of the illuminated screens ($L_1$ and $L_2$) relative to the distance between them ($D_{1,2}$), the modulation index of the relatively broad scale of scintillation decreases following 
\begin{equation}
    m = \frac{1}{\sqrt{1+\frac{\pi^2}{4^3}\left(\frac{L_1L_2}{\lambda D_{1,2}}\right)^2}},
\end{equation}
where $\lambda$ is the wavelength. Commensurate with this change in modulation index, the relatively broad scale of scintillation associated with a screen at $D_1$ is expected to widen following
\begin{equation}
    \nu_{\text{DC}} = \frac{cD_{1}}{\pi L_1^2}\sqrt{1+\frac{\pi^2}{4^3}\left(\frac{L_1L_2}{\lambda D_{1,2}}\right)^2}.
\end{equation}
Given our modulation index of $m=0.4$, we can use the observed scattering times to constrain the minimum distance between these two screens for some reasonable assumption of their effective distances ($D_s D_{sf}/D_f$, where each are angular diameter distances to screen, between screen, and the FRB and to the FRB, respectively). We assume one screen is located at the peak of turbulence strength along this line of site at a distance $D_{MW}=0.25\,$kpc inferred from the \texttt{ne2001} profile \citep{ne2001,KochOcker_2022}, where this source is at a high Galactic latitude of $b=57.46\degree$. As a result, the distance between the screens must be greater than 1\,Mpc for all second screens at effective distances greater than 3\,pc. We therefore conclude that one of the screens must be extragalactic.

The product of separations of extragalactic and Galactic screens from host and observer, respectively, may then be constrained, following the updated method outlined by \cite{t2025scintillometryfastradiobursts}, using the scintillation and the larger of the two scattering times. We constrain the product of the angular diameter distances between observer and Galactic scattering screen ($D_g$) and source and host galaxy scattering screen ($D_{x,s}$) to be $D_gD_{x,s}\lesssim0.015(4)\,$kpc$^2$. We describe this constraint conservatively here as a limit, but highlight that in reality it should be an equality provided the systematics are well understood. This is orders of magnitude more constraining than typical \citep{sammons_two_screen_2023}, due primarily to the proximity of this source. Assuming again that our Galactic scattering screen lies at the peak of turbulence strength for this sight-line of $250\,$pc, the extragalactic scattering screen must be within approximately $60\pm20$\,pc of the source, the tightest constraint yet.

\subsection{Limits on FRB Repetition Rate}
\label{sec:rate}
CHIME/FRB's 269.77 hours of exposure towards \frb\, is distributed in 8--11 minute\footnote{The daily transit time has steadily increased as the precession of the Earth moves the source transit more centrally through its corresponding row of formed beams.} intervals over our six years of operation. CHIME/FRB has not detected any other candidate bursts with S/N $\geq$ 8.5 and a position and DM consistent with \frb\, (within 10 pc cm$^{-3}$ and an angular separation of one degree). CHIME/FRB has some sensitivity and much more exposure to very bright bursts in our side-lobes, where we have detected at least 10 FRBs with \frb-level fluences many degrees off meridian \citep{2020Natur.582..351C,2024ApJ...975...75L}. Thus, we also search positions that are consistent with the time-dependent apparent position of \frb\ as it transits through the side-lobes up to $60\degree$ from the telescope's meridian, but find no convincing burst candidates. 

The HyperFlash observations (\S\ref{sec:hyperflash}, the timeline of which is visualized in Figure \ref{fig:hyperflash}) targeted the source shortly after discovery and began with a particularly constraining 33-hour L-band exposure just 14 hours after the initial detection of \frb. No bursts were detected. Repeater bursts are known to cluster in time, with waiting distributions well-described by a Weibull distribution \citep{2018MNRAS.475.5109O}, but see also \cite{2023ApJ...944...70G}. The Weibull distribution describes the intervals between events, and it is a generalization of the Poisson distribution described by two parameters. In this analysis, we adopt the parameterization of \cite{2018MNRAS.475.5109O} such that the first Weibull parameter, $r \in \mathbb{R}^+$, can still be interpreted as a rate and the additional shape parameter $k \in \mathbb{R}^+$ describes the extent of clustering. When $k=1$, the Weibull distribution reduces to the Poissonian case, and when $k<1$, small intervals are favored compared to the Poissonian case.

\begin{figure}
    \centering
    \includegraphics[width=\linewidth]{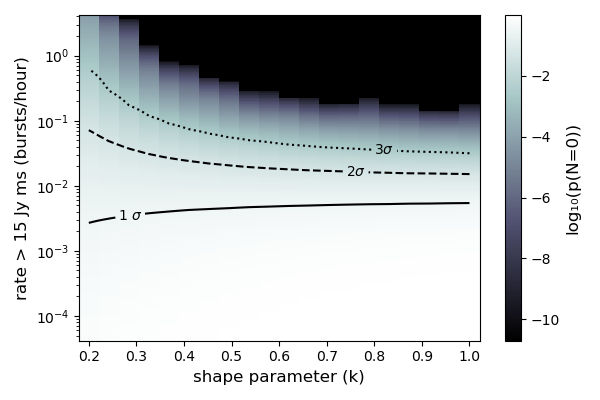}
    \caption{Probability of detecting no bursts in the HyperFlash observations (color map) as a function of assumed shape parameter (x-axis) and burst rate above 15 Jy ms (y-axis). Contours corresponding to the 1-, 2-, and 3-$\sigma$ confidence upper limits are superimposed.}
    \label{fig:weibull_const}
\end{figure}

Using the HyperFlash L-band observations, in Figure~\ref{fig:weibull_const} we show the region of parameter space that can be ruled out at different confidence levels given our observations. This is done via simulation: for different values of $k$ and $r$, we randomly sample waiting times from the corresponding Weibull distribution, and check how frequently zero bursts occur during any of our observing windows. Assuming Poisson statistics ($k=1$), we can rule out rates greater than 2.4$\times10^{-2}$ bursts per hour above the sensitivity threshold, 15 Jy ms, at the 99.7\% level. 

When additionally considering CHIME/FRB's 269.77 hours of exposure over the six years prior to the discovery of \frb, Poisson rates above 1.5$\times10^{-2}$ bursts per hour are ruled out at the 99.7\% confidence level above 15 Jy ms. These rate upper limits are only valid under the assumption that the source was in an active state with constant rate for the entirety of the observations, which we caution is a much more reasonable assumption for the HyperFlash observations that took place in the weeks following our detection than it is for the six previous years of CHIME/FRB observations. %$ given observed repeater `lifetimes'. 

\subsection{Burst Energy Distribution Constraints}

The more puzzling aspect of \frb, if it were to be a repeater, is its high S/N, which is a reflection of its relative proximity. Given its measured fluence of 1.7(2) kJy ms and NGC 4141's $\sim$ 40 Mpc distance, we derive a burst spectral energy of of $(3.1 \pm 0.7)\times10^{30}$ erg Hz$^{-1}$. 

\begin{figure}
    \centering
    \includegraphics[width=\linewidth]{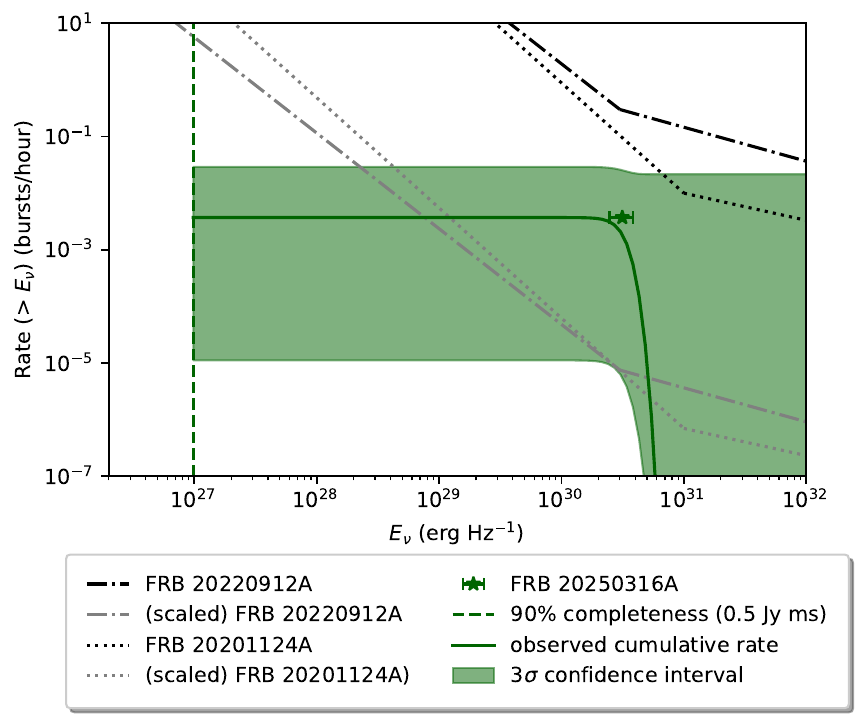}
    \caption{Cumulative rate as a function of spectral energy. The empirical/observed cumulative rate for \frb is shown as a solid green line. We show the 99.7\% confidence intervals around the observed cumulative rate assuming Poisson statistics (green region). The single observed burst from \frb is marked at its measured spectral energy (green star). Our sensitivity threshold (0.5 Jy ms) is shown with a green, vertical, dashed line.  For comparison, we show the observed best-fit spectral energy function for FRB 20201124A \citep{2024NatAs...8..337K} and FRB 20220912A \citep{2024arXiv241017024O}, with black-dotted and dot-dashed lines, respectively. Even scaling these observed overall rates down orders of magnitude, essentially as low as possible without making the observation of the first burst improbable (gray-dotted and dot-dashed lines for scaled-down cumulative rate distributions of FRB 20201124A and FRB 20220912A, respectively), reveals a significant dearth of lower-energy bursts compared to the expected rates, illustrating the tension between this burst and well-studied repeating FRBs. 
    \label{fig:energyrate}}
\end{figure}

Known repeaters are observed to have power-law fluence distributions, with higher repetition rates at lower fluence. If \frb were a repeater, CHIME/FRB should be sensitive to those lower-fluence bursts, and a lack of detection allows us to constrain the likely burst luminosity function. Assuming a cumulative rate distribution as a function of burst spectral energy with a power-law index of $\alpha = -1.5$,
the probability that we detect a burst of this spectral energy and nothing less energetic, considering our fluence threshold, is less than $2\times10^{-6}$ (4.7$\sigma$ Gaussian equivalent) for this source alone (i.e., even maximizing among all possible overall burst rates). If one controls for the family-wise error rate using the Bonferroni correction \citep{bonferroni} considering our ($\sim 100$) total confirmed repeaters, this observation is inconsistent with a $\alpha = -1.5$ power-law index at a 3.7-$\sigma$ Gaussian equivalent level ($p < 2\times10^{-4}$). Indeed, even assuming that every unique FRB source that we have detected to date is a repeater (using a trials factor extrapolated from the source count at the cutoff of CHIME/FRB's Second Catalog to $\sim 5000$), there is a $<1\%$ chance of such an observation per CHIME/FRB experiment, i.e., in the first 5000 detected FRB sources. This tension is illustrated in the empirical cumulative rate as a function of spectral energy in Figure \ref{fig:energyrate}, where even scaling the overall rate of the repeaters down orders of magnitude, essentially as low as would still statistically allow for the detection of the first burst, reveals a significant dearth of lower energy bursts. 

 Assuming the measured power-law indices of known repeaters FRB 20220912A and FRB 20201124A -- $-1.69$ and $-1.94$, respectively; relevant for spectral energies below $3\times 10^{30}$ and $1\times10^{31}$ erg Hz$^{-1}$, respectively \citep{2024NatAs...8..337K,2024arXiv241017024O} -- the probability of detecting only one burst with the spectral energy $(3.1\pm 0.7)\times10^{30}$ erg Hz$^{-1}$ and no lower-energy bursts above our $1\times10^{27}$ erg Hz$^{-1}$ sensitivity threshold
 decreases by one and two orders of magnitude (to $5.1,\, 5.4 \sigma$ Gaussian equivalent). 
 Considering the non-detection by FAST \citep{ATel17126} in the week following \frb's detection, this probability decreases by an additional order of magnitude ($5.2.\, 6.1 \sigma$ assuming $\Gamma=-1.5, -1.96$, respectively). However, this number should be considered with the caveat that both the HyperFLASH and FAST observations took place at a higher radio observing frequency, and while FRBs discovered in the CHIME band are commonly detected at $\sim 1400$MHz (L-band), ultimately the average repeater burst spectral energy distribution is not well-characterized. Similarly, while the FAST observations took place at L-band promptly after established source activity, the comparatively short exposure time of $13$ hours in six exposures spread over a week suggests that the constraining power on the rate will be dominated by the assumed clustering, which is not accounted for in this probability. 
\subsection{V/\texorpdfstring{V\textsubscript{max}}{V max} Analysis}
V/\texorpdfstring{V\textsubscript{max}}{V max} serves as a useful statistic to highlight the uniqueness of \frb. V/\texorpdfstring{V\textsubscript{max}}{V max} is defined as the comoving volume interior to a source's distance, divided by the comoving volume interior to the maximal distance at which that source could be detected \citep[$V/V_\text{max}=D_c^3(z)/D_c^3(z_\text{max})$; ][]{1970MNRAS.151...45L}. For a survey of sources distributed uniformly in comoving space, which is complete above some threshold, this statistic is also expected to be uniformly distributed between zero and one. For the case of FRBs, $D_c(z_\text{max})$ corresponds to the distance where the fluence threshold of detection ($F_{\nu,\text{thresh}}$) corresponds to the energy of the burst $E$ following \citep{2018MNRAS.480.4211M}:
\begin{equation}
     E = \frac{4\pi d_L^2(z)\Delta\nu F_{\nu,\text{thresh}} \sqrt{w_\text{eff}(z)}}{(1+z)^{2+\beta}},
\end{equation}
where $\beta$ is the spectral index, $d_L$ is the luminosity distance and $\Delta\nu$ is the burst bandwidth. We include an additional factor representing the relative change in burst width as a function of redshift ($w_\text{eff}(z)$) to capture how increased width from DM and cosmological time dilation reduce the S/N of a detection at a given fluence, following the formalism in \cite{2003ApJ...596.1142C}
\begin{equation}
    w_\text{eff} = \frac{\sqrt{(w_\text{m}(1+z))^2+w_{{\text{DM}(z)}}^2+\tau^2}}{w_m},
\end{equation}
where $w_m$ is the width measured in the observed burst, $(1+z)$ is the cosmological time dilation factor, $w_{\text{DM}(z)}$ is the contribution to the observed width from dispersive smearing as a function of redshift and $\tau$ is an additional scattering component from intervening structures at larger distances which we do not take to have a strict functional form with redshift. 

To calculate V/\texorpdfstring{V\textsubscript{max}}{V max} we use \frb's measured burst fluence of $1700\,$Jy\,ms, a fluence threshold of $0.5\,$Jy\,ms (see Appendix~\ref{app:sens_exp}), a fiducial burst bandwidth of $1\,$GHz, a spectral index of $\beta=-1.39$ \citep{2023ApJ...944..105S}, and assume a Planck cosmology \citep{2020A&A...641A...6P}. To scale the width we use a fiducial measured width of $w_m=1$\,ms, calculate the dispersive smearing at the central observing frequency of CHIME, $600\,$MHz, using the 24.4 kHz channels in the CHIME/FRB detection system (\citetalias{CHIME+2021} \citeyear{CHIME+2021}), and include a flat $\tau=1\,$ms scattering time as a conservative estimate of the contribution to scattering from intervening structures \citep{2019Sci...366..231P}. To maintain this conservative approach we deliberately overestimate the excess DM acquired at greater distances by using the 3-$\sigma$ excess from the Macquart relation \citep{2020Natur.581..391M} for the highest DM variance scenario (F $\approx0.25$) seen among the CAMELS suite of simulations \citep{2024ApJ...967...32M}. Under these assumptions \frb\, is detectable out to $z=0.42$ ($d_L=2370\,$Mpc), at which point the burst has accrued an additional $555\,$pc cm$^{-3}$ of DM, $1\,$ms of additional scattering and is $61\%$ of its original energy following the spectral index and emission in a higher frequency rest frame. This maximum redshift yields a V/\texorpdfstring{V\textsubscript{max}}{V max} $=1.4\times10^{-5}$. To contextualise the scale of this value, if we integrate the comoving volume over the whole sky out to infinite redshift it converges to a value $\approx3800\,$Gpc$^3$. Taking this value as V\textsubscript{max}, a V/\texorpdfstring{V\textsubscript{max}}{V max} $\leq 1.4\times10^{-5}$ is only possible for sources interior to $z\geq0.079$ ($D_c=340\,$Mpc), for all other sources there is insufficient volume in the observable Universe. We highlight that several other nearby, high-fluence bursts have been detected by CHIME/FRB in its broad side-lobes \citep{2024ApJ...975...75L}. In these cases, however, the associated detection thresholds are much higher than for \frb, substantially reducing their expected V/\texorpdfstring{V\textsubscript{max}}{V max} values.

\begin{figure}
    \centering
    \includegraphics[width=\linewidth]{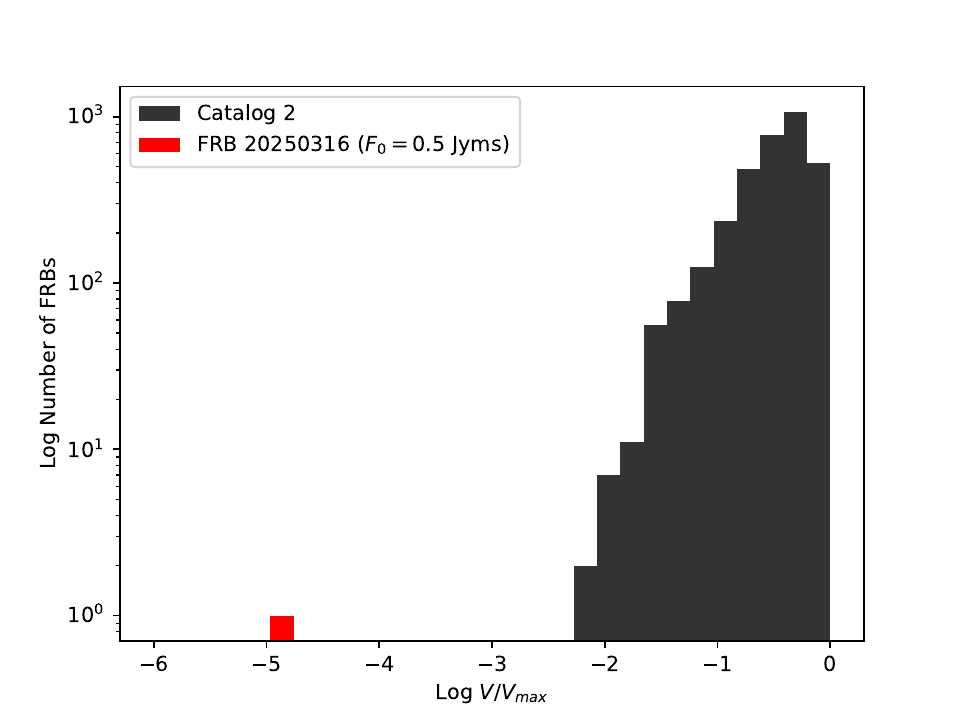}
    \caption{Comparison of the V/\texorpdfstring{V\textsubscript{max}}{V max} statistic for \frb, in red, against a simulated distribution for the one-off bursts in the Second CHIME/FRB Catalog, using DM as a proxy for distance under the z-DM distribution derived by \cite{2023ApJ...944..105S}.}
    \label{fig:VonVmax}
\end{figure}

Using the above method we can compare this result to V/\texorpdfstring{V\textsubscript{max}}{V max} for the 3375 one-off FRB sources with measured fluence in the Second CHIME/FRB Catalog (\citetalias{catIIsubmitted} \citeyear{catIIsubmitted}). To estimate the true redshift of each FRB, we translate observed dispersion measures, using the $p(z|DM)$ relationship modelled by \cite{2023ApJ...944..105S}. The resulting V/\texorpdfstring{V\textsubscript{max}}{V max} values are shown in Figure \ref{fig:VonVmax}. Notably, the observed bursts do not populate to V/\texorpdfstring{V\textsubscript{max}}{V max} = $1/3375=10^{-3.53}$ level as expected for a uniform population of sources in comoving space which should yield a correspondingly uniform distribution of V/\texorpdfstring{V\textsubscript{max}}{V max} values. We expect this is due to instrumental selection bias introduced when high S/N events are flagged as RFI. If we assume the selection-corrected distribution is uniform, a fiducial CHIME-like sample of 5000\footnote{This burst was observed subsequently to the cutoff date of CHIME/FRB's Second Catalog and therefore belongs to a larger sample whose size we estimate through extrapolation.} FRBs has a 7.0\% chance of containing a V/\texorpdfstring{V\textsubscript{max}}{V max} on this scale, and therefore \frb\ is potentially consistent with the observed population of one-off CHIME bursts. 

In the nearby Universe, where cosmological populations do not undergo significant redshift evolution of their ensemble density and luminosity functions, the assumption of a uniform V/\texorpdfstring{V\textsubscript{max}}{V max} distribution is well justified. Conversely, as the next generation of FRB surveys like Canadian Hydrogen Observatory and Radio Transient Detector and the Deep Synoptic Array 2000 \citep{2019clrp.2020...28V, 2019BAAS...51g.255H} begin to detect more distant FRBs, the evolution of these properties is expected to impact the observed V/\texorpdfstring{V\textsubscript{max}}{V max} distributions. If FRBs trace star formation, as expected for a young magnetar progenitor model, the increase in cosmic star formation toward redshift 2 will bias observations toward higher V/\texorpdfstring{V\textsubscript{max}}{V max} values. Whereas progenitor channels involving long delay time distributions, such as neutron star mergers \citep{2022ApJ...940L..18Z}, are expected to have a flatter redshift distribution between $z=0-2$, yielding a more uniform distribution of V/\texorpdfstring{V\textsubscript{max}}{V max}. The measurement of this cosmological statistic therefore represents a useful tool for understanding FRBs and should be considered in the future.

\section{Host Galaxy NGC 4141 and FRB Environment}
\label{sec:environment}
\subsection{Host Identification and Galactocentric Offset}\label{ss:PATH}
 We used the VLBI localization to associate a host candidate for \frb using the Probabilistic Association of Transients to their Hosts (PATH) analysis, as described by \citet{PATH_2021}. We identified host candidates using galaxy angular sizes and magnitudes from the DECaLS DR10 imaging catalog \citep{DECALs_2019}. \frb is located within $\sim0.5\arcmin$ of NGC 4141. However, the imaging catalog also identifies star-forming regions within $\approx40\arcsec$ from the galactic center of NGC 4141 as individual objects. We therefore use {\tt photutils} to estimate the half-light radius of NGC 4141 and update the candidate list by removing these regions. We assume an exponential offset prior and a prior probability of an unseen host, $P(U|x)$, of 0.10 (Anderson et al. \textit{in prep.}). The PATH analysis identifies NGC 4141 as the top candidate for the host of \frb with a posterior probability, $P(O|x)$, of 99.95\%. 

 From our MMT imaging (\S\ref{sec:MMT}), we use {\tt Source Extractor} \citep{SourceExtractor} to derive a light-weighted center for NGC~4141 of R.A. (ICRS) = 12$^\text{h}$09$^\text{m}$47$\dotsec$347 and Decl.(ICRS) = 58$^{\circ}$50$\arcmin$56.841$\arcsec$. Incorporating the FRB positional, host positional, and astrometric tie uncertainties, we calculate a projected offset between the FRB and the host center of $23.50 \pm 0.11\arcsec$ or $4.49 \pm 0.02$~kpc at 40~Mpc\footnote{This value is the straight line distance between the centre of the FRB localization and the host galaxy with 1-$\sigma$ uncertainties added in quadrature.}.

\subsection{Limits on Emission at FRB Location}
\paragraph{Radio} Our continuum radio observations (\S\ref{sec:cradio} and Appendix~\ref{app:crcal}) identified no evidence of compact radio emission associated with the FRB position, placing a 5-$\sigma$ upper limit on any associated counterpart of $\leq 22~\mu\mathrm{Jy}$ at $4.87~\mathrm{GHz}$ and $\leq11~\mu\mathrm{Jy}$ at $9.9~\mathrm{GHz}$. At a distance of $40~\mathrm{Mpc}$, this corresponds to an upper limit on the specific luminosity of $L_{\mathrm{4.93~\mathrm{GHz}}}\leq4.2\times10^{25}~\mathrm{erg~s^{-1}~Hz^{-1}}$ and $L_{\mathrm{9.9~\mathrm{GHz}}}\leq2.1\times10^{25}~\mathrm{erg~s^{-1}~Hz^{-1}}$. Previous limits reported using archival VLA data performed at X-band placed a 5-$\sigma$ upper limit on any associated compact radio emission of $4.8\times10^{25}~\mathrm{erg~s^{-1}~Hz^{-1}}$ \citep{2025ATel17111....1G}. VLBI measurements performed using eMERLIN achieved an upper limit of $2.2\times10^{26}~\mathrm{erg~s^{-1}~Hz^{-1}}$ \citep{2025ATel17120....1B}.  A summary of upper limits on any persistent radio counterpart are provided in Table \ref{tab:radiofluxes}.

\paragraph{Optical} We can also place constraints on transient optical emission prior to and immediately following the FRB. Prior to the FRB, we use historical exposures of this field with the 0.76-m Katzman Automatic Imaging Telescope (KAIT) at Lick Observatory \citep{2001ASPC..246..121F} and the Coddenham Observatory, which have a cadence of $\sim$10 exposures per year in 2001--2009 and 3 exposures per year thereafter. 
There is no evidence for point-like optical transient emission down to $M_\mathrm{clear} \approx -13.5$ mag ($1\sigma$). These exposures are deep enough to detect luminous optical transients including core-collapse and Type Ia SNe as well as optical transients associated with the final stages of stellar evolution in some massive stars~\citep[such as those reported by][]{pastorello2007giant}. However, the observing cadence of both KAIT and the Coddenham Observatory are sensitive to only half of each year of observation, allowing for the possibility that a weeks-long optical transient was simply missed.  We expect the observations would have been sensitive to typical 1--2 month old supernovae had one occurred in the gaps between observations.

From our MMT imaging (\S~\ref{sec:opt}), we detect no apparent unresolved optical emission at the FRB position to $r \gtrsim 25.6$~mag or $M_r \gtrsim -7.4$~mag. At the distance to NGC 4141 ($d \approx 40$~Mpc), this upper limit corresponds to an in-band luminosity of $1.9 \times 10^{38}$ erg~s$^{-1}$. With this limit, a comparison to single-star evolutionary tracks \citep{Choi+2016} suggests that we can rule out the presence of massive main-sequence stars with initial masses $\gtrsim 40\,M_{\odot}$, as well as evolved stars (i.e., red supergiants) with masses $\gtrsim 20\,M_{\odot}$. Similarly, our limit can be used to place constraints on slowly evolving transients. For instance, we can rule out the presence of a young ($\lesssim 1$~yr) core-collapse supernova (CCSN) at the FRB position based on CCSN light curves \citep{JacobsonGalan+2025}. After the FRB occurred, our MMT imaging indicates no new point source relative to archival imaging (taken in 2016--2017), and no transient emission to $M_{r} \approx -8$~mag on timescales of $\sim 8.2$--9.5 days after the FRB. This limit is sensitive to rapidly evolving transients such as gamma-ray burst afterglows, kilonovae, and fast blue optical transients \citep{Villar+2017,Ho+2023}.

If the FRB progenitor is a magnetar formed through the core-collapse of a massive star, a past SN or historical optical transient may be spatially coincident with the FRB. To this end, we search for cataloged optical transients from the Transient Name Server at the FRB position following the method described in \cite{Dong25}. We find no significant associations within the 5-$\sigma$ localization uncertainty of \frb. Within NGC 4141, there are two reported Type II SNe, SN\,2008X and SN\,2009E \citep{2008CBET.1239....1B,2012A&A...537A.141P} which lie 31.5\arcsec~and 41.2\arcsec~away from \frb, respectively. More recently, an unclassified optical transient, AT\,2025erx, was discovered in the same galaxy prior to the FRB on 2025 January 17 \citep{Andreoni25}, and lies 57.7\arcsec~away from the FRB. Given such large positional separations, we conclude that a physical association to either the SNe or the optical transient is ruled out. 

\subsection{Radio Star Formation Rate}\label{ss:sfr_radio}
Following \cite{greiner2016}, we estimate the radio-inferred star formation rate, $\mathrm{SFR}_\mathrm{radio}$, of NGC 4141 to compare to the optically inferred value. The radio-inferred SFR is given by 
\begin{equation}
    \mathrm{SFR_\mathrm{radio}} = 0.059~M_\odot~\mathrm{yr^{-1}} F_{\nu,\mu\mathrm{Jy}} d_{L,\mathrm{Gpc}}^2\nu_\mathrm{GHz}^{\gamma}(1 + z)^{(\gamma-1)},
\end{equation}
where $F_\nu$ is the observed flux density at a frequency $\nu$ and $d_L$ is the luminosity distance at a redshift $z$. We adopt $\gamma = -0.75$ \citep{Condon1992} for the spectral index, as the measured flux becomes increasingly resolved out at higher frequencies, leading to unreliable estimates of the source flux and hence spectral index. Using the VLA S-band flux density (see Table \ref{tab:radiofluxes}), we infer a 3.2-GHz radio SFR for the host galaxy of $\mathrm{SFR_\mathrm{radio}} = 0.2 ~\rm M_\odot~\mathrm{yr^{-1}}$. This is comparable to, albeit lower than, the SFR as inferred from \halpha\, of $0.6 \ \rm M_\odot~\mathrm{yr^{-1}}$ \citep{2005PASP..117..227K}. This minor discrepancy may be attributed to variations in the radio-infrared correlation across spiral arms and interarm regions due to differences in the interstellar medium properties that complicate the use of non-thermal radio emission as an SFR indicator (e.g., \citealt{Dumas2011}). Alternatively, the larger \halpha\, SFR may point to a recent enhancement in star formation, given the typical timescales of 10 Myr probed by line indicators compared to the 100 Myr timescales associated with non-thermal synchrotron emission.

\subsection{Local Environment}
\label{sec:clump}

\begin{figure*}[ht]
    \centering
    \includegraphics[width=0.8\linewidth]{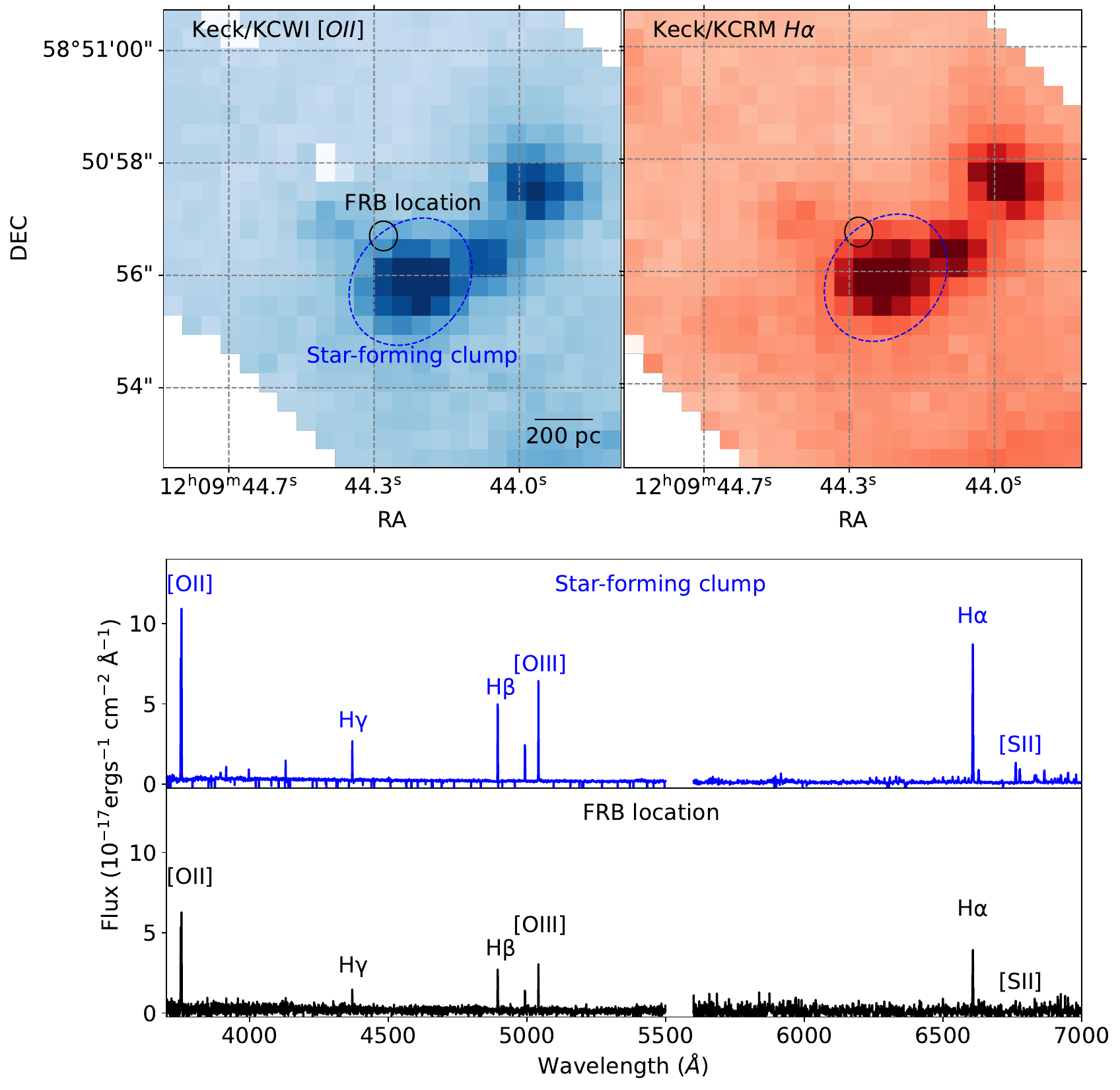}
    \caption{KCWI narrow-band images of the burst location. The top panels show [OII]3726-3729~\AA\ and \halpha~images extracted from the co-added cubes of the blue and red arm of KCWI respectively. The solid black ellipse shows the 2-$\sigma$ localization region of the FRB, placing its center at 190 pc projected separation ($0.9^{\prime\prime}$ angular offset) from the emission peak of the nearby star-forming region, which is marked by the dashed blue ellipse. The middle panel shows the combined blue and red side 1D spectrum of the star-forming clump extracted within the dashed ellipse. The bottom panel shows the spectrum extracted from the FRB localization ellipse.  The gap in the spectra between 5500~\AA\ and 5600~\AA\ masks out the region where the dichroic splits the blue and red light. Both spectra show narrow nebular emission features and we do not see any transient features at the location of the FRB.}
    \label{fig:kcwi_mosaic}
\end{figure*}

The co-added KCWI data cubes reveal several compact star-forming clumps with strong nebular emission features. Indeed, consistent with the MMT imaging, the FRB localization centroid is offset from the centroid of the nearest star-forming clump by $0.9\arcsec\pm0.1\arcsec$ \footnote{As mentioned in \S\ref{sec:kcwi}, the localization error ellipses include the radio localization error, astrometric uncertainty in the MMT-Gaia tie-in and the astrometric uncertainty in aligning the KCWI cubes to the MMT image.}, which corresponds to $190\pm20$ pc in projected separation at the host distance. In Figure~\ref{fig:kcwi_mosaic}, we show the combined spectrum (from the red and blue arms) of the clump extracted within a $1.5\arcsec$ aperture. Also shown is the spectrum of the region extracted from an elliptical aperture corresponding to the 2-$\sigma$ localization region of the FRB. Comparing the two spectra, there is no strong evidence of any emission lines associated with a potential transient, i.e., no significant difference in line-widths or the detection of features atypical of galactic nebular emission.

In addition to investigating the presence of transient optical emission, we leverage our spectroscopic data cubes to characterize the local gas environment in the vicinity of the FRB. To this end, we produce emission line maps from the cube by extracting spectral slabs covering the emission lines of interest and then fitting a flat continuum model with Gaussian line profiles for individual lines for each spaxel. Instead of fitting the entire wavelength range, we divide our cubes into three slabs covering (1) the [OII] 3726-3729~\AA\ doublet, (2) H$\beta$ and the [OIII] 4959-5007~\AA\ doublet, and (3) \halpha, the [NII] 6548-6584~\AA\ doublet and the [SII] 6718-6733~\AA\ doublet. This simplifies our computation and also avoids the need for reprojecting the blue cube to the red WCS system, as there is a slight offset between the cubes. With the line fluxes extracted at each spaxel ($0.34^{\prime\prime}\times0.34^{\prime\prime}$), we employ standard emission line diagnostics (that we shall describe below) to infer gas properties of the host ISM.

\subsubsection{ISM Ionization State}

The Baldwin-Phillip-Terelevich (BPT) diagram \citep{BPT} is a line-ratio diagnostic of excitation mechanism driving gas ionization of ISM gas. Since we detect the four lines \halpha, $\rm [NII]6584$, $\rm [OIII]5007$ and \hbeta, we can place the emitting clump near the FRB location on the diagram to ascertain the nature of the radiation. We do not perform a per-spaxel BPT analysis since there is a very slight offset between the red and blue arm data cubes that precludes a one-to-one mapping between the spaxels of the two cubes. Instead, we opt to extract the line fluxes from the spectra of the clump and the FRB location (middle and bottom panels of Figure \ref{fig:kcwi_mosaic}) and use these to determine the gas properties at the two locations.

\begin{figure}[ht]
    \centering
    \includegraphics[width=\linewidth]{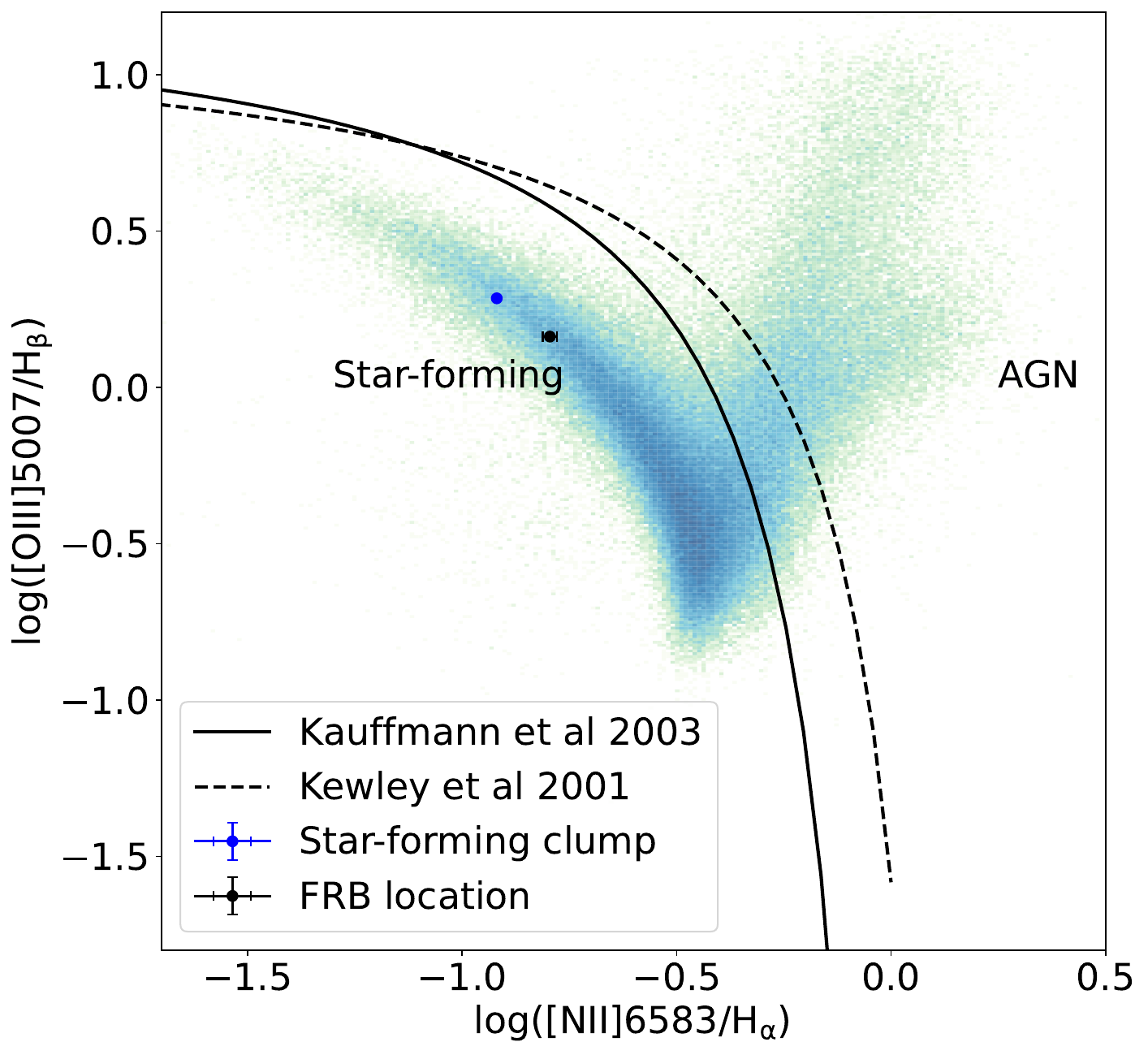}
    \caption{BPT diagram showing the clump and the emission at the FRB location compared to the field galaxies (blue-green colormap). Both locations show ionization consistent with star-forming activity, i.e. to the left of the \citet{Kauffmann+2003} demarcation (black solid line) and the \citet{Kewley+2001} demarcation (black dashed line).}
    \label{fig:bpt_kcwi}
\end{figure}

Figure~\ref{fig:bpt_kcwi} shows the BPT diagram with our data points overlaid. Both the clump and the FRB location are consistent with gas ionization from star-forming activity within \ion{H}{2} regions. 

\subsubsection{Star-formation rate and dust extinction}
\label{sec:optical_sfr}

The \halpha~emission flux is a well-known indicator of present star formation rate (SFR) through the \citet{Kennicutt1998} relation. It is strongly detected throughout the cube footprint (see Figure \ref{fig:emha} and \S~\ref{sec:ism} for the spatially resolved emission map). However, one must account for dust extinction from the Milky Way and the host galaxy ISM before converting the line flux to the SFR. To correct the photometry for Galactic extinction, we query the IRSA dust extinction database\footnote{https://irsa.ipac.caltech.edu/applications/DUST/} at the FRB position and find $\rm E(B-V) = 0.05$ \citep{SchlaflyFinkbeiner2011}.  We rely on the measured flux ratio \halpha/\hbeta~to first estimate the dust extinction and subsequently correct the \halpha~flux. We assume $R_V = 3.1$ and subsequently apply the extinction correction to the line fluxes. Then, to correct for the host galaxy dust extinction, we assume that the intrinsic ratio of \halpha/\hbeta~fluxes is 2.87 \citep{OsterbrockFerland2006} and attribute the deviation from this ratio to dust extinction, allowing us to compute the visual extinction $\rm A_V$ from the spectra (we also assume $\rm R_V = 3.1$ here). To this end, we assume the G23 extinction model \citep{G23} from the \texttt{dust\_extinction} Python package \citep{gordon_2024}.  After de-reddening, we apply the Kennicutt relation and multiply the result by 0.63 \citep{Conroy+2009} to report the SFR with the \citet{Chabrier03} Initial Mass Function (IMF). The error is largely from the scatter in the Kennicutt relation ($30\%$). As in the case of the BPT diagram, we do not provide maps of the either star-formation or dust extinction for the entire cube footprint due to the slight offset between the red and blue cubes. We again adopt the spectra extracted at the FRB location and the star forming clump and tabulate our measurements in Table \ref{tab:sfr}. We estimate that the star-forming clump is roughly producing 5 $\times 10^{-4} \rm M_\odot/yr$. 

\begin{deluxetable}{lcc}
\tablecaption{Dust extinction and the star-formation rate of the host galaxy in the vicinity of the FRB. We note here that the reported error in $\rm A_V$ is purely from propagating the line flux measurement uncertainties and do not include the systematic error in assuming the G23 dust extinction model.\label{tab:sfr}}
\tablehead{
\colhead{Location} & \colhead{A$_V$} & \colhead{SFR} \\
\colhead{} & \colhead{(mag)} & \colhead{($10^{-4}\ M_\odot\,\mathrm{yr}^{-1}$)}
}
\startdata
Star-forming clump & $1.15\pm0.02$ & $5\pm 2$  \\
FRB location       & $0.32\pm0.09$ & $1.4\pm 0.4$ \\
\enddata
\end{deluxetable}

\subsubsection{ISM metallicity}
\label{sec:optical_met}

Of the various probes of ISM metallicity, the flux ratio $\rm R_{23}$ is known to have the least scatter \citep[0.14 dex;][]{Nagao+2006, Nakajima+2022}, where
\begin{equation}
    \rm R_{23} = \frac{[OII]3729+[OIII]4959+[OIII]5007}{\hbeta}.
\end{equation}

\begin{figure*}
    \centering
    \includegraphics[width=\textwidth]{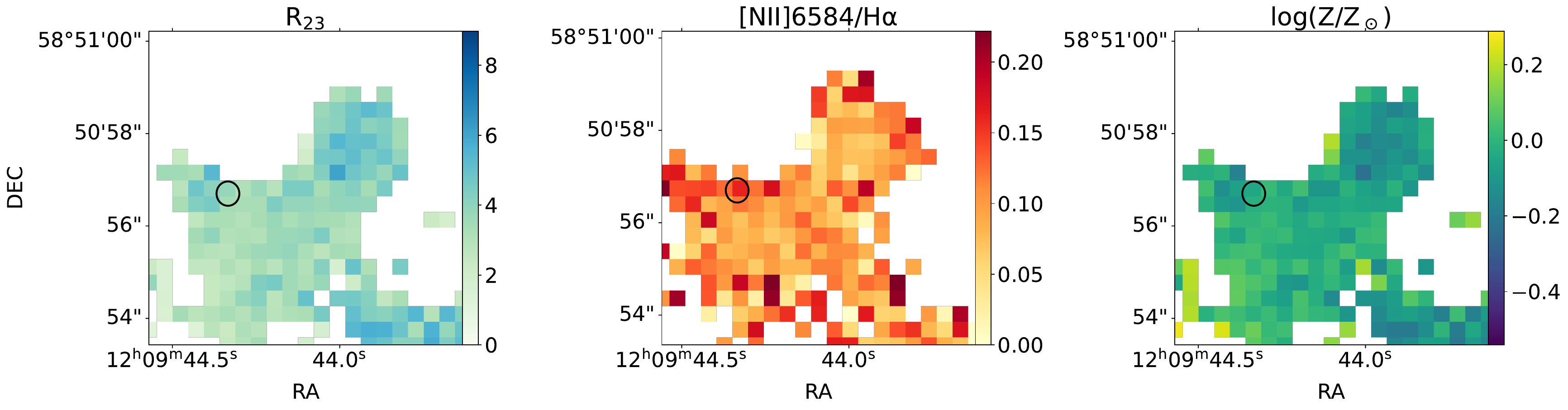}
    \caption{Strong-line metallicity diagnostics. Left: A map of $\rm R_{23}$ computed from the blue arm data cube of KCWI. Middle: The [N II]6584/\halpha~flux ratio estimated from the red arm data cube. Given that most spaxels near the FRB location show the $\rm [N II]6584/\halpha>0.1$, we can safely assume the metallicity is higher than 8.1, and thus break the degeneracy in $\rm R_{23}$. Right: Subsequently, we can invert the best-fit $\rm R_{23}$--$\rm \log (Z/Z_\odot)$ relation on the high-metallicity branch of the curve to produce a metallicity map near the FRB (2-$\sigma$ localization shown by the black ellipse). We note that the FRB environment's metallicity is only weakly sub-solar.}
    \label{fig:met}
\end{figure*}

However, converting $\rm R_{23}$ to metallicity requires an additional line-diagnostic as it is a double-valued, i.e., both low and high metallicity gas can produce similar values of $\rm R_{23}$. Therefore, we show our maps of $\rm R_{23}$ and the flux ratio [NII]6584/\halpha~in the left and middle panels of Figure \ref{fig:met}. Indeed, despite the noisier [NII]6584/\halpha~map, we note that most of the spaxels show that flux ratio is $\gtrsim0.1$, thus implying a metallicity $\rm \log(Z/Z_\odot) > -0.5$ \citep{Nakajima+2022}. Subsequently, we invert $\rm R_{23}$ in the high-metallicity regime to map $\rm \log(Z/Z_\odot)$, shown in the right panel of Figure \ref{fig:met}. The FRB location exhibits a weakly sub-solar metallicity. This is consistent with the distribution of metallicity estimated by \citet{Sharma+2024} for their host galaxy sample. We recognize that their measurements are for global host properties and thus more IFU data on nearby FRB hosts is necessary to determine the distribution of ISM metallicity of FRB environments.         

\subsubsection{ISM Electron Density}
\label{sec:ism}
The [O~II] 3726,3729 \AA\ doublet is collisionally excited and the line flux ratio [O~II] 3729/[O~II] 3726 can serve as a probe of the ISM gas density assuming a typical ISM temperature ($\sim 10^4$~K; \citealt{OsterbrockFerland2006}). A similar analysis can be performed using the [S~II] 6717, 6731 \AA\ doublet, but we opt to use [O~II] as it has much higher detection S/N throughout our data. We employ the line fluxes extracted on a per-spaxel basis (as described previously) to generate a map of the flux ratio over our full KCWI footprint. We only estimate the ratio in spaxels where the combined [O~II] doublet emission is detected at a significant level ($3\sigma$) over the background noise. To select these pixels, we first extract a narrow-band image encompassing the wavelength range of the combined [O~II] doublet emission (3746--3760~\AA). Then, we sigma-clip iteratively to identify the background spaxels and estimate the spaxel-to-spaxel standard deviation in flux. Finally, we select spaxels for our ratio estimation by requiring their flux to be above the 3-$\sigma$ background limit. In addition to the flux selection above, we also require the line flux ratios only be estimated where both emission lines have been fit with a Gaussian with positive amplitude.

\begin{figure*}
    \centering
    \includegraphics[width=\linewidth]{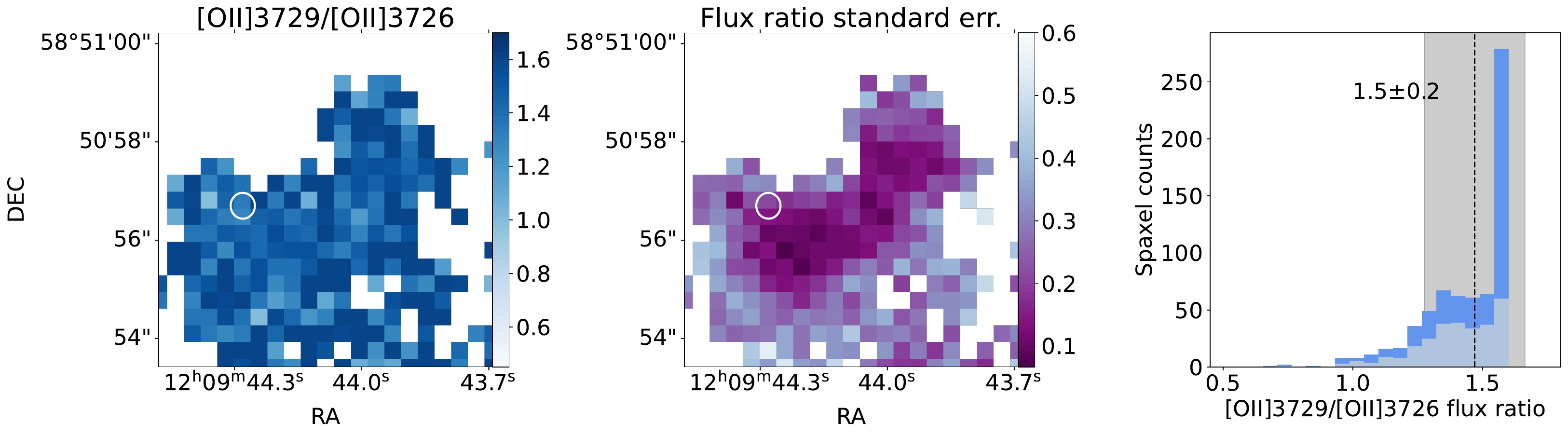}
    \caption{Left: [O~II] 3729/[O~II] 3726 line-ratio map over the KCWI footprint. The flux ratios are only computed where the combined [O~II] doublet emission is over the 3-$\sigma$ limit set by the background and where the line fluxes are measured (i.e., positive amplitude Gaussian fit to the emission line). Other spaxels are blanked out. The 2-$\sigma$ FRB localization is marked by the white ellipse. Middle: The per spaxel error in the flux-ratio measurement estimated from propagating the curve fitting errors. Right: A histogram of line-ratio values from all the colored spaxels in the left frame (deep blue). The median value is 1.5 with a standard deviation of 0.2, although the distribution is clearly non-Gaussian and most spaxels have line ratio of 1.6. If we limit histogram to spaxels where the flux error is below 0.25 (i.e., the blue-green regions in the middle panel), then the peak at 1.6 is no longer as prominent but most spaxels still have a flux ratio consistent with low-density gas ($\lesssim 1$\,e$^-\,$cm$^{-3}$). The FRB line ratio at the FRB location could be slightly lower, implying an enhanced density, but our measurement errors preclude a definitive gas-density estimate.}
    \label{fig:oii_ratio}
\end{figure*}

The left panel of Figure \ref{fig:oii_ratio} shows the [O~II] doublet emission-line ratio for our footprint. Two spaxels coincide with the 2-$\sigma$ localization region of the FRB (see left panel of Figure \ref{fig:oii_ratio}). Both the top and bottom spaxels indicate a flux ratio between 1.3 and 1.4, which could correspond to an electron density of $\sim 100$~cm$^{-3}$. However, we estimate a standard error of $\sim0.2$ for both spaxels (middle panel of Figure \ref{fig:oii_ratio}). The error is estimated by propagating the curve-fitting error for the two emission lines. Thus, the flux ratios of both spaxels could very well be $\gtrsim1.45$, at the low-density limit for the line ratio ($\lesssim 1$\,e$^-\,$cm$^{-3}$). Thus, while it appears that the electron density could be enhanced at the FRB location, we cannot claim a statistically significant detection of overdensity. The right panel of Figure \ref{fig:oii_ratio} shows a histogram of flux-ratio measurements for all colored spaxels in the left panel. Most spaxels show a doublet ratio at or beyond the low-density limit. It is therefore possible that the ISM at the FRB location is also at similarly low density.

\subsubsection{\dmhism}
\label{sec:dmhism}
In addition to assessing the electron density of the ISM from the doublet, we also leverage the \halpha\ emission measure (\emha) to constrain the DM contribution from the host ISM. This method was first described in the context of pulsars in the Milky Way \citep{Reynolds+1977} but was subsequently adopted by the FRB community \citep{cordes+2016,Tendulkar+2017,Bernales-Cortes+2025} to derive empirical upper bounds on \dmhism. Similar to our [O~II] analysis, we start with the spaxel-by-spaxel \halpha\ line-flux estimate from our fits to the data. Here too, we restrict our analysis to spaxels which correspond to \halpha~emission above the 3-$\sigma$ background threshold (wavelength range 6602--6612 \AA). The \dmhism~estimate from \emha~is given by
\begin{equation}
    \begin{aligned}
        \dmhism = 387~\dmunits&\left(\frac{\emha}{600~\rm pc~cm^{-6}}\right)^{0.5}\\
        &\times\left(\frac{L}{1~\rm kpc}\right)^{0.5}\left(\frac{F}{1}\right)^{0.5}\, ,
    \end{aligned}
\end{equation}

\begin{figure}
    \centering
    \includegraphics[width=\linewidth]{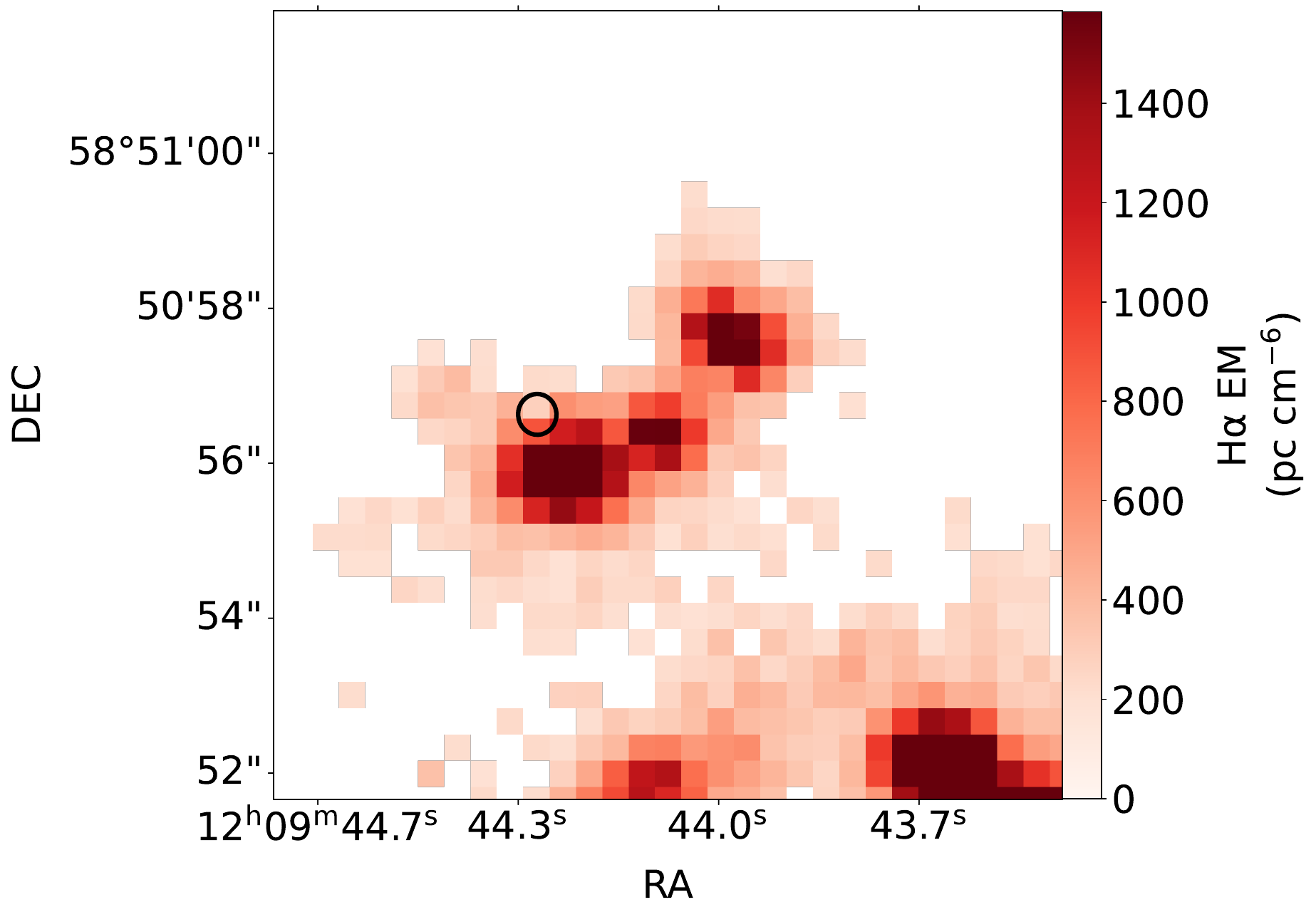}
    \caption{\emha~ estimated over the KCWI footprint. The 2-$\sigma$ localization regions of the FRB are shown by the solid black ellipse. Here, we are showing the \emha~map without applying any reddening correction and assuming that the gas temperature is $10^4$ K throughout.}
    \label{fig:emha}
\end{figure}

\noindent
where $F = 4f_f/\zeta(1+\epsilon^2)$ is a term that describes the properties of the emitting \halpha~clouds and encompasses cloud properties such as the volume filling factor of emitting clouds ($f_f$), turbulence-based fractional density variation in each cloud ($\epsilon$), and density fluctuation between individual clouds ($\zeta$) \citep{cordes+2016}. However, most FRB studies choose the value of $F=1$ \citep[the latest being ][]{Bernales-Cortes+2025}, which shall be our assumption as well. $L$ is the column length of the emitting gas that the FRB traverses. We compute the \halpha~surface brightness (in photon flux per unit solid angle measured in rayleighs; $I_R$) from our red-arm data cube and convert this to an emission-measure map using the equation $\emha=2.75$~pc~cm$^{-6}~T_4^{0.9}~I_R$ \citep{Reynolds+1977}. Here, $T_4$ is the gas temperature in units of $10^4$ K. In the absence of accurate temperature diagnostics, we assume $T_4 = 1$, following the references mentioned above. Figure \ref{fig:emha} shows our \emha~map without any host galaxy dust extinction correction applied. Within the 2-$\sigma$ localization region, $\rm EM = 282-987 ~pc~cm^{-6}$, the  range arising from the two KCWI pixels the uncertainty ellipse spans. From Table \ref{tab:sfr}, we can apply the estimated dust correction at the FRB location. Again, as done in \S\ref{sec:optical_sfr}, we use the extinction model from \citet{G23} and $R_V = 3.1$ and estimate that the intrinsic \halpha~surface brightness at the FRB location is 6\% higher than the observed value, i.e. $\rm EM_{dereddened} = 300-1050 ~pc~cm^{-6}$.  This implies $\dmhism=61-115\dmunits(L/50~\rm pc)^{0.5}$ at the FRB location. We choose 50 pc as a representative length scale here but emphasize that $L$ is not known {\it a priori}. 

One could, however, place constraints on $L$ using DM-budget arguments. Accounting for the DM from all other foreground components (the Milky Way, cosmic web, and host-galaxy halo), one can place bounds on \dmhism{} which subsequently constrains $L$. Using models of the Milky Way ISM and halo \citep{ne2001, ProchaskaZheng2019,2023ApJ...946...58C}, we can estimate that the Milky Way contribution to the FRB DM is $\sim70~\dmunits$. If all of the remaining 90 \dmunits{} were attributed to the ISM alone, the gas column length allowed by our \emha-based \dmhism~estimate is 30--110 pc. Stricter bounds can be placed by accounting for the host halo and the foreground cosmic web. The stellar mass of the host is estimated to be $\rm 1.7\times10^9~M_\odot$ 
in the NED Local-volume sample\footnote{We converted the NED estimate to the Chabrier initial mass function by dividing the listed value by 1.7.} 
\citep{Cook+2023_NED}. Assuming the stellar-to-halo mass ratio of \citet{Moster+2013} and applying the Modified Navarro-Frenk-White (NFW) profile halo gas model prescribed by \citet{ProchaskaZheng2019} (profile parameters $\alpha=2$, $y_0=2$, $f_{hot}=0.75$ extending to 2 virial radii), roughly 20 \dmunits~can be  attributed to the host halo. The cosmic web could contribute another 6 \dmunits\ to the DM budget \citep{Huang+2025}. Thus, if only 64 \dmunits~arise from the host ISM, the gas column further reduces to 17--60 pc. Thus, we can conclude that the FRB is not deeply embedded in any \halpha~emitting gas region. Interestingly, the scintillation and scattering properties of the burst (\S\ref{sec:scint}) indicate a small separation ($\lesssim60\,$pc) between the dominant scattering screen in the host and the FRB source. This coincidence suggests that the same \halpha~emitting gas region could also be responsible for the large-scale scattering ($\tau_2$) observed in the burst profile.

We must further stress an important point: our KCWI data are seeing limited. The star-forming clump is roughly $1.2^{\prime\prime}$ in diameter, comparable to the PSF size at the \halpha\ wavelength ($\sim0.6^{\prime\prime}$ FWHM, estimated from our standard-star observation). Given that the FRB is located at the edge of this emission, it is likely the clump is much more compact and thus, the true \halpha\ flux is even lower at the FRB location and commensurately our \dmhism\ estimate is lower too. This could imply a significant circumburst contribution to the FRB DM. Indeed, space-based imaging with  {\it JWST} obtained by \cite{Blanchard25} revealed that the FRB localization region is consistent with a faint point-like source ($\gtrsim30$ mag in F150W2), and the higher resolution data confirm the 1-$\sigma$ localizations are well outside the nearby star-forming clump. This observation could even imply a negligible \dmhism~with all of the remaining DM in the budget arising from the circumburst medium.

\section{Discussion}
\label{sec:discussion}
\subsection{Repetition} 

No repeat bursts from \frb have been detected in 269.77 hours of CHIME/FRB's exposure towards its sky position, and the morphology of the original detection (broadband emission spanning the entire observable 400 MHz and only $\lesssim$ 1\,ms in intrinsic duration, see Figure \ref{fig:pol_summary}) is similar to that of other one-off events.

Our implied 99.7\% upper limit rate ($0.015$ bursts per hour above 15 Jy ms; \S\ref{sec:rate}) is not in tension with repeater rates seen from CHIME/FRB's sample. The measured burst rates in \cite{CHIME+2023} span $0.001$ to $3$ bursts per hour above 5 Jy ms for established repeaters. To compare the rate upper limits derived from this work more directly, we can recalculate this rate using the same assumptions as used by \cite{CHIME+2023}: assuming burst arrival times are Poisson and scaling the upper limit on the rate we derive to $5$ Jy ms assuming a power-law index of $-1.5$. Based on these assumptions, and combining the CHIME/FRB data and the HyperFlash L-band data, we place a 99.7\% Poisson upper limit on the repetition rate of $\sim 0.001$ bursts per hour above 5 Jy ms. Considering only the HyperFlash L-band data, scaled with the assumptions above, we derive a 99.7\% upper limit on the repetition rate of $0.1$ bursts per hour above 5 Jy ms. 

In addition to being highly clustered in time, evidence of months- to years- long repeater `lifetimes' has begun to emerge; for example, FRB 20201124A,  FRB 20220912A, and FRB 20240209A were not detected for years by CHIME/FRB until their initial discovery, at which point they were in hyperactive phases (\citealt{2022ApJ...927...59L};  \citealt{Shah+2025}; T. C. Abbott et. al. \textit{in prep.}). The former two sources have not been detected by CHIME/FRB for years after their initial months-long hyperactive phases. The third known repeater, FRB 20180916B, was discovered by CHIME and later was found to be emitting only in approximately four day windows every 16.3 days \citep{2020Natur.582..351C}, the first periodic behavior confirmed from an FRB source. For these reasons, the detection of only one burst by CHIME/FRB from \frb is insufficient evidence that it is non-repeating.

The implied upper-limit burst rate at low-spectral energies relative to the implied burst rate at high-spectral energies for \frb, is, however, in significant tension with the known population of repeaters. The proximity of \frb means that CHIME/FRB would likely have been sensitive to bursts about one thousand times less energetic than \frb. Well-studied repeaters have burst spectral energy distributions with steep power-law burst cumulative rates with indices ranging from $-1.9$ to $-1.6$ \citep{2023ApJ...955..142Z,2024NatAs...8..337K,2024MNRAS.534.3331K,2024arXiv241017024O,2025MNRAS.tmp..751T} at spectral energies comparable to \frb, and below. The steep  $-1.9$ to $-1.6$ power-law rate distributions suggest that we would expect to detect these less energetic bursts $> 10^5$ more frequently than bursts with the spectral energy of \frb. We find the non-detection of bursts from \frb with lower spectral energies incompatible with the properties of known repeaters at a minimum level of $3.7\sigma$. Flatter spectral indices are required to relieve this tension: we find that only $\alpha > -0.6$ is consistent at the 3-$\sigma$ level. Shallow spectral indices have been measured for the higher-energy part of the broken power laws of FRB 20201124A and FRB 20220912A, but the spectral energy corresponding to the transition point between steep and shallow spectral indices for \frb would need to be approximately four orders of magnitude lower than observed for these sources to be consistent \citep{2024NatAs...8..337K,2024arXiv241017024O}. The FAST observations \citep{ATel17126}, which are sensitive to bursts 100 times less energetic than those detectable by CHIME/FRB, demonstrate that \frb is not highly active just below the sensitivity limit of CHIME/FRB. 

\frb is in tension with the properties of all well-studied repeaters to date. We do not detect FRBs often within the volume of \frb, and \frb is the highest-S/N event among the FRBs for which CHIME/FRB captured raw voltage data between the years 2019 and 2023\footnote{We search this sample because the real-time detection S/N is not an accurate predictor of the true S/N, see \S\ref{sec:CHIMEFRB}.}. The position of \frb transits through our main-lobe $<10\degree$ from the zenith of CHIME/FRB, and centrally through corresponding row of E--W synthesized beams, which means CHIME/FRB would have been particularly sensitive to bursts from this location.

Higher-fluence events than \frb are detected far into CHIME/FRB's side-lobes a few times a year \citep{2024ApJ...975...75L}. Despite comparable fluences, the non-detection of lower-fluence bursts from main-lobe transits of \frb's position represent a substantially stronger constraint on the repetition than for main-lobe transits of the high-fluence side-lobe sources. This is because the rate of low-fluence bursts expected for the side-lobe detections is smaller on account of the much larger exposure time over which the high-fluence detection was made. In comparison, the rate of high-fluence bursts required to explain the detection of \frb\ during its short main-lobe exposure implies a higher expected rate of low-fluence bursts. This difference in exposure leads to \frb\ having a much greater inconsistency with the known repeater population than the sources with high-fluence side-lobe bursts.

\subsection{Host Galaxy NGC 4141}
The host galaxy NGC 4141 is a relatively nearby SBc galaxy with a distance of 37--44 Mpc, notable for its abundance of \ion{H}{2} regions and for having hosted two recent Type II supernovae, SN2008X and SN2009E \citep{2008CBET.1239....1B,2009CBET.1648....1B,2008CBET.1239....2M,2012A&A...537A.141P} and unclassified optical transient, AT\,2025erx \citep{Andreoni25}. While the CHIME/FRB Outrigger localization is not positionally coincident with either of these supernovae and rules them out as related to the FRB, the occurrence of two supernovae within a 15-year span may point to an elevated SFR in the recent past \citep{Strolger2015,Michalowski2020}, and demonstrates a global environment with active star formation. However, the ongoing SFR in the host is not elevated, at $0.6 \ \rm M_\odot$ yr$^{-1}$ \citep{2005PASP..117..227K}. Combined with its inferred stellar mass of $1.7\times10^{9}$ M$_\odot$ \citep{Cook+2023_NED}, the specific star-formation rate (sSFR) is $\sim 3.5 \times 10^{-10}$~yr$^{-1}$, at the median value for the host galaxies of non-repeating FRBs \citep{Gordon+2023}. Moreover, the projected physical offset of the FRB of $\sim 4.5$~kpc from its host galaxy is also an average value compared to the population of FRBs \citep{Gordon2025}. Overall, NGC 4141 lies on the `star forming main sequence' of galaxies at comparable redshifts \citep{Whitaker2012} and is unremarkable given the typical ranges of such properties observed for FRB host galaxies \citep{Gordon+2023,Sharma+2024}.

\subsection{Environment}

Our results paint a picture of an FRB on the outskirts of a star-forming environment of slightly sub-solar metallicity. The DM budget allows for $\sim$ 40 \dmunits{} to be contributed locally from the ISM and circum-burst material. This local DM  is small compared to the estimated hundreds of \dmunits{} local contributions of FRB 20121102A and FRB 20190520B \citep{Tendulkar+2017, KochOcker_2022}. FRB 20121102A and FRB 20190520B additionally show evidence of having extreme magneto-ionic environments \citep{2018Natur.553..182M,2012ApJ...744..108B,2023Sci...380..599A}, characterized by large excess rotation measures, whereas \frb's measured local $|\text{RM}| \lesssim 10$ rad m$^{-2}$ is consistent with zero given the level of uncertainty on the foreground Galactic contribution to the RM. 

Our scintillation results (\S\ref{sec:scint}) and \halpha\, measurements (\S\ref{sec:dmhism}) are both consistent with a relatively dense\footnote{when inferring \dmhism{} from the DM budgeting arguments and accounting for uncertainty} (EM/\dmhism{} $\gg 0.1$~cm$^{-3}$; cf.\ \citealt{2008PASA...25..184G}) ionized cloud or screen that is $\lesssim$ 60 pc in front of the FRB, over and above any intervening diffuse ionized gas in NGC~4141. While we cannot rule out such a discrete source at our limited spatial resolution, these properties are consistent with the FRB being inside or immediately behind the outskirts of the adjacent star-forming region, whose center is separated in projection from the FRB by $190\pm20$ pc. We therefore propose that the FRB is physically associated with this region, and that the true offset of the FRB from the region's center is not too different from its projected offset. 
The observed $190\pm20$ pc offset between this star-forming region is similar to the case for FRB 20180916B, which is offset by 250 pc from the center of a star-forming complex. If we follow the logic from \cite{Tendulkar+2021} for FRB 20180916B, assuming that \frb's progenitor was formed near the center of the star-formation region, we can estimate the minimum age of the FRB source given the observed offset and likely kick velocities by progenitors of interest. This is particularly constraining for the young magnetar scenario of FRBs, in which, if one assumes typical velocities from magnetars and pulsars in our own Galaxy (between 60 and 750 km s$^{-1}$ for 90\% of observed pulsars, magnetars, and neutron stars in binaries; \citealt{2005MNRAS.360..974H,2020MNRAS.498.3736D}), implies ages of $2\times10^5 - 3\times10^6$ years for the source, much longer than typical active lifetimes of magnetars in our own Galaxy \citep{2014ApJS..212....6O,2017ARA&A..55..261K}. However, using the fit from \cite{2025arXiv250522102D} of the distribution of kick velocities for $<10$ Myr isolated neutron stars, which spans a wider range of velocities than those considered in \cite{Tendulkar+2021}, and accounting for uncertainties in the measured offset and best-fit parameters, we estimate a $5\%$ probability of observing such an offset for source ages less than 10 kyr. Even this slight tension between the offsets of \frb and FRB 20180916B and the age of young magnetars can be relieved if one assumes that the neutron star/magnetar is created {\it in situ} via a runaway OB star. 

Importantly, however, from JWST near-infrared (NIR) imaging, \cite{Blanchard25} find stars with similar population age to those in the star-formation region co-located with the FRB localization, indicative that the stellar population associated with the \ion{H}{2} region extends beyond its apparent size in the KCWI images.  Therefore, the authors conclude that a neutron star could have formed {\it in situ} and as a result, one cannot place a meaningful constraint on its age based on the observed offset from the center of the star-forming region.

Other similarities between \frb and FRB 20180916B are their expected DM$_{\text{host}}$\ contributions (40 and 70 \dmunits, respectively; \citealt{Marcote+2020}) and that both lack compact persistent radio counterparts. Of course, \frb does not seem to be a repeater, whereas FRB 20180916B exhibits a $\sim$16.3-day periodic activity cycle. The local environment of FRB 20180916B has an estimated metallicity of $12+\log(\text{O/H})= 8.4, \text{ or }\rm \log(Z/Z_\odot)\sim -0.3$, which is consistent with the metallicity of its host, as well as with NGC 4141, the host of \frb ($12+\log(\text{O/H})\sim 8.2$; \citealt{2013A&A...558A.143T}). We estimate a slightly higher gas-phase metallicity of $\sim8.6$ or $\rm \log(Z/Z_\odot) \sim -0.1$ for the local environment of \frb (\S\ref{sec:optical_met}) compared to the average value for NGC 4141, and substantially higher than the mean values at de-projected normalized distances similar to the measured offset of \frb $\rm \log(Z/Z_\odot) \sim -0.4$ \citep{2013A&A...558A.143T}. 

The local metallicity of the FRB environment is consistent with the distribution of global metallicity measurements of \cite{Sharma+2024}, who interpret the observed preferential occurrence of FRBs in high-mass galaxies as a preference for high-metallicity environments and hence favor progenitors like CCSNe of stellar
merger remnants. This disfavors transient progenitors which are in preferentially low-metallicity environments including superluminous supernovae \citep{2017MNRAS.470.3566C} and long gamma-ray bursts \citep{2024NatAs...8..774B}. Our VLBI observations (\S\ref{sec:cradio}) also rule out afterglow emission like that associated with a long gamma-ray burst that went off within the last $\sim 10$ years\footnote{We infer that the luminosity limits from our VLBI observations would be sensitive at the 3-$\sigma$ level to a typical long GRB afterglow on a time scale of 10 years assuming a power-law drop off in luminosity and extrapolating to 1.3 GHz \citep{2002ApJ...568..820G}.} \citep{2020MNRAS.496.3326R}. Short GRB radio afterglows have been detected up to $\sim$30 days post burst \citep{2024ApJ...970..139S} with $L_\nu\gtrsim 10^{39}$ erg\,s$^{-1}$\,Hz$^{-1}$ at $\sim$ GHz frequencies \citep{2015ApJ...815..102F}, which is a thousand times brighter than what our EVN limits would have been sensitive to nine days after \frb. 

In fact, the proximity of \frb\ allows us to place the deepest specific luminosity limit on any associated PRS with a non-repeater to-date, with $L_\nu < 4.2\times10^{25}~\mathrm{erg~s^{-1}~Hz^{-1}}$ at $4.9~\mathrm{GHz}$ and $L_\nu < 2.2\times10^{25}~\mathrm{erg~s^{-1}~Hz^{-1}}$ at $9.9~\mathrm{GHz}$. Prior to \frb, the deepest limit on a non-repeater was set by FRB 20220319D with an upper limit of $1.8\times10^{26}~\mathrm{erg~s^{-1}~Hz^{-1}}$ \citep{2024ApJ...967...29L}. The limits imposed by our observations point to a radio environment more in-line with that of FRB\,20200120E, the repeater localized to a globular cluster in M81, with a limit on a radio counterpart of $L_\nu \leq 3.1\times10^{23}~\mathrm{erg~s^{-1}~Hz^{-1}}$ at $1.5~\mathrm{GHz}$ \citep{Kirsten+2022}. Much like FRB \, 20200120E, \frb\ stands in contrast to the specific luminosities of the known sample of PRSs associated with some repeating FRBs, interpreted as dense magnetospheric nebular environments with specific luminosities ranging from $10^{27}-10^{29}~\mathrm{erg~s^{-1}~Hz^{-1}}$ \citep{Marcote_2017,Niu_2022,Bhandari_2023,bruni2024nebularoriginpersistentradio,Bruni_2025}. Additionally, we can rule out a pulsar wind-like nebula that is $\gtrsim20\times$ more luminous than the Crab, with a present-day specific luminosity of $\sim 2\times10^{24}~\mathrm{erg~s^{-1}~Hz^{-1}} $ at $4.93~\mathrm{GHz}$ \citep{1968AJ.....73..535T,Perley_2017}.  Our observations cannot rule out emission like that of the supernova remnant surrounding SGR\,1935+2154, which is about a thousand times less luminous than our specific luminosity upper limits \citep{2018ApJ...852...54K}. 

Given \frb's apparent tension with the spectral-energy distributions of the repeating population and nearby nature of the burst it is interesting to consider the gravitational wave (GW) counterpart which would be expected along with a cataclysmic merger scenario. Two sites, Hanford and Livingston, of the Laser Interferometer Gravitational-Wave Observatory (LIGO), were operating in the five hours before and two hours after the arrival of \frb and had an average sensitivity at S/N = 8 to binary neutron star inspirals to distances out to about $150$ Mpc, about three times more distant than \frb \footnote{Sensitivity according to the detector status summary page \url{https://gwosc.org/detector_status/day/20250316/}, accessed 2025 May 19.}, but no alerts were issued.  Generally it is expected that in these near-simultaneous windows around a BNS merger the ejecta environment would be opaque to the radio emission corresponding to what is observed at 400--800MHz for \frb \citep{2024ApJ...977..122B}, although there are some models that predict near simultaneous counterparts \citep[e.g.,][]{2014ApJ...780L..21Z}. We cannot, however, rule out a GW counterpart that occurred decades or centuries before \frb.

There is no significant X-ray emission detected and no evidence for X-ray variability from this source (see \S\ref{sec:xray}). The 0.5--10.0 keV, 99.7\% upper limits on the unabsorbed flux are of order $10^{-13}$ erg\,cm$^{-2}$\,s$^{-1}$, or $10^{40}$ erg\,s$^{-1}$ assuming a source distance of 40 Mpc, both limits assuming an isotropic emission with $\Gamma = 2 $ power-law spectrum, absorbed by a $N_H = 1.4\times10^{20}$ cm$^{-2}$ neutral hydrogen column. Higher absorptive columns, like those expected if the source was located in a decades-old supernova remnant \citep{2017ApJ...841...14M}, could decrease the constraining power of these X-ray upper limits considerably, 
 but neither our historical SN data nor the JWST observations \citep{Blanchard25} support the presence of very young supernova remnant. The persistent X-ray limits rule out most ultraluminous X-ray sources at the position of \frb \citep{2022A&A...659A.188B}, in line with luminosities that have been previously ruled out for other FRBs \citep{2023ApJ...958...66E, 2025NatAs...9..111P}. This persistent X-ray limit cannot, however, rule out the presence of a Crab-like supernova remnant or pulsar wind nebula, which is about 1000 times less luminous than the X-ray limit \citep{2025NatAs...9..111P}. 

\section{Conclusion}

The VLBI localization capabilities of the newly operational CHIME Outriggers array have enabled $\sim 50-100$ mas localizations of FRBs detected by CHIME/FRB's wide field of view. The continental baselines achieve a precision well matched to the highest-resolution instruments at other wavelengths (e.g., JWST; \citealt{Blanchard25}), allowing for unambiguous host associations and opening the door to spatially-resolved, multi-wavelength studies that will address key questions about the sources of FRBs and maximize the utility of FRBs as cosmological probes.

The full Outrigger array localization of \frb, paired with its proximity, allowed for the 13-parsec precision localization and deep multiwavelength environment study of a non-repeating FRB. 
Rapid MMT and Gemini follow-up observations allowed us to rule out transient optical emission above $M_r \sim -8$ mag nine days after the event. Regular cadence Coddenhiem Observatory and KAIT observations disfavor luminous optical transients like core-collapse and Type Ia SNe which could be consistent with the location of \frb for nearly 25~yr before the radio burst.
The spectral energy of \frb, particularly given our sensitivity toward this line of sight, was found to be inconsistent at a $>3.7$-$\sigma$ level with having arisen from the burst energy distributions of known repeaters. We expect this source to remain the observational limit of certainty that a source is a non-repeater given the rarity of detections this high above CHIME/FRB's fluence threshold. If there is a distinct population of non-repeating FRBs, given this apparent tension with the known repeater population, our study of \frb provides unique insight into the local environment of a source most likely from that non-repeating population.

KCWI data allowed us to probe the gas density, metallicity, nature of gas ionization, dust extinction and star-formation rate in the environment of the FRB via emission line fluxes, and revealed that \frb is offset by $190\pm20$ pc from the center of the most proximal star-formation region. The observed scintillation and scattering timescales additionally place a scattering screen within about 60 pc in front of \frb, and we propose that the scattering screen is likely physically associated with the \halpha/star-forming region seen by KCWI. In this case, the true offset of the FRB from the region's center is similar to its projected offset.

By itself, the lack of strong apparent star formation at the location of \frb, coupled with limits at the FRB position on stars with M $\gtrsim 20 \ M_{\odot}$ \citep{Blanchard25}, could have been interpreted as being in tension with a young (10--100 Myr) compact object like a magnetar. However, \cite{Blanchard25} show that the location of young stars in the vicinity of \frb indicate that the observations of \frb are not inconsistent with a compact object formed {\it in situ}. 

One can imagine needing to consider many possible conclusions about the source of \frb if we only had access to a CHIME/FRB-core-level localization $(\sim$ arcmin) or even a localization using only the first CHIME/FRB Outrigger ($\sim $ tens of arcsec), especially given the recent supernovae in NGC~4141. For example, the local metallicity has implications for comparisons to known transient populations. Yet, the measured local gas-phase metallicity is higher than the average host value and even the metallicity extrapolated to the measured offset of the FRB position. % This highlights the power of precision localizations in population studies used to make inferences on the validity of candidate progenitor(s) of FRBs.

Instead, zooming into the $\sim$13-pc environment tells a much richer story: we find a clean radio environment with relatively small DM contribution for host and local environment compared to other FRBs, slightly sub-solar gas-phase metallicity, and 190$\pm20$ pc offset of the source from the nearest knot of star formation. This highlights the constraining power that large numbers of sub-arcsecond precision localized FRBs, particularly one-off FRBs, will represent in the future. 
The incoming wealth of precisely localized FRBs from the now-operational CHIME Outriggers (\citetalias{outriggers_2025} \citeyear{outriggers_2025}), and eventually also from planned telescopes like the Canadian Hydrogen Observatory and Radio-transient Detector (CHORD), and the Deep Synoptic Array (DSA-2000), will be revolutionary for disentangling multiple populations of FRBs and unveiling their sources \citep{2019MNRAS.489..919K,2019BAAS...51g.255H,2019clrp.2020...28V}. 
\begin{acknowledgements}

We thank Lauren Rhodes for inspiring the RBFLOAT name and for helpful comments that increased the quality of the draft. We also thank Reshma Anna Thomas for inspecting burst candidates from \textit{realfast}.  We are deeply grateful to Keith Gendreau, Zaven Arzoumanian, and Elizabeth Ferrera for promptly scheduling the NICER observations and for their support of our work. This work benefited from early access to the scintillometry methods described in \cite{t2025scintillometryfastradiobursts}, and we are grateful to Sachin Pradeep E. T. and co-authors for sharing it. We also thank Jillian Rastinejad, who kindly triggered the first epoch of the MMT observations under her ToO program.

M.B. is a McWilliams fellow and an International Astronomical Union Gruber fellow. M.B. also receives support from the McWilliams seed grants.
A.M.C. is a Banting Postdoctoral Researcher.
A.P.C. is a Vanier Canada Graduate Scholar.
M.D. is supported by a CRC Chair, Natural Sciences and Engineering Research Council of Canada (NSERC) Discovery Grant, CIFAR, and by the FRQNT Centre de Recherche en Astrophysique du Qu\'ebec (CRAQ).
Y.D. is supported by the National Science Foundation Graduate Research Fellowship under grant No. DGE-2234667.
G.M.E. acknowledges funding from NSERC through Discovery Grant RGPIN-2020-04554. 
W.F. gratefully acknowledges support by the National Science Foundation (NSF) under grants AST-2206494, AST-2308182, and CAREER grant AST-2047919, the David and Lucile Packard Foundation, the Alfred P. Sloan Foundation, and the Research Corporation for Science Advancement through Cottrell Scholar Award \#28284.
E.F. and S.S.P. are supported by the National Science Foundation (NSF) under grant number AST-2407399. %
J.W.T.H. and the AstroFlash research group acknowledge support from a Canada Excellence Research Chair in Transient Astrophysics (CERC-2022-00009); an Advanced Grant from the European Research Council (ERC) under the European Union's Horizon 2020 research and innovation programme (``EuroFlash''; Grant agreement \#101098079); and an NWO-Vici grant (``AstroFlash''; VI.C.192.045).
V.M.K. holds the Lorne Trottier Chair in Astrophysics \& Cosmology, a Distinguished James McGill Professorship, and receives support from an NSERC Discovery grant (RGPIN 228738-13).
C.L. acknowledges support from the Miller Institute for Basic Research in Science at UC Berkeley.
K.W.M. holds the Adam J. Burgasser Chair in Astrophysics and received support from NSF grant AST-2018490.
K.T.M. is supported by a FRQNT Master's Research Scholarship.
J.M.P. acknowledges the support of an NSERC Discovery Grant (RGPIN-2023-05373).
D.M. acknowledges support from the French government under the France 2030 investment plan, as part of the Initiative d'Excellence d'Aix-Marseille Universit\'e -- A*MIDEX (AMX-23-CEI-088).
M.N. is a Fonds de Recherche du Quebec –- Nature et Technologies (FRQNT) postdoctoral fellow.
K.N. is an MIT Kavli Fellow.
A.P. is funded by the NSERC Canada Graduate Scholarships -- Doctoral program.
A.B.P. is a Banting Fellow, a McGill Space Institute~(MSI) Fellow, and a FRQNT postdoctoral fellow.
U.P. is supported by NSERC (funding reference number RGPIN-2019-06770, ALLRP 586559-23), Canadian Institute for Advanced Research (CIFAR), AMD AI Quantum Astro.
Z.P. is supported by an NWO Veni fellowship (VI.Veni.222.295).
M.W.S. acknowledges support from the Trottier Space Institute Fellowship program.
K.R.S. is supported by a Fonds de Recherche du Quebec—Nature et Technologies (FRQNT) Doctoral Research Award.
S.M.R. is a CIFAR Fellow and is supported by the NSF Physics Frontiers Center award 2020265.
P.S. acknowledges the support of an NSERC Discovery Grant (RGPIN-2024-06266).
V.S. is supported by a Fonds de Recherche du Quebec—Nature et Technologies (FRQNT) Doctoral Research Award.

K.S. is supported by the NSF Graduate Research Fellowship Program.
S.S. is supported by the joint Northwestern University and University of Chicago Brinson Fellowship.
C.D.K. gratefully acknowledges support from the NSF through AST-2432037, the HST Guest Observer Program through HST-SNAP-17070 and HST-GO-17706, and from JWST Archival Research through JWST-AR-6241 and JWST-AR-5441.
I.H.S. is supported by an NSERC Discovery Grant and by the Canadian Institute for Advanced Research.
D.C.S. is supported by an NSERC Discovery Grant (RGPIN-2021-03985) and by a Canadian Statistical Sciences Institute (CANSSI) Collaborative Research Team Grant.
F.K. acknowledges support from Onsala Space Observatory for  the  provisioning of its facilities/observational support. The Onsala Space Observatory national research infrastructure is funded through Swedish Research Council grant No 2017-00648.
A.V.F. and W.Z. are grateful for financial support from the Christopher R. Redlich Fund, Gary and Cynthia Bengier, Clark and Sharon Winslow, and Alan Eustace and Kathy Kwan (W.Z. is a Bengier-Winslow-Eustace Specialist in Astronomy).

We acknowledge that CHIME and the \kkonamecaps{} Outrigger (KKO) are built on the traditional, ancestral, and unceded territory of the Syilx Okanagan people. \kkonamecaps{}  is situated on land leased from the Imperial Metals Corporation. We are grateful to the staff of the Dominion Radio Astrophysical Observatory, which is operated by the National Research Council of Canada. CHIME operations are funded by a grant from the NSERC Alliance Program and by support from McGill University, University of British Columbia, and University of Toronto. CHIME/FRB Outriggers are funded by a grant from the Gordon \& Betty Moore Foundation. We are grateful to Robert Kirshner for early support and encouragement of the CHIME/FRB Outriggers Project, and to Du{\v{s}}an Pejakovi\'c of the Moore Foundation for continued support. CHIME was funded by a grant from the Canada Foundation for Innovation (CFI) 2012 Leading Edge Fund (Project 31170) and by contributions from the provinces of British Columbia, Qu\'ebec, and Ontario. The CHIME/FRB Project was funded by a grant from the CFI 2015 Innovation Fund (Project 33213) and by contributions from the provinces of British Columbia and Qu\'ebec, and by the Dunlap Institute for Astronomy and Astrophysics at the University of Toronto. Additional support was provided by CIFAR, the Trottier Space Institute at McGill University, and the University of British Columbia. The CHIME/FRB baseband recording system is funded in part by a CFI John R. Evans Leaders Fund award to IHS. The National Radio Astronomy Observatory is a facility of the NSF operated under cooperative agreement by Associated Universities, Inc. The Fast and Fortunate for FRB Follow-up team acknowledges support from NSF grants AST-1911140, AST-1910471, and AST-2206490. FRB research at WVU is supported by NSF grants AST-2006548 and AST-2018490.

MMT Observatory access was supported by Northwestern University and the Center for Interdisciplinary Exploration and Research in Astrophysics (CIERA). Observations reported here were obtained at the MMT Observatory, a joint facility of the University of Arizona and the Smithsonian Institution. Some of the data presented herein were obtained at the W. M. Keck Observatory, which is operated as a scientific partnership among the California Institute of Technology, the University of California, and the NASA. The Observatory was made possible by the generous financial support of the W. M. Keck Foundation. The authors wish to recognize and acknowledge the very significant cultural role and reverence that the summit of Maunakea has always had within the indigenous Hawaiian community. We are most fortunate to have the opportunity to conduct observations from this mountain.

This work is based in part on observations carried out using the $32$\,m radio telescope operated by the Institute of Astronomy of the Nicolaus Copernicus University in Toru\'n (Poland) and supported by a Polish Ministry of Science and Higher Education SpUB grant. 
This work makes use of data from the Westerbork Synthesis Radio Telescope owned by ASTRON. ASTRON, the Netherlands Institute for Radio Astronomy is an institute of the Dutch Scientific Research Council NWO (Nederlandse Organisatie voor Wetenschappelijk Onderzoek). We thank the Westerbork operators Richard Blaauw, Jurjen Sluman, and Henk Mulders for scheduling and supporting observations.     
The Nan\c{c}ay Radio Observatory is operated by the Paris Observatory, associated with the French Centre National de la Recherche Scientifique (CNRS), and partially supported by the Region Centre in France. We acknowledge financial support from ``Programme National de Cosmologie and Galaxies'' (PNCG), and ``Programme National Hautes Energies'' (PNHE) funded by CNRS/INSU-IN2P3-INP, CEA and CNES, France.

KAIT and its ongoing operation were made possible by donations from Sun Microsystems, Inc., the Hewlett-Packard Company, AutoScope Corporation, Lick Observatory, the NSF, the University of California, the Sylvia \& Jim Katzman Foundation, and the TABASGO Foundation. Research at Lick Observatory is partially supported by a generous gift from Google, Inc. 

The European VLBI Network is a joint facility of independent European, African, Asian, and North American radio astronomy institutes.  Scientific results from data presented in this publication are derived from the following EVN project code: EN006 \& EL071. 

The National Radio Astronomy Observatory is a facility of the National Science Foundation operated under cooperative agreement by Associated Universities, Inc. This work made use of the Swinburne University of Technology software correlator, developed as part of the Australian Major National Research Facilities Programme and operated under licence.
The NICER mission and portions of the NICER science team activities are funded by the National Aeronautics and Space Administration (NASA).

We thank the staff of the GMRT that made these observations possible. GMRT is run by the National Centre for Radio Astrophysics of the Tata Institute of Fundamental Research.

\end{acknowledgements}

\facilities{CHIME(FRB), DRT, EVN, Gemini:North(GMOS), GMRT, KAIT, Keck(KCWI), MMT, NRT, Swift(XRT), VLA, VLBA, WSRT}

%\software{astropy \citep{2013A&A...558A..33A,2018AJ....156..123A,2022ApJ...935..167A}}
\software{photutils \citep{larry_bradley_photutils_2024}}

\newpage

\appendix

\section{CHIME/FRB\textquotesingle s sensitivity and exposure towards FRB 20250316A}
\label{app:sens_exp}
The position of \frb\ transits through all four E--W formed beams in its corresponding beam row at near-peak sensitivity. We define its exposure as the total amount of time that the position of \frb\ was within the FWHM of a CHIME-formed beam evaluated at 600 MHz (the center of the observing band) and both the telescope and real-time pipeline were operational and performing at an acceptable level of sensitivity. 
 Our regular procedure for estimating our sensitivity or fluence threshold for a given source leverages the detection S/N, detection S/N threshold, estimated fluence of a burst from that source, and a primary beam correction. The primary beam correction is dependent on the position of the source, and thus the sensitivity is averaged over the localization uncertainty region of the source weighted by the probability density of each position. This is necessary because the system response can vary significantly over the non-negligible positional uncertainty for bursts where CHIME/FRB captured total intensity only.  

For \frb, there is negligible positional uncertainty, owing to the VLBI localization, but we do not know the true detection S/N given the significant masking of this bright burst (as discussed above). Thus, instead of the regular procedure, we estimate the fluence threshold to be 0.5 Jy ms, which corresponds to the 5th percentile of the 68\% upper limit values of fluence thresholds from CHIME/FRB's Second Catalog of FRBs detected at zenith angles within one degree of \frb's (\citetalias{catIIsubmitted} \citeyear{catIIsubmitted}). We excluded sensitivity thresholds from repeat bursts from this distribution as they have a lower S/N callback threshold in the real-time pipeline. We motivate the choice of the 5th percentile given the negligible positional uncertainty of \frb and its central transit through a formed beam row.  

\section{Continuum Radio Calibration \& Images}
\label{app:crcal}
For both EVN and HSA data, we performed phase-referenced observations, interleaving observations of FRB\,20250316A and  NVSS J121710+583526 on 5 minute intervals: 3.5 minutes on target and 1.5 minutes on the phase calibrator. Every 4 cycles, we interleaved observations of a check source, NVSS J120303+603119  for HSA and  IVS B1212+602 for EVN. For the HSA observations, an intermediate scan was inserted every $\sim20$ minutes to phase-up the VLA (i.e., aligning the phases of the individual VLA dishes). We observed both NVSS J092703+390220 and  NVSS J180045+782804 to serve as fringe finders. The visibilities for both EVN and HSA were reduced using standard procedures in the Astronomical Image Processing System (AIPS; \citealt{greisen_2003}). {\it A-priori} system temperature and gain corrections were applied, followed by deriving instrumental delay corrections using a $1.5$-minute interval on NVSS J092703+390220. The visibilities were manually inspected for RFI and/or periods when an antenna was not on source and subsequently flagged. Ionospheric corrections using the \texttt{vlbatecr} function were applied to the visibilities. Multi-band delay solutions were derived using the phase calibrator on $20$-second intervals with either Effelsberg (EVN observations) or the VLA (HSA observations) as the reference antenna. Bandpass solutions were derived using NVSS J092703+390220. Once calibrated, the visibilities of the phase calibrator were exported for image processing using DIFMAP \citep{1994BAAS...26..987S}. Calibration solutions were improved via self-calibration of the phase calibrator (both phase and amplitude) and the corrections applied to the target visibilities. We validated our calibration techniques by imaging the respective check sources prior to imaging \frb. The final set of calibrated visibilities were exported to search for compact, persistent radio emission. Assuming natural weighting, the resulting EVN and HSA dirty images had root-mean-squares of $9.2~\mu\mathrm{Jy~beam}^{-1}$ and $4.4~\mu\mathrm{Jy~beam}^{-1}$, respectively, in-line with the expected thermal noise sensitivity limit. We detect maximum peaks in the dirty images at the $\sim4.5$-$\sigma$ levels; however, we identify no evidence of fringing and thus deem it unlikely to be of astrophysical origin. We place a 5-$\sigma$ upper limit on any radio flux associated with compact radio emission of $\leq22~\mu\mathrm{Jy}$ and $\leq46~\mu\mathrm{Jy}$ with the HSA and EVN, respectively. 

\startlongtable
\begin{deluxetable}{cccccc}
\tabletypesize{\small}
\tablecaption{Summary of multi-band, multi-epoch continuum radio observations. \label{tab:radiofluxes}}
\tablehead{\colhead{Date} & \colhead{Array$^e$} & \colhead{$\nu^a$} & \colhead{$\sigma^b$} & \colhead{$F_{\nu, \mathrm{NGC~4141}}$$^c$} & \colhead{$L_{\nu,\mathrm{PRS}}$$^d$} \\
\colhead{(UTC)} & \colhead{} & \colhead{(GHz)} & \colhead{$(\mu\mathrm{Jy~beam}^{-1})$} & \colhead{$(\mathrm{mJy})$} & \colhead{$(\mathrm{erg~s^{-1}~Hz^{-1}})$}
}
\startdata  
    2025 Mar. 25 & EVN & $4.9$ & $9.2$ & $-$ & $<8.8\times10^{25}$ \\
    2025 Mar. 30 & uGMRT & $0.65$ & $20$ & $-$ & $\leq1.9\times10^{26}$ \\
    2025 Mar. 31 & uGMRT & $1.4$ & $14$ & $-$ & $\leq1.3\times10^{26}$ \\
    2025 Apr. 01 & HSA & $4.9$ & $4.4$ & $-$ & $\leq4.2\times10^{25}$ \\
    2025 Apr. 04 & VLA & $3.2$ & $13.6$ & $1.31\pm0.07$ & $\leq1.3\times10^{26}$ \\
    2025 Apr. 04 & VLA & $6.1$ & $5.6$ & $-$ & $\leq5.4\times10^{25}$ \\
    2025 Apr. 04 & VLA & $9.9$ & $5.2$ & $-$ & $\leq5.0\times10^{25}$ \\
    2025 Apr. 04 & VLA & $21.8$ & $6.9$ & $-$ & $\leq6.6\times10^{25}$ \\
    2025 May 10 & VLA & $3.2$ & $13.5$ & $1.31\pm0.07$ & $\leq1.3\times10^{26}$ \\
    2025 May 10 & VLA & $6.1$ & $5.8$ & $-$ & $\leq5.6\times10^{25}$ \\
    2025 May 10 & VLA & $9.9$ & $5.6$ & $-$ & $\leq5.4\times10^{25}$ \\
    2025 May 10 & VLA & $21.8$ & $7.8$ & $-$ & $\leq7.5\times10^{25}$ \\
    2025 May 30 & ~~VLA$^f$ & $9.9$ & $2.2$ & $-$ & $\leq2.1\times10^{25}$ \\
\enddata
\hspace{1mm} \\
%\textbf{Notes.} \\
$^a$ Central observing frequency. \\
$^b$ Measured noise (root-mean-square of the residuals). \\
$^c$ Integrated flux density associated with NGC 4141 used to estimate the SFR. Quoted values correspond to the flux density contained within 5-$\sigma$ contours. Uncertainties include both statistical and a flat $5\%$ systematic uncertainty, added in quadrature. \\
$^d$ Limits on any associated persistent radio counterpart. Upper limits on specific luminosity are quoted at the 5-$\sigma$ level assuming a luminosity distance of $40~\mathrm{Mpc}$. \\
$^e$ VLA = Very Large Array (D-configuration). HSA = High Sensitivity Array (VLBA + phased VLA). EVN = European VLBI Network. uGMRT = upgraded Giant Metrewave Radio Telescope. \\
$^f$ VLA observations were performed in C-configuration. 
\end{deluxetable}

\begin{figure}[htbp]
  \centering
  \begin{minipage}[b]{0.48\textwidth}
    \centering
    \includegraphics[width=\textwidth]{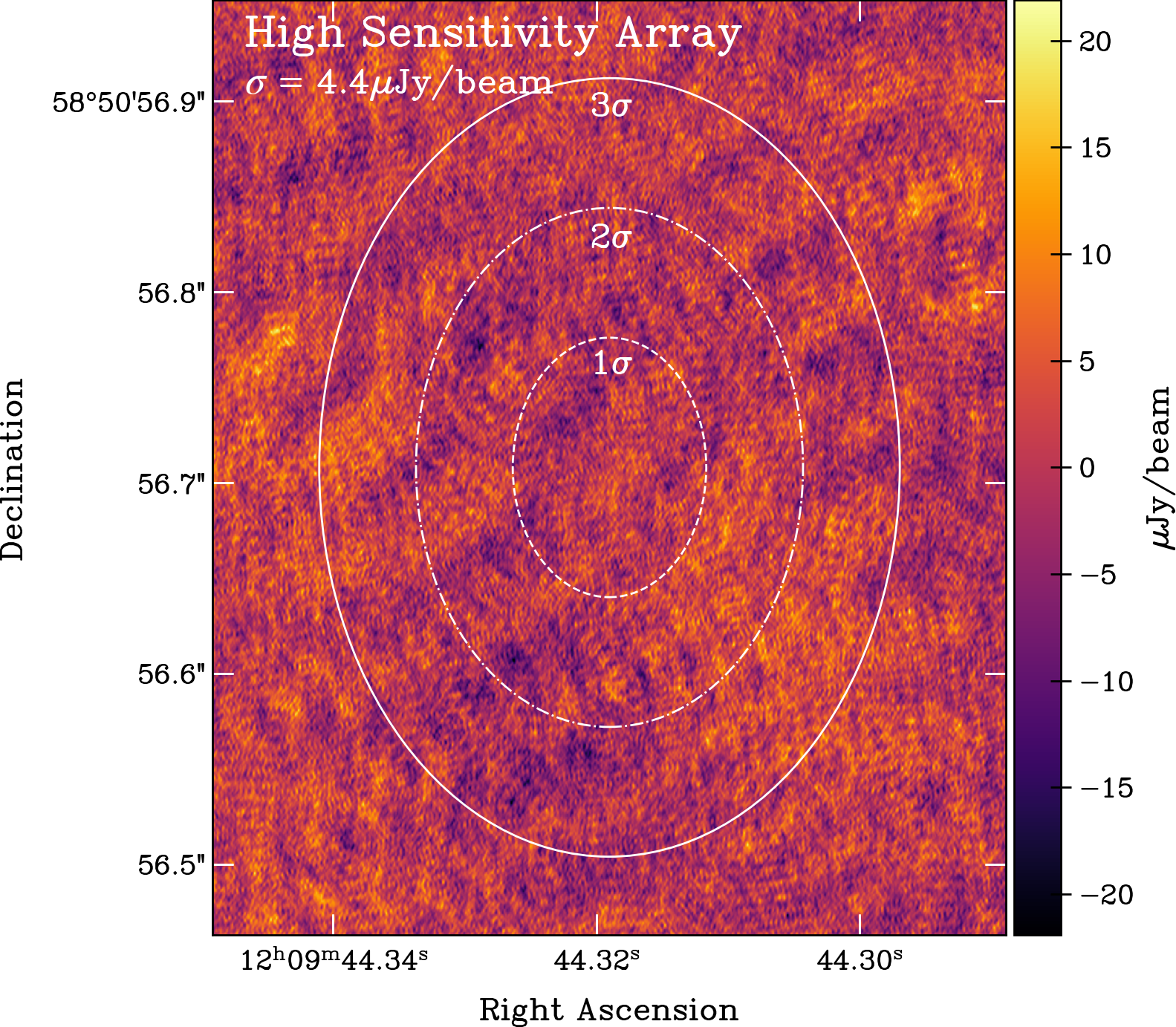}
  \end{minipage}
  \hfill
  \begin{minipage}[b]{0.48\textwidth}
    \centering
    \includegraphics[width=\textwidth]{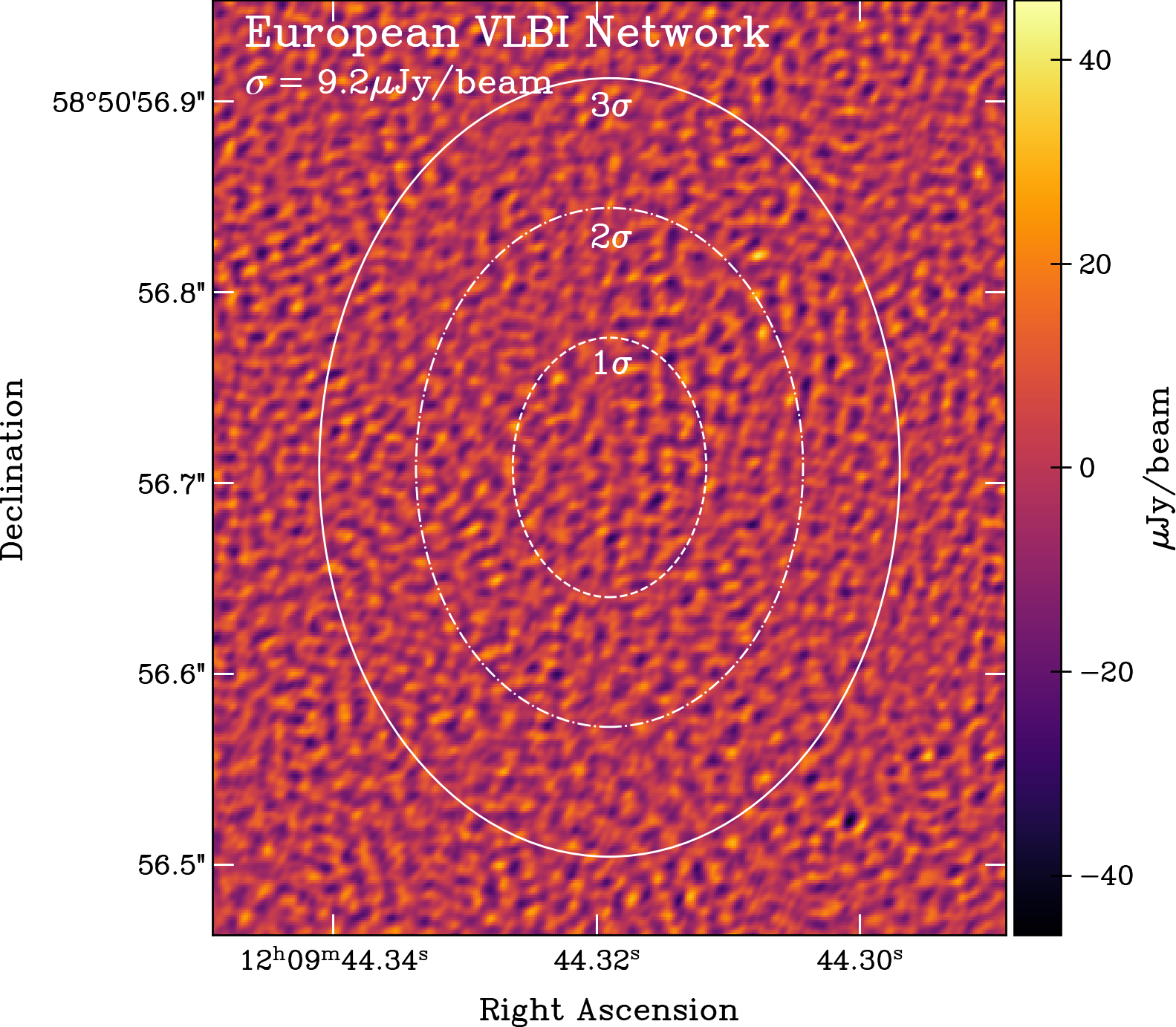}
  \end{minipage}

  \vspace{1em}

  \caption{Residual dirty images of \frb using the High Sensitivity Array (HSA; left) and European VLBI Network (EVN; right) at $4.9~\mathrm{GHz}$. The root-mean-square of each image is $\sigma =4.4~\mu\mathrm{Jy/beam}$ and $\sigma = 9.2~\mu\mathrm{Jy/beam}$, respectively. Images are saturated to $\pm5\sigma$. We overplot the 1-, 2-, and 3-$\sigma$ localization contours of the FRB as white ellipses, the levels indicated on the image. We identify no evidence for compact radio emission in the field of \frb, placing a 5-$\sigma$ upper limit on any PRS emission of $\le 22~\mu\mathrm{Jy}$. }  
  \label{fig:vlbi_dirtyimages}
\end{figure}

\begin{figure}[htbp]
  \centering
  \begin{minipage}[b]{0.42\textwidth}
    \centering
    \includegraphics[width=\textwidth]{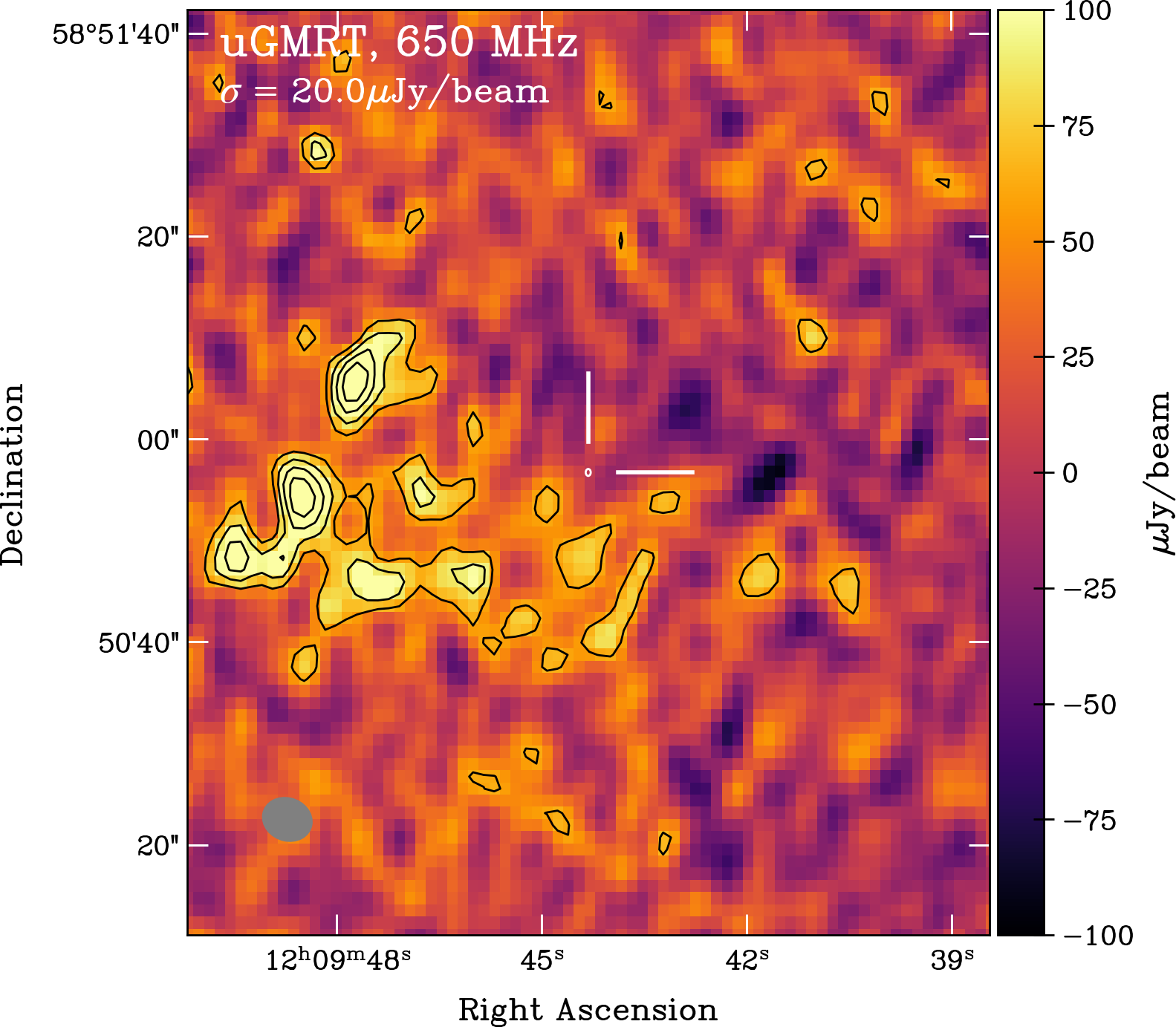}
  \end{minipage}
  \hspace{2mm}
  \begin{minipage}[b]{0.42\textwidth}
    \centering
    \includegraphics[width=\textwidth]{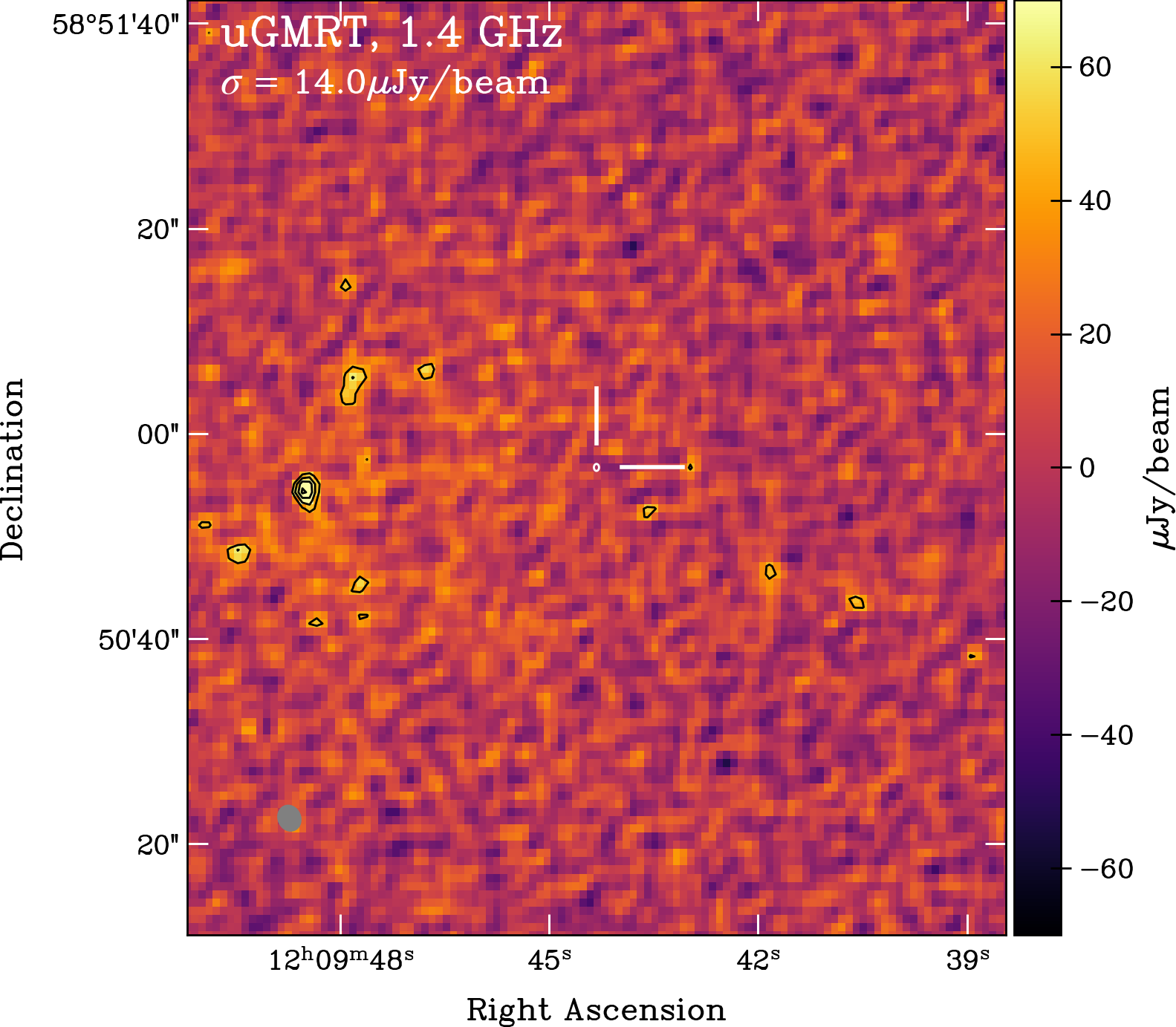}
  \end{minipage}

  \vspace{1em}
  
  \begin{minipage}[b]{0.42\textwidth}
    \centering
    \includegraphics[width=\textwidth]{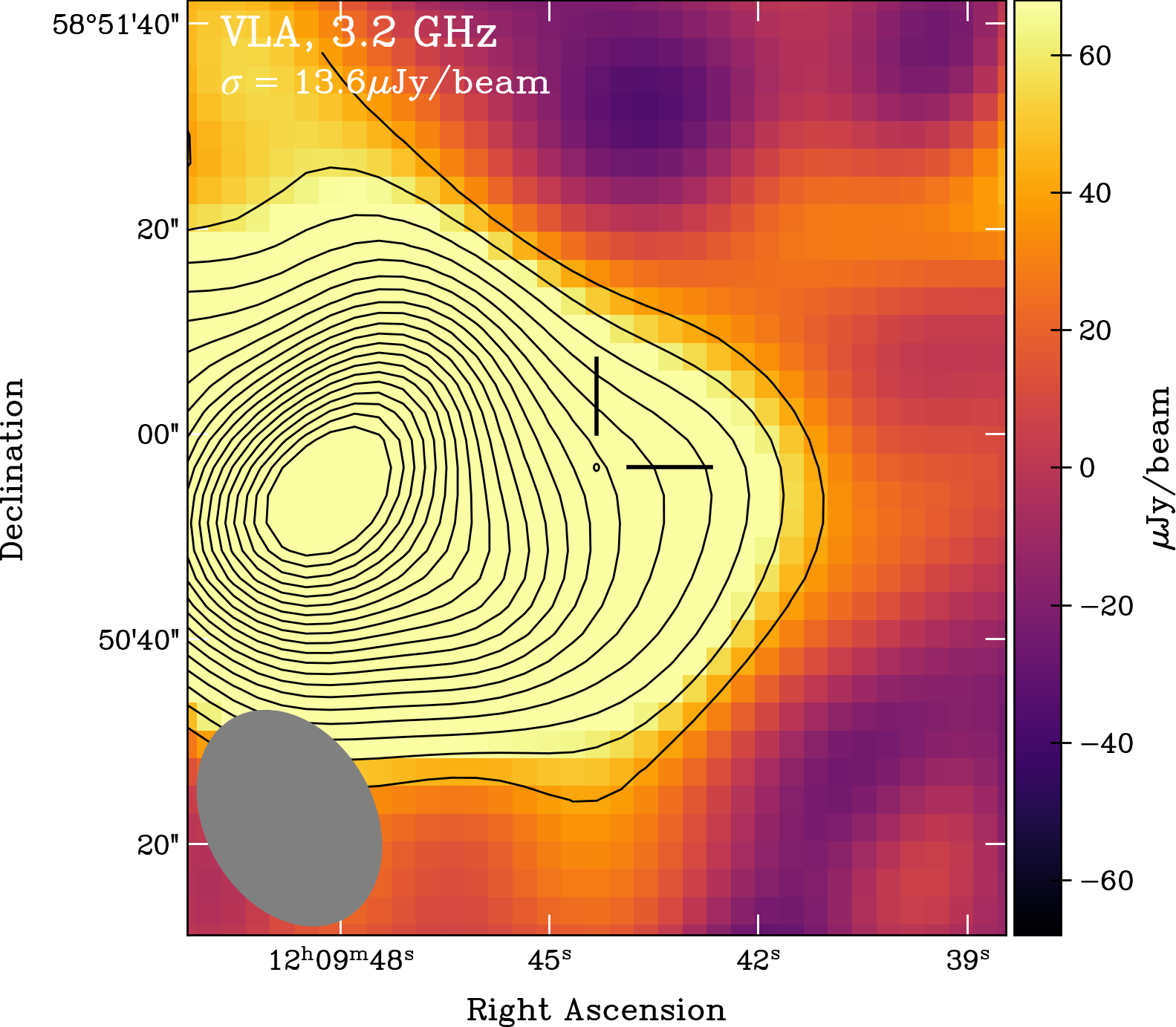}
  \end{minipage}
  \hspace{2mm}
  \begin{minipage}[b]{0.42\textwidth}
    \centering
    \includegraphics[width=\textwidth]{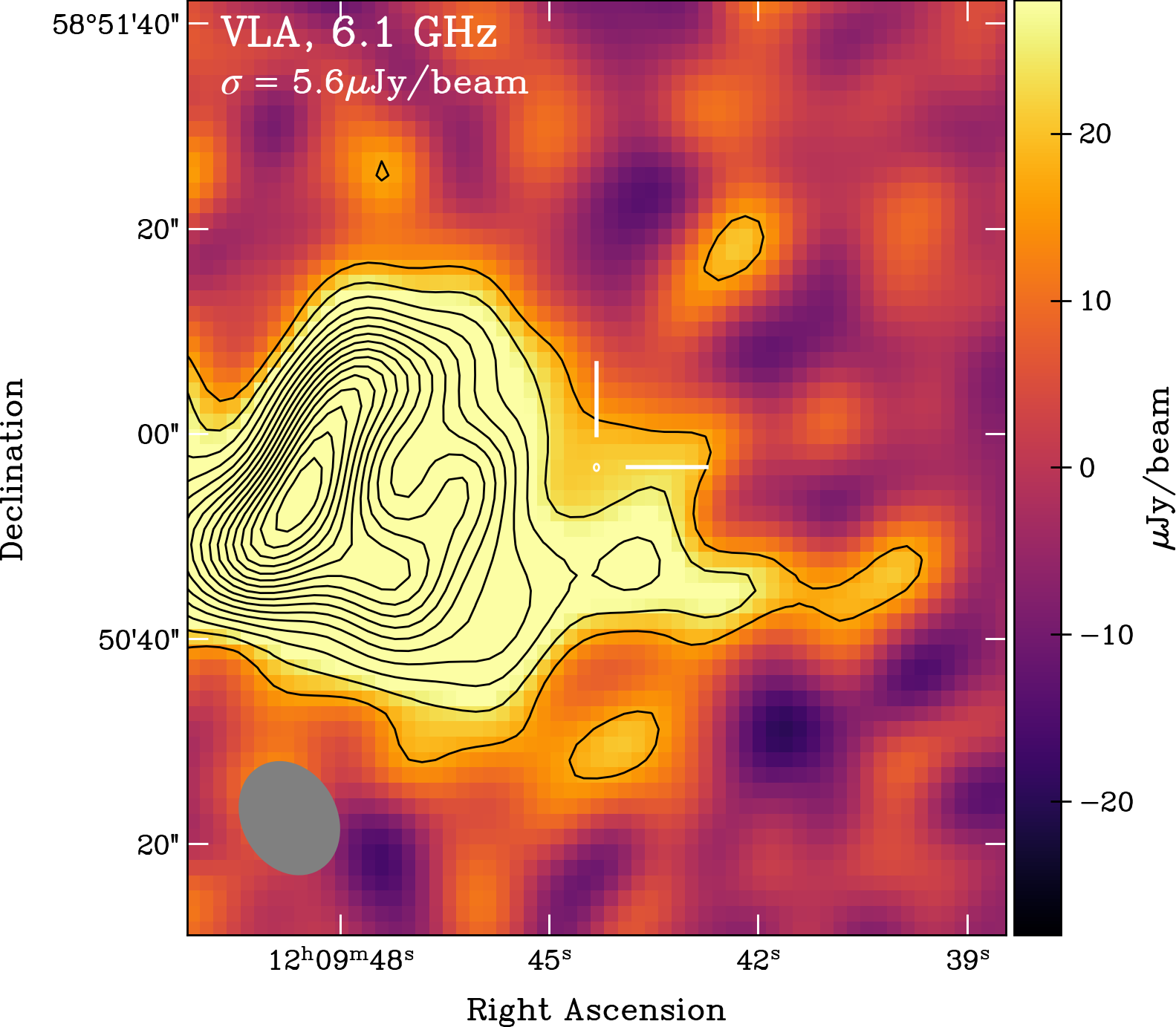}
  \end{minipage}

  \vspace{1em}

  \begin{minipage}[b]{0.42\textwidth}
    \centering
    \includegraphics[width=\textwidth]{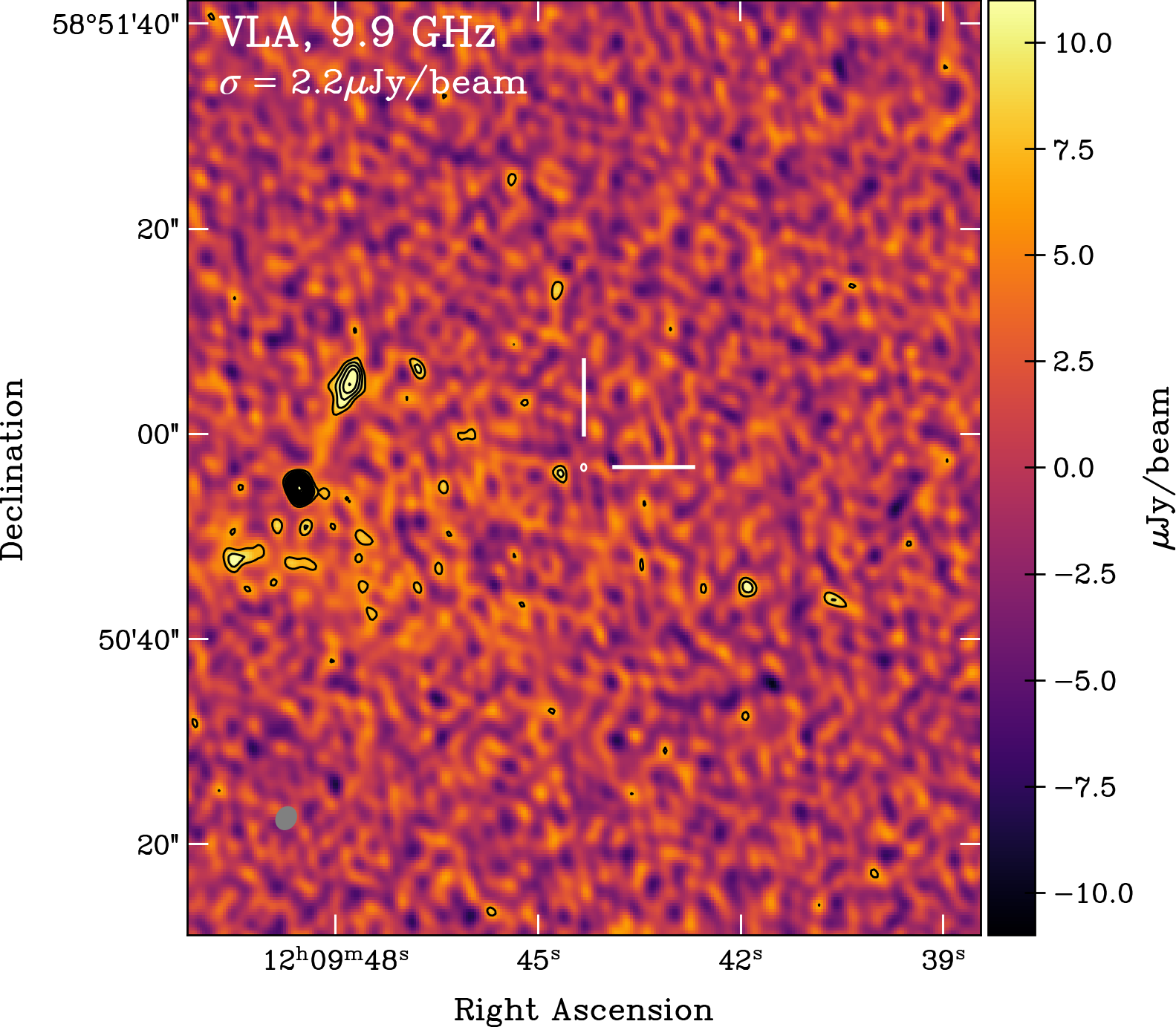}
  \end{minipage}
  \hspace{2mm}
  \begin{minipage}[b]{0.42\textwidth}
    \centering
    \includegraphics[width=\textwidth]{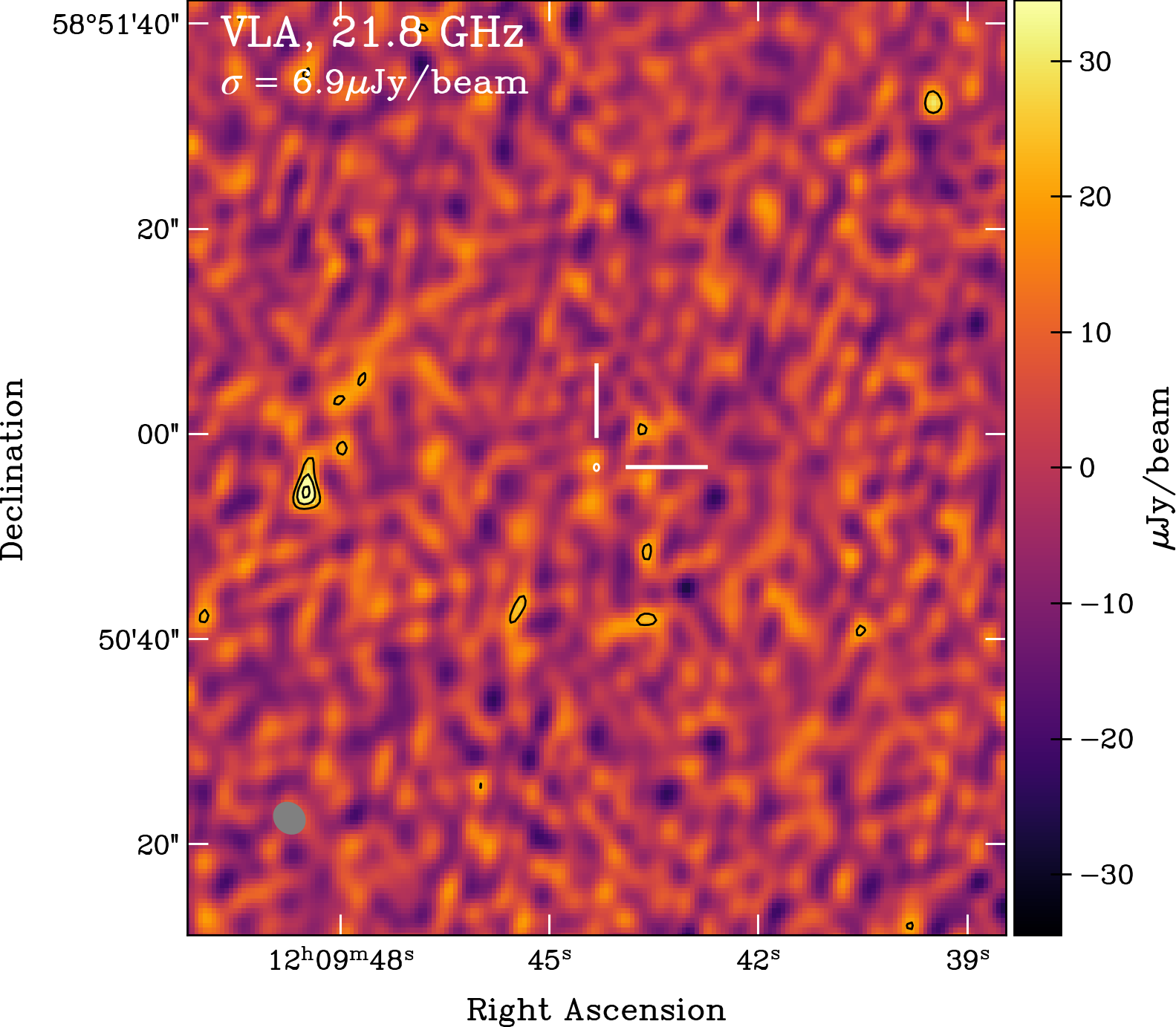}
  \end{minipage}

  \caption{Multi-band radio images of NGC 4141 using the VLA and the uGMRT. The FRB position is indicated by cross-hairs, with the 5-$\sigma$ localization ellipse drawn at the center. Contours begin at $3\sigma$ and increase by factors of $\sqrt{2}$ where $\sigma$ corresponds to the root-mean-square of the residuals. Images are saturated $\pm5\sigma$. The synthesized beam is indicated in the bottom left of each image by a gray ellipse. Host galaxy emission dominates at low angular resolution and low frequencies, gradually fading as the angular resolution increases. }  
  \label{fig:2x2grid}
\end{figure}

\section{Radio Burst Morphology Modeling}
\label{app:morphfit}
We model the burst as two Gaussian components characterized by start times and widths ($t_1,\,t_2,\,w_1,\,w_2$), convolved with two thin-screen scattering kernels characterised by scattering times referenced to $600\,$MHz and power-law scaling indices ($\tau_1,\, \tau_2,\, \alpha_1,\,\alpha_2$). We perform a joint maximum likelihood fit of these parameters using the \texttt{Bilby} implementation of the Dynesty nested-sampling library to minimize the square of the residuals between our model and the observed dynamic spectra at 0.390625\,MHz spectral resolution and 25.6-$\mu s$ temporal resolution \citep{2019ApJS..241...27A,2020MNRAS.493.3132S}. We assume prior independence among the model parameters and specify the joint prior distribution as the product of the following univariate distributions:
\begin{align*}
    \log_{10}(\tau_1/[\text{s}])   &  \sim \mathcal{U}[10^{-5},\, 2 \times 10^{-4}], &  \log_{10}(w_1/[\text{s}])              & \sim \mathcal{U}[10^{-5},\, 5 \times 10^{-5}],  \\
\tau_2  & \sim \mathcal{U}[2.1 \times 10^{-4},\, 5 \times 10^{-3}]\text{ s}, & \log_{10}(w_2/[\text{s}])              & \sim \mathcal{U}[10^{-4},\, 3 \times 10^{-4}],  \\
\alpha_1    & \sim \mathcal{U}[-6,\, -2], & t_1             & \sim \mathcal{U}[1.3,\, 2.0] \text{ ms}, \\
\alpha_2    & \sim \mathcal{U}[-6,\, -2], & t_2              & \sim \mathcal{U}[2.2,\, 3.0] \text{ ms},  \\
\log_{10}(\sigma)             &  \sim \mathcal{U}[5,\, 50],
\end{align*}
where $\mathcal{U}[a,b]$ represents the uniform distribution on closed interval $[a,b]$.

\begin{figure}
    \centering
    \includegraphics[width=\linewidth]{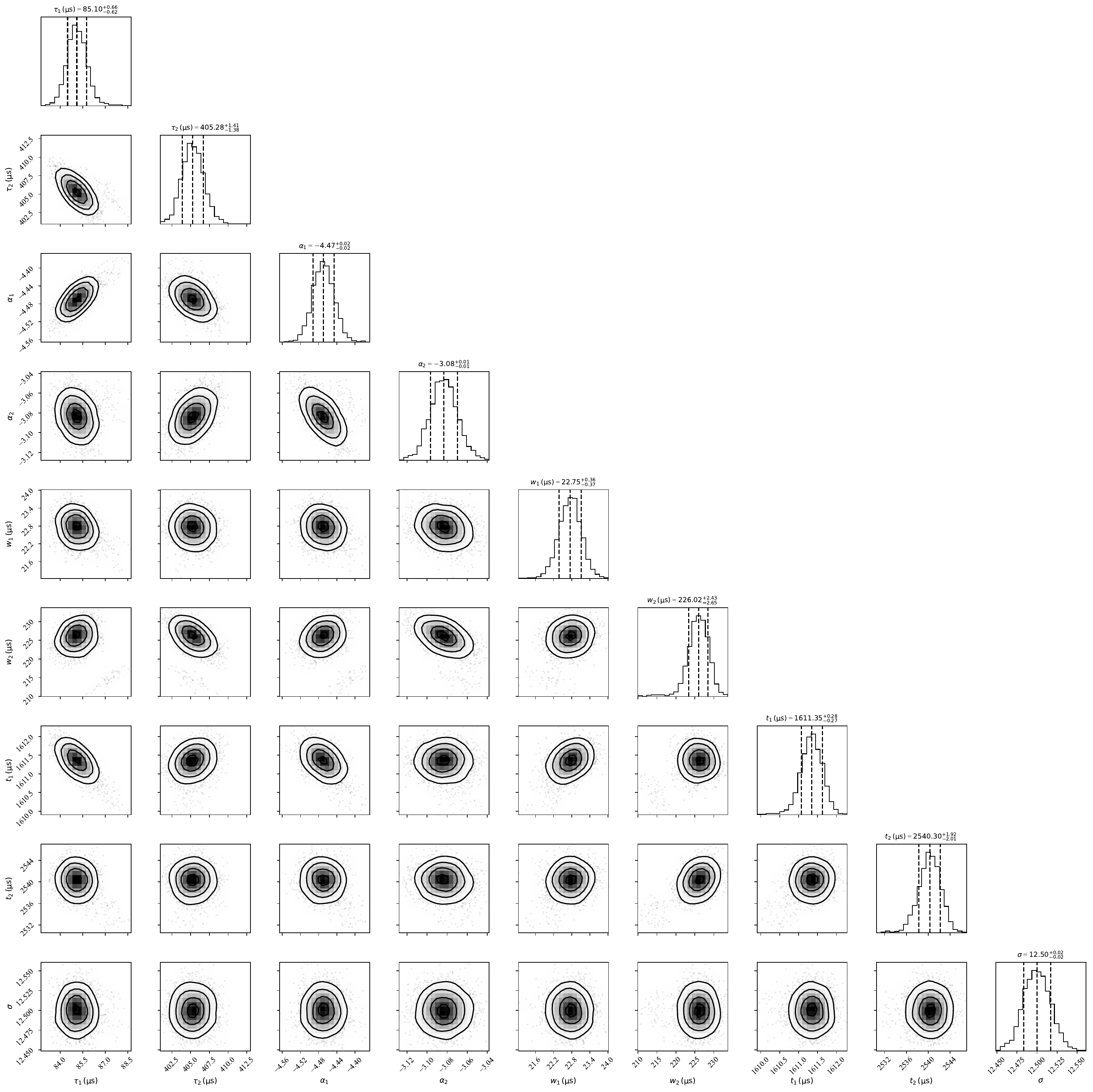}
    \caption{Best-fit solutions and marginalized errors for the two-screen, two-component model described in the text. Gaussianity in each parameter is consistent with a converged solution with some covariance between parameters indicating that our model is mildly degenerate but mutually constrained.}
    \label{fig:morph-corner}
\end{figure}

\begin{figure}
    \centering
    \includegraphics[width=\linewidth]{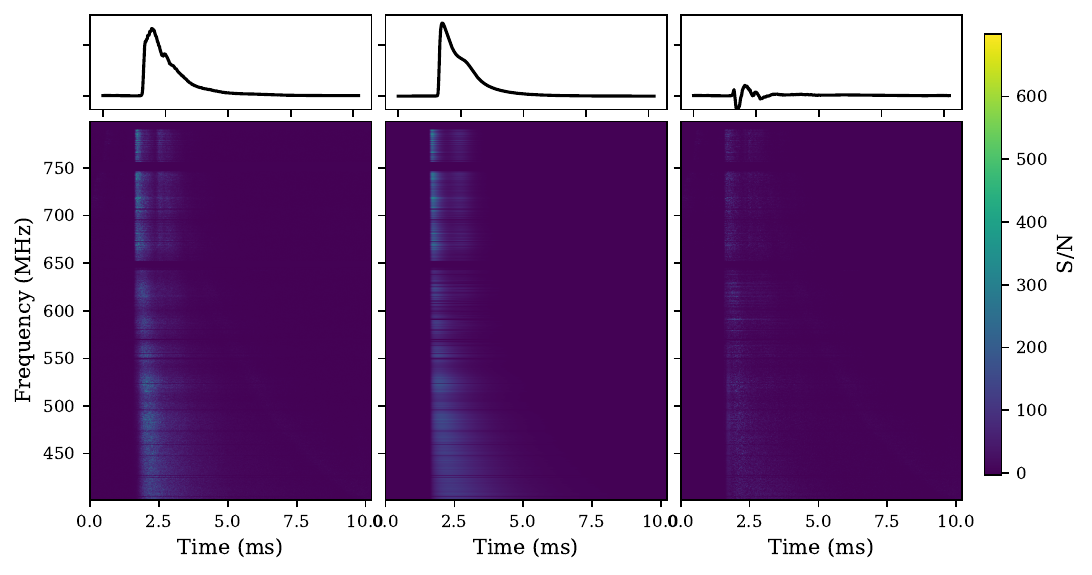}
    \caption{Comparison of the observed dynamic spectra of \frb (\textit{left}) with the best-fit model (\textit{middle}) and squared residuals (\textit{right}).}
    \label{fig:morph-residual}
\end{figure}

To account for the instrumental spectral ripple and any underlying spectral index of the intrinsic components we normalize our model to the bursts time-integrated spectra and thereby fit only for the temporal shape of the burst as a function of frequency and time. Figure \ref{fig:morph-corner} shows the corner plot for our best-fit solution with the associated model and residuals compared against the data shown in Figure \ref{fig:morph-residual}. 
 We do not attempt to evaluate our goodness of fit using a $\chi^2$ as the noise over the burst duration is dominated by physical source noise \citep[the random fluctuations in the intrinsic signal, see][for an in-depth discussion]{Morgan_Ekers_2021} which is not captured by our smooth model.
\begin{figure}
    \centering
    \includegraphics[width=\linewidth]{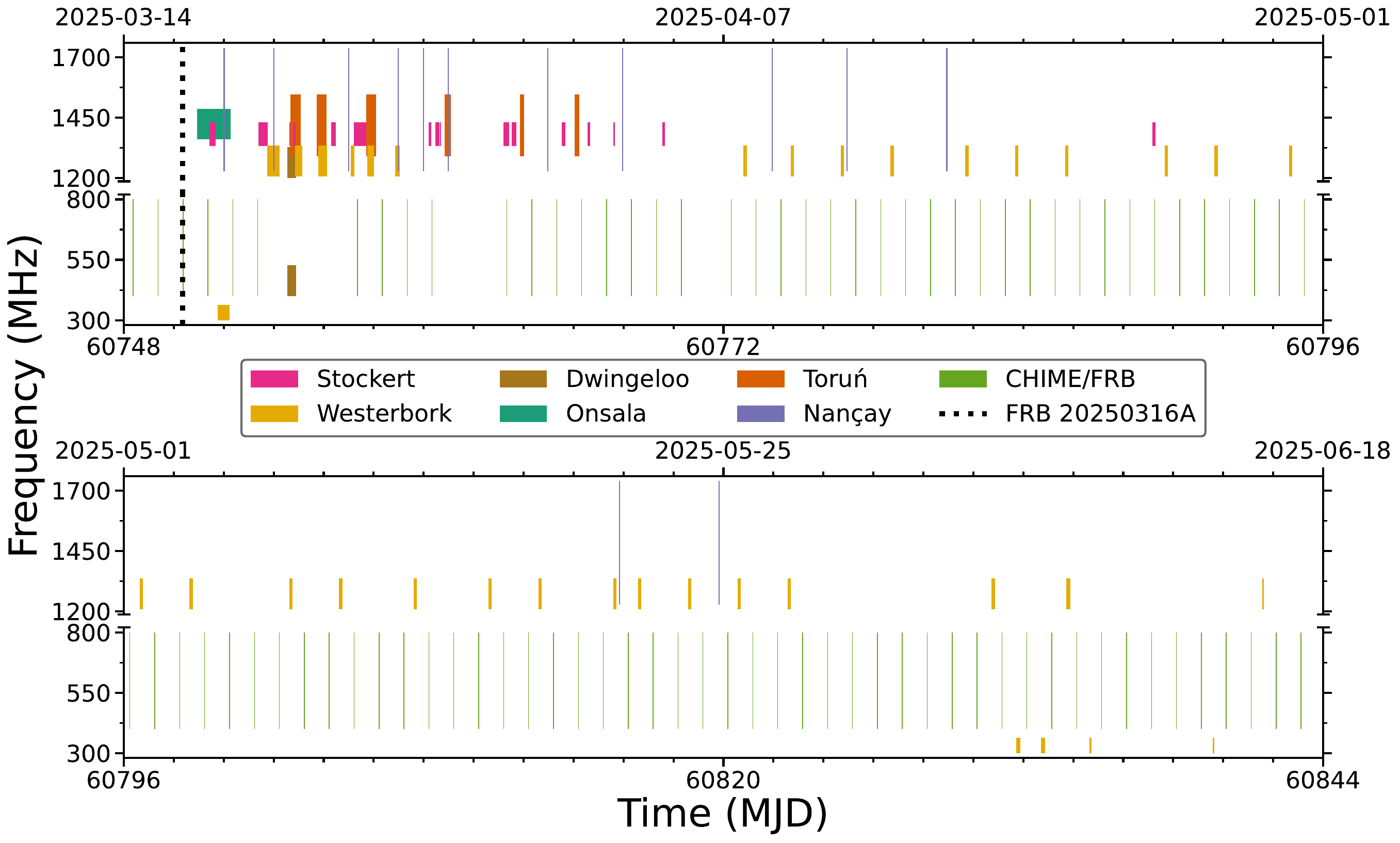}
    \caption{Overview of the \frb observing campaign for repeat bursts by HyperFlash, \'ECLAT, and CHIME/FRB for two months following the discovery of the as-yet non-repeating burst. \frb's time of arrival is denoted with a dotted black vertical line. Each colored block represents an observation with a telescope at a given frequency (y-axis). Prior to the period shown above, CHIME/FRB additionally observed the position of \frb nearly each sidereal day for $\sim 11$ minutes per transit since it commenced operations in 2018.}
    \label{fig:hyperflash}
\end{figure}


\bibliography{bib}{}

\begin{thebibliography}{}
\expandafter\ifx\csname natexlab\endcsname\relax\def\natexlab#1{#1}\fi
\providecommand{\url}[1]{\href{#1}{#1}}
\providecommand{\dodoi}[1]{doi:~\href{http://doi.org/#1}{\nolinkurl{#1}}}
\providecommand{\doeprint}[1]{\href{http://ascl.net/#1}{\nolinkurl{http://ascl.net/#1}}}
\providecommand{\doarXiv}[1]{\href{https://arxiv.org/abs/#1}{\nolinkurl{https://arxiv.org/abs/#1}}}

\bibitem[{T.~C. {Abbott} {et~al.}(2025){Abbott}, {Zwaniga}, {Brar}, {Kaspi}, {Petroff}, {Bhardwaj}, {Boyle}, {Cook}, {Joseph}, {Masui}, {Pandhi}, {Pleunis}, {Scholz}, {Shin}, \& {Tendulkar}}]{2025AJ....169...39A}
{Abbott}, T.~C., {Zwaniga}, A.~V., {Brar}, C., {et~al.} 2025, \bibinfo{title}{{frb-voe: A Real-time Virtual Observatory Event Alert Service for Fast Radio Bursts},} \aj, 169, 39, \dodoi{10.3847/1538-3881/ad9451}

\bibitem[{K. {Aggarwal} {et~al.}(2021){Aggarwal}, {Budav{\'a}ri}, {Deller}, {Eftekhari}, {James}, {Prochaska}, \& {Tendulkar}}]{PATH_2021}
{Aggarwal}, K., {Budav{\'a}ri}, T., {Deller}, A.~T., {et~al.} 2021, \bibinfo{title}{{Probabilistic Association of Transients to their Hosts (PATH)},} \apj, 911, 95, \dodoi{10.3847/1538-4357/abe8d2}

\bibitem[{I. {Andreoni} \& A. {Palmese}(2025){Andreoni} \& {Palmese}}]{Andreoni25}
{Andreoni}, I., \& {Palmese}, A. 2025, \bibinfo{title}{{ZTF Transient Discovery Report for 2025-03-17},} Transient Name Server Discovery Report, 2025-1016, 1

\bibitem[{S. {Andrew} \&  {CHIME/FRB Collaboration}(2025){Andrew} \& {CHIME/FRB Collaboration}}]{ATel17114}
{Andrew}, S., \& {CHIME/FRB Collaboration}. 2025, \bibinfo{title}{{The sub-arcsecond localization of FRB 20250316A using CHIME/FRB Outriggers coincides with reported X-ray counterparts},} The Astronomer's Telegram, 17114, 1

\bibitem[{S. {Andrew} {et~al.}(2025){Andrew}, {Leung}, {Li}, {Masui}, {Andersen}, {Bandura}, {Curtin}, {Kaczmarek}, {Lanman}, {Lazda}, {Mena-Parra}, {Michilli}, {Nimmo}, {Pearlman}, {Rahman}, {Shah}, {Shin}, {Wang}, \& {CHIME/FRB Collaboration}}]{Andrew+2025}
{Andrew}, S., {Leung}, C., {Li}, A., {et~al.} 2025, \bibinfo{title}{{A Very Long Baseline Interferometry Calibrator Grid at 600 MHz for Fast Radio Transient Localizations with CHIME/FRB Outriggers},} \apj, 981, 39, \dodoi{10.3847/1538-4357/adaf8d}

\bibitem[{R. {Anna-Thomas} {et~al.}(2023){Anna-Thomas}, {Connor}, {Dai}, {Feng}, {Burke-Spolaor}, {Beniamini}, {Yang}, {Zhang}, {Aggarwal}, {Law}, {Li}, {Niu}, {Chatterjee}, {Cruces}, {Duan}, {Filipovic}, {Hobbs}, {Lynch}, {Miao}, {Niu}, {Ocker}, {Tsai}, {Wang}, {Xue}, {Yao}, {Yu}, {Zhang}, {Zhang}, {Zhu}, \& {Zhu}}]{2023Sci...380..599A}
{Anna-Thomas}, R., {Connor}, L., {Dai}, S., {et~al.} 2023, \bibinfo{title}{{Magnetic field reversal in the turbulent environment around a repeating fast radio burst},} Science, 380, 599, \dodoi{10.1126/science.abo6526}

\bibitem[{G. Ashton {et~al.}(2019)Ashton {et~al.}}]{bilby_paper}
Ashton, G., {et~al.} 2019, \bibinfo{title}{{BILBY: A user-friendly Bayesian inference library for gravitational-wave astronomy},} Astrophys. J. Suppl., 241, 27, \dodoi{10.3847/1538-4365/ab06fc}

\bibitem[{G. {Ashton} {et~al.}(2019){Ashton}, {H{\"u}bner}, {Lasky}, {Talbot}, {Ackley}, {Biscoveanu}, {Chu}, {Divakarla}, {Easter}, {Goncharov}, {Hernandez Vivanco}, {Harms}, {Lower}, {Meadors}, {Melchor}, {Payne}, {Pitkin}, {Powell}, {Sarin}, {Smith}, \& {Thrane}}]{2019ApJS..241...27A}
{Ashton}, G., {H{\"u}bner}, M., {Lasky}, P.~D., {et~al.} 2019, \bibinfo{title}{{BILBY: A User-friendly Bayesian Inference Library for Gravitational-wave Astronomy},} \apjs, 241, 27, \dodoi{10.3847/1538-4365/ab06fc}

\bibitem[{J.~A. {Baldwin} {et~al.}(1981){Baldwin}, {Phillips}, \& {Terlevich}}]{BPT}
{Baldwin}, J.~A., {Phillips}, M.~M., \& {Terlevich}, R. 1981, \bibinfo{title}{{Classification parameters for the emission-line spectra of extragalactic objects.},} \pasp, 93, 5, \dodoi{10.1086/130766}

\bibitem[{C.~G. {Bassa} {et~al.}(2017){Bassa}, {Tendulkar}, {Adams}, {Maddox}, {Bogdanov}, {Bower}, {Burke-Spolaor}, {Butler}, {Chatterjee}, {Cordes}, {Hessels}, {Kaspi}, {Law}, {Marcote}, {Paragi}, {Ransom}, {Scholz}, {Spitler}, \& {van Langevelde}}]{2017ApJ...843L...8B}
{Bassa}, C.~G., {Tendulkar}, S.~P., {Adams}, E.~A.~K., {et~al.} 2017, \bibinfo{title}{{FRB 121102 Is Coincident with a Star-forming Region in Its Host Galaxy},} \apjl, 843, L8, \dodoi{10.3847/2041-8213/aa7a0c}

\bibitem[{A. Becker(2015)Becker}]{HOTPANTS}
Becker, A. 2015, \bibinfo{title}{{HOTPANTS: High Order Transform of PSF ANd Template Subtraction},}, Astrophysics Source Code Library, record ascl:1504.004 \url{https://ui.adsabs.harvard.edu/abs/2015ascl.soft04004B}

\bibitem[{M.~C.~i. {Bernadich} {et~al.}(2022){Bernadich}, {Schwope}, {Kovlakas}, {Zezas}, \& {Traulsen}}]{2022A&A...659A.188B}
{Bernadich}, M.~C.~i., {Schwope}, A.~D., {Kovlakas}, K., {Zezas}, A., \& {Traulsen}, I. 2022, \bibinfo{title}{{An expanded ultraluminous X-ray source catalogue},} \aap, 659, A188, \dodoi{10.1051/0004-6361/202141560}

\bibitem[{L. {Bernales-Cortes} {et~al.}(2025){Bernales-Cortes}, {Tejos}, {Prochaska}, {Khrykin}, {Marnoch}, {Ryder}, \& {Shannon}}]{Bernales-Cortes+2025}
{Bernales-Cortes}, L., {Tejos}, N., {Prochaska}, J.~X., {et~al.} 2025, \bibinfo{title}{{Empirical estimation of host galaxy dispersion measure toward well-localized fast radio bursts},} \aap, 696, A81, \dodoi{10.1051/0004-6361/202452026}

\bibitem[{E. {Bertin} \& S. {Arnouts}(1996){Bertin} \& {Arnouts}}]{SourceExtractor}
{Bertin}, E., \& {Arnouts}, S. 1996, \bibinfo{title}{{SExtractor: Software for source extraction.},} \aaps, 117, 393, \dodoi{10.1051/aas:1996164}

\bibitem[{S. {Bhandari} {et~al.}(2022){Bhandari}, {Heintz}, {Aggarwal}, {Marnoch}, {Day}, {Sydnor}, {Burke-Spolaor}, {Law}, {Xavier Prochaska}, {Tejos}, {Bannister}, {Butler}, {Deller}, {Ekers}, {Flynn}, {Fong}, {James}, {Lazio}, {Luo}, {Mahony}, {Ryder}, {Sadler}, {Shannon}, {Han}, {Lee}, \& {Zhang}}]{Bhandari+2022}
{Bhandari}, S., {Heintz}, K.~E., {Aggarwal}, K., {et~al.} 2022, \bibinfo{title}{{Characterizing the Fast Radio Burst Host Galaxy Population and its Connection to Transients in the Local and Extragalactic Universe},} \aj, 163, 69, \dodoi{10.3847/1538-3881/ac3aec}

\bibitem[{S. Bhandari {et~al.}(2023)Bhandari, Marcote, Sridhar, Eftekhari, Hessels, Hewitt, Kirsten, Ould-Boukattine, Paragi, \& Snelders}]{Bhandari_2023}
Bhandari, S., Marcote, B., Sridhar, N., {et~al.} 2023, \bibinfo{title}{Constraints on the Persistent Radio Source Associated with FRB 20190520B Using the European VLBI Network,} The Astrophysical Journal Letters, 958, L19, \dodoi{10.3847/2041-8213/ad083f}

\bibitem[{M. {Bhardwaj} {et~al.}(2024{\natexlab{a}}){Bhardwaj}, {Palmese}, {Maga{\~n}a Hernandez}, {D'Emilio}, \& {Morisaki}}]{2024ApJ...977..122B}
{Bhardwaj}, M., {Palmese}, A., {Maga{\~n}a Hernandez}, I., {D'Emilio}, V., \& {Morisaki}, S. 2024{\natexlab{a}}, \bibinfo{title}{{Challenges for Fast Radio Bursts as Multimessenger Sources from Binary Neutron Star Mergers},} \apj, 977, 122, \dodoi{10.3847/1538-4357/ad9023}

\bibitem[{M. {Bhardwaj} {et~al.}(2021){Bhardwaj}, {Gaensler}, {Kaspi}, {Landecker}, {Mckinven}, {Michilli}, {Pleunis}, {Tendulkar}, {Andersen}, {Boyle}, {Cassanelli}, {Chawla}, {Cook}, {Dobbs}, {Fonseca}, {Kaczmarek}, {Leung}, {Masui}, {Mnchmeyer}, {Ng}, {Rafiei-Ravandi}, {Scholz}, {Shin}, {Smith}, {Stairs}, \& {Zwaniga}}]{Bhardwaj+2021}
{Bhardwaj}, M., {Gaensler}, B.~M., {Kaspi}, V.~M., {et~al.} 2021, \bibinfo{title}{{A Nearby Repeating Fast Radio Burst in the Direction of M81},} \apjl, 910, L18, \dodoi{10.3847/2041-8213/abeaa6}

\bibitem[{M. {Bhardwaj} {et~al.}(2024{\natexlab{b}}){Bhardwaj}, {Michilli}, {Kirichenko}, {Modilim}, {Shin}, {Kaspi}, {Andersen}, {Cassanelli}, {Brar}, {Chatterjee}, {Cook}, {Dong}, {Fonseca}, {Gaensler}, {Ibik}, {Kaczmarek}, {Lanman}, {Leung}, {Masui}, {Pandhi}, {Pearlman}, {Petroff}, {Pleunis}, {Prochaska}, {Rafiei-Ravandi}, {Sand}, {Scholz}, \& {Smith}}]{2024ApJ...971L..51B}
{Bhardwaj}, M., {Michilli}, D., {Kirichenko}, A.~Y., {et~al.} 2024{\natexlab{b}}, \bibinfo{title}{{Host Galaxies for Four Nearby CHIME/FRB Sources and the Local Universe FRB Host Galaxy Population},} \apjl, 971, L51, \dodoi{10.3847/2041-8213/ad64d1}

\bibitem[{M. {Bhardwaj} {et~al.}(2025){Bhardwaj}, {Snelders}, {Hessels}, {Gil de Paz}, {Bhandari}, {Marcote}, {Kirichenko}, {Ould-Boukattine}, {Kirsten}, {Bempong-Manful}, {Bezrukovs}, {Bray}, {Buttaccio}, {Corongiu}, {Feiler}, {Gawronski}, {Giroletti}, {Hewitt}, {Lindqvist}, {Maccaferri}, {Moroianu}, {Nimmo}, {Paragi}, {Puchalska}, {Wang}, {Williams-Baldwin}, \& {Yuan}}]{2025arXiv250611915B}
{Bhardwaj}, M., {Snelders}, M.~P., {Hessels}, J.~W.~T., {et~al.} 2025, \bibinfo{title}{{A Hyperactive FRB Pinpointed in an SMC-Like Satellite Host Galaxy},} arXiv e-prints, arXiv:2506.11915, \dodoi{10.48550/arXiv.2506.11915}

\bibitem[{P.~K. {Blanchard} {et~al.}(2025){Blanchard}, {Berger}, {Andrew}, {Aswin}, {Uno}, \& {Kilpatrick}}]{Blanchard25}
{Blanchard}, P.~K., {Berger}, E., {Andrew}, S.~E., {et~al.} 2025, \bibinfo{title}{{},} Submitted to ApJ Letters

\bibitem[{P.~K. {Blanchard} {et~al.}(2024){Blanchard}, {Villar}, {Chornock}, {Laskar}, {Li}, {Leja}, {Pierel}, {Berger}, {Margutti}, {Alexander}, {Barnes}, {Cendes}, {Eftekhari}, {Kasen}, {LeBaron}, {Metzger}, {Muzerolle Page}, {Rest}, {Sears}, {Siegel}, \& {Yadavalli}}]{2024NatAs...8..774B}
{Blanchard}, P.~K., {Villar}, V.~A., {Chornock}, R., {et~al.} 2024, \bibinfo{title}{{JWST detection of a supernova associated with GRB 221009A without an r-process signature},} Nature Astronomy, 8, 774, \dodoi{10.1038/s41550-024-02237-4}

\bibitem[{A. {Bodaghee} {et~al.}(2012){Bodaghee}, {Tomsick}, {Rodriguez}, \& {James}}]{2012ApJ...744..108B}
{Bodaghee}, A., {Tomsick}, J.~A., {Rodriguez}, J., \& {James}, J.~B. 2012, \bibinfo{title}{{Clustering between High-mass X-Ray Binaries and OB Associations in the Milky Way},} \apj, 744, 108, \dodoi{10.1088/0004-637X/744/2/108}

\bibitem[{R.~C. {Bohlin} {et~al.}(2025){Bohlin}, {Deustua}, {Narayan}, {Saha}, {Calamida}, {Gordon}, {Holberg}, {Hubeny}, {Matheson}, \& {Rest}}]{Bohlin+2025}
{Bohlin}, R.~C., {Deustua}, S., {Narayan}, G., {et~al.} 2025, \bibinfo{title}{{Faint White Dwarf Flux Standards: Data and Models},} \aj, 169, 40, \dodoi{10.3847/1538-3881/ad93d8}

\bibitem[{T. {Boles}(2008){Boles}}]{2008CBET.1239....1B}
{Boles}, T. 2008, \bibinfo{title}{{Supernova 2008X in NGC 4141},} Central Bureau Electronic Telegrams, 1239, 1

\bibitem[{T. {Boles}(2009){Boles}}]{2009CBET.1648....1B}
{Boles}, T. 2009, \bibinfo{title}{{Supernova 2009E in NGC 4141},} Central Bureau Electronic Telegrams, 1648, 1

\bibitem[{L. Bradley {et~al.}(2024)Bradley, Sip{\H o}cz, Robitaille, Tollerud, Vin{\'{\i}}cius, Deil, Barbary, Wilson, Busko, Donath, G{\"u}nther, Cara, Lim, Me{\ss}linger, Conseil, Burnett, Bostroem, Droettboom, Bray, Bratholm, Ginsburg, Jamieson, Barentsen, Craig, Morris, Perrin, Rathi, Pascual, \& Georgiev}]{larry_bradley_photutils_2024}
Bradley, L., Sip{\H o}cz, B., Robitaille, T., {et~al.} 2024, \bibinfo{title}{{astropy/photutils: 2.0.2},}, Zenodo \dodoi{10.5281/zenodo.13989456}

\bibitem[{M.~A. {Brentjens} \& A.~G. {de Bruyn}(2005){Brentjens} \& {de Bruyn}}]{brentjens_deBruyn_2005_RM_synth}
{Brentjens}, M.~A., \& {de Bruyn}, A.~G. 2005, \bibinfo{title}{{Faraday rotation measure synthesis},} \aap, 441, 1217, \dodoi{10.1051/0004-6361:20052990}

\bibitem[{G. Bruni {et~al.}(2024)Bruni, Piro, Yang, Quai, Zhang, Palazzi, Nicastro, Feruglio, Tripodi, O'Connor, Gardini, Savaglio, Rossi, Guelbenzu, \& Paladino}]{bruni2024nebularoriginpersistentradio}
Bruni, G., Piro, L., Yang, Y.-P., {et~al.} 2024, \bibinfo{title}{A nebular origin for the persistent radio emission of fast radio bursts,} \doarXiv{2312.15296}

\bibitem[{G. Bruni {et~al.}(2025)Bruni, Piro, Yang, Palazzi, Nicastro, Rossi, Savaglio, Maiorano, \& Zhang}]{Bruni_2025}
Bruni, G., Piro, L., Yang, Y.-P., {et~al.} 2025, \bibinfo{title}{Discovery of a persistent radio source associated with FRB 20240114A,} Astronomy \&; Astrophysics, 695, L12, \dodoi{10.1051/0004-6361/202453233}

\bibitem[{G. {Bruni} {et~al.}(2025){Bruni}, {Piro}, {Thakur}, {Zhang}, {Sun}, {Yuan}, {Liu}, {Wu}, {Nicastro}, \& {Palazzi}}]{2025ATel17120....1B}
{Bruni}, G., {Piro}, L., {Thakur}, A.~L., {et~al.} 2025, \bibinfo{title}{{e-MERLIN follow-up of FRB 20250316A and its candidate X-ray counterpart},} The Astronomer's Telegram, 17120, 1

\bibitem[{B.~J. {Burn}(1966){Burn}}]{burn_1966}
{Burn}, B.~J. 1966, \bibinfo{title}{{On the depolarization of discrete radio sources by Faraday dispersion},} \mnras, 133, 67, \dodoi{10.1093/mnras/133.1.67}

\bibitem[{G. {Chabrier}(2003){Chabrier}}]{Chabrier03}
{Chabrier}, G. 2003, \bibinfo{title}{{Galactic Stellar and Substellar Initial Mass Function},} \pasp, 115, 763, \dodoi{10.1086/376392}

\bibitem[{T.-W. {Chen} {et~al.}(2017){Chen}, {Smartt}, {Yates}, {Nicholl}, {Kr{\"u}hler}, {Schady}, {Dennefeld}, \& {Inserra}}]{2017MNRAS.470.3566C}
{Chen}, T.-W., {Smartt}, S.~J., {Yates}, R.~M., {et~al.} 2017, \bibinfo{title}{{Superluminous supernova progenitors have a half-solar metallicity threshold},} \mnras, 470, 3566, \dodoi{10.1093/mnras/stx1428}

\bibitem[{ {\!\!CHIME Collaboration} {et~al.}(2022){\!\!CHIME Collaboration}, {Amiri}, {Bandura}, {Boskovic}, {Chen}, {Cliche}, {Deng}, {Denman}, {Dobbs}, {Fandino}, {Foreman}, {Halpern}, {Hanna}, {Hill}, {Hinshaw}, {H{\"o}fer}, {Kania}, {Klages}, {Landecker}, {MacEachern}, {Masui}, {Mena-Parra}, {Milutinovic}, {Mirhosseini}, {Newburgh}, {Nitsche}, {Ordog}, {Pen}, {Pinsonneault-Marotte}, {Polzin}, {Reda}, {Renard}, {Shaw}, {Siegel}, {Singh}, {Smegal}, {Tretyakov}, {van Gassen}, {Vanderlinde}, {Wang}, {Wiebe}, {Willis}, \& {Wulf}}]{chimeoverview}
{\!\!CHIME Collaboration}, {Amiri}, M., {Bandura}, K., {et~al.} 2022, \bibinfo{title}{{An Overview of CHIME, the Canadian Hydrogen Intensity Mapping Experiment},} \apjs, 261, 29, \dodoi{10.3847/1538-4365/ac6fd9}

\bibitem[{ {\!\!CHIME/FRB Collaboration} {et~al.}(2018){\!\!CHIME/FRB Collaboration}, {Amiri}, {Bandura}, {Berger}, {Bhardwaj}, {Boyce}, {Boyle}, {Brar}, {Burhanpurkar}, {Chawla}, {Chowdhury}, {Cliche}, {Cranmer}, {Cubranic}, {Deng}, {Denman}, {Dobbs}, {Fandino}, {Fonseca}, {Gaensler}, {Giri}, {Gilbert}, {Good}, {Guliani}, {Halpern}, {Hinshaw}, {H{\"o}fer}, {Josephy}, {Kaspi}, {Landecker}, {Lang}, {Liao}, {Masui}, {Mena-Parra}, {Naidu}, {Newburgh}, {Ng}, {Patel}, {Pen}, {Pinsonneault-Marotte}, {Pleunis}, {Rafiei Ravandi}, {Ransom}, {Renard}, {Scholz}, {Sigurdson}, {Siegel}, {Smith}, {Stairs}, {Tendulkar}, {Vanderlinde}, \& {Wiebe}}]{chimefrboverview}
{\!\!CHIME/FRB Collaboration}, {Amiri}, M., {Bandura}, K., {et~al.} 2018, \bibinfo{title}{{The CHIME Fast Radio Burst Project: System Overview},} \apj, 863, 48, \dodoi{10.3847/1538-4357/aad188}

\bibitem[{ {\!\!CHIME/FRB Collaboration} {et~al.}(2020{\natexlab{a}}){\!\!CHIME/FRB Collaboration}, {Andersen}, {Bandura}, {Bhardwaj}, {Bij}, {Boyce}, {Boyle}, {Brar}, {Cassanelli}, {Chawla}, {Chen}, {Cliche}, {Cook}, {Cubranic}, {Curtin}, {Denman}, {Dobbs}, {Dong}, {Fandino}, {Fonseca}, {Gaensler}, {Giri}, {Good}, {Halpern}, {Hill}, {Hinshaw}, {H{\"o}fer}, {Josephy}, {Kania}, {Kaspi}, {Landecker}, {Leung}, {Li}, {Lin}, {Masui}, {McKinven}, {Mena-Parra}, {Merryfield}, {Meyers}, {Michilli}, {Milutinovic}, {Mirhosseini}, {M{\"u}nchmeyer}, {Naidu}, {Newburgh}, {Ng}, {Patel}, {Pen}, {Pinsonneault-Marotte}, {Pleunis}, {Quine}, {Rafiei-Ravandi}, {Rahman}, {Ransom}, {Renard}, {Sanghavi}, {Scholz}, {Shaw}, {Shin}, {Siegel}, {Singh}, {Smegal}, {Smith}, {Stairs}, {Tan}, {Tendulkar}, {Tretyakov}, {Vanderlinde}, {Wang}, {Wulf}, \& {Zwaniga}}]{CHIME+2020}
{\!\!CHIME/FRB Collaboration}, {Andersen}, B.~C., {Bandura}, K.~M., {et~al.} 2020{\natexlab{a}}, \bibinfo{title}{{A bright millisecond-duration radio burst from a Galactic magnetar},} \nat, 587, 54, \dodoi{10.1038/s41586-020-2863-y}

\bibitem[{ {\!\!CHIME/FRB Collaboration} {et~al.}(2020{\natexlab{b}}){\!\!CHIME/FRB Collaboration}, {Amiri}, {Andersen}, {Bandura}, {Bhardwaj}, {Boyle}, {Brar}, {Chawla}, {Chen}, {Cliche}, {Cubranic}, {Deng}, {Denman}, {Dobbs}, {Dong}, {Fandino}, {Fonseca}, {Gaensler}, {Giri}, {Good}, {Halpern}, {Hessels}, {Hill}, {H{\"o}fer}, {Josephy}, {Kania}, {Karuppusamy}, {Kaspi}, {Keimpema}, {Kirsten}, {Landecker}, {Lang}, {Leung}, {Li}, {Lin}, {Marcote}, {Masui}, {McKinven}, {Mena-Parra}, {Merryfield}, {Michilli}, {Milutinovic}, {Mirhosseini}, {Naidu}, {Newburgh}, {Ng}, {Nimmo}, {Paragi}, {Patel}, {Pen}, {Pinsonneault-Marotte}, {Pleunis}, {Rafiei-Ravandi}, {Rahman}, {Ransom}, {Renard}, {Sanghavi}, {Scholz}, {Shaw}, {Shin}, {Siegel}, {Singh}, {Smegal}, {Smith}, {Stairs}, {Tendulkar}, {Tretyakov}, {Vanderlinde}, {Wang}, {Wang}, {Wulf}, {Yadav}, \& {Zwaniga}}]{2020Natur.582..351C}
{\!\!CHIME/FRB Collaboration}, {Amiri}, M., {Andersen}, B.~C., {et~al.} 2020{\natexlab{b}}, \bibinfo{title}{{Periodic activity from a fast radio burst source},} \nat, 582, 351, \dodoi{10.1038/s41586-020-2398-2}

\bibitem[{ {\!\!CHIME/FRB Collaboration} {et~al.}(2021){\!\!CHIME/FRB Collaboration}, {Amiri}, {Andersen}, {Bandura}, {Berger}, {Bhardwaj}, {Boyce}, {Boyle}, {Brar}, {Breitman}, {Cassanelli}, {Chawla}, {Chen}, {Cliche}, {Cook}, {Cubranic}, {Curtin}, {Deng}, {Dobbs}, {Dong}, {Eadie}, {Fandino}, {Fonseca}, {Gaensler}, {Giri}, {Good}, {Halpern}, {Hill}, {Hinshaw}, {Josephy}, {Kaczmarek}, {Kader}, {Kania}, {Kaspi}, {Landecker}, {Lang}, {Leung}, {Li}, {Lin}, {Masui}, {McKinven}, {Mena-Parra}, {Merryfield}, {Meyers}, {Michilli}, {Milutinovic}, {Mirhosseini}, {M{\"u}nchmeyer}, {Naidu}, {Newburgh}, {Ng}, {Patel}, {Pen}, {Petroff}, {Pinsonneault-Marotte}, {Pleunis}, {Rafiei-Ravandi}, {Rahman}, {Ransom}, {Renard}, {Sanghavi}, {Scholz}, {Shaw}, {Shin}, {Siegel}, {Sikora}, {Singh}, {Smith}, {Stairs}, {Tan}, {Tendulkar}, {Vanderlinde}, {Wang}, {Wulf}, \& {Zwaniga}}]{CHIME+2021}
{\!\!CHIME/FRB Collaboration}, {Amiri}, M., {Andersen}, B.~C., {et~al.} 2021, \bibinfo{title}{{The First CHIME/FRB Fast Radio Burst Catalog},} \apjs, 257, 59, \dodoi{10.3847/1538-4365/ac33ab}

\bibitem[{ {\!\!CHIME/FRB Collaboration} {et~al.}(2023){\!\!CHIME/FRB Collaboration}, {Andersen}, {Bandura}, {Bhardwaj}, {Boyle}, {Brar}, {Cassanelli}, {Chatterjee}, {Chawla}, {Cook}, {Curtin}, {Dobbs}, {Dong}, {Faber}, {Fandino}, {Fonseca}, {Gaensler}, {Giri}, {Herrera-Martin}, {Hill}, {Ibik}, {Josephy}, {Kaczmarek}, {Kader}, {Kaspi}, {Landecker}, {Lanman}, {Lazda}, {Leung}, {Lin}, {Masui}, {McKinven}, {Mena-Parra}, {Meyers}, {Michilli}, {Ng}, {Pandhi}, {Pearlman}, {Pen}, {Petroff}, {Pleunis}, {Rafiei-Ravandi}, {Rahman}, {Ransom}, {Renard}, {Sand}, {Sanghavi}, {Scholz}, {Shah}, {Shin}, {Siegel}, {Smith}, {Stairs}, {Su}, {Tendulkar}, {Vanderlinde}, {Wang}, {Wulf}, \& {Zwaniga}}]{CHIME+2023}
{\!\!CHIME/FRB Collaboration}, {Andersen}, B.~C., {Bandura}, K., {et~al.} 2023, \bibinfo{title}{{CHIME/FRB Discovery of 25 Repeating Fast Radio Burst Sources},} \apj, 947, 83, \dodoi{10.3847/1538-4357/acc6c1}

\bibitem[{ {\!\!CHIME/FRB Collaboration} {et~al.}(2025){\!\!CHIME/FRB Collaboration}, {Amiri}, {Andersen}, {Andrew}, {Bandura}, {Bhardwaj}, {Bhopi}, {Bidula}, {Boyle}, {Brar}, {Carlson}, {Cassanelli}, {Cassity}, {Chatterjee}, {Cliche}, {Curtin}, {Darlinger}, {DeBoer}, {Dobbs}, {Dong}, {Eadie}, {Fonseca}, {Gaensler}, {Gusinskaia}, {Halpern}, {Hendricksen}, {Hessels}, {Joseph}, {Kaczmarek}, {Kaspi}, {Khairy}, {Landecker}, {Lanman}, {Kit Lau}, {Lazda}, {Leung}, {Main}, {Masui}, {Mckinven}, {Mena-Parra}, {Meyers}, {Michilli}, {Milutinovic}, {Nimmo}, {Noble}, {Pandhi}, {Pearlman}, {Peterson}, {Petroff}, {Pleunis}, {Pollak}, {Rafiei-Ravandi}, {Renard}, {Sammons}, {Sand}, {Sanghavi}, {Scholz}, {Shah}, {Shin}, {Siegel}, {Siemion}, {Sievers}, {Smith}, {Spear}, {Stairs}, {Vanderlinde}, {Wang}, {Willis}, \& {Zegmott}}]{outriggers_2025}
{\!\!CHIME/FRB Collaboration}, {Amiri}, M., {Andersen}, B.~C., {et~al.} 2025, \bibinfo{title}{{CHIME/FRB Outriggers: Design Overview},} arXiv e-prints, arXiv:2504.05192, \dodoi{10.48550/arXiv.2504.05192}

\bibitem[{ {\!\!CHIME/FRB Collaboration et al.}(2025){\!\!CHIME/FRB Collaboration et al.}}]{catIIsubmitted}
{\!\!CHIME/FRB Collaboration et al.} 2025, \bibinfo{title}{The Second CHIME/FRB Catalog of Fast Radio Bursts,} Submitted to ApJS

\bibitem[{J. {Choi} {et~al.}(2016){Choi}, {Dotter}, {Conroy}, {Cantiello}, {Paxton}, \& {Johnson}}]{Choi+2016}
{Choi}, J., {Dotter}, A., {Conroy}, C., {et~al.} 2016, \bibinfo{title}{{Mesa Isochrones and Stellar Tracks (MIST). I. Solar-scaled Models},} \apj, 823, 102, \dodoi{10.3847/0004-637X/823/2/102}

\bibitem[{J.~J. {Condon}(1992){Condon}}]{Condon1992}
{Condon}, J.~J. 1992, \bibinfo{title}{{Radio emission from normal galaxies.},} \araa, 30, 575, \dodoi{10.1146/annurev.aa.30.090192.003043}

\bibitem[{L. {Connor} {et~al.}(2025){Connor}, {Athukoralalage}, {Bieryla}, {Lin}, {Miller}, {Oliva}, \& {Patton}}]{2025ATel17091....1C}
{Connor}, L., {Athukoralalage}, W., {Bieryla}, A., {et~al.} 2025, \bibinfo{title}{{Optical spectrum of the host galaxy of ultra-bright fast radio burst source FRB 20250316A},} The Astronomer's Telegram, 17091, 1

\bibitem[{C. {Conroy} {et~al.}(2009){Conroy}, {Gunn}, \& {White}}]{Conroy+2009}
{Conroy}, C., {Gunn}, J.~E., \& {White}, M. 2009, \bibinfo{title}{{The Propagation of Uncertainties in Stellar Population Synthesis Modeling. I. The Relevance of Uncertain Aspects of Stellar Evolution and the Initial Mass Function to the Derived Physical Properties of Galaxies},} \apj, 699, 486, \dodoi{10.1088/0004-637X/699/1/486}

\bibitem[{A.~M. {Cook} {et~al.}(2023){Cook}, {Bhardwaj}, {Gaensler}, {Scholz}, {Eadie}, {Hill}, {Kaspi}, {Masui}, {Curtin}, {Dong}, {Fonseca}, {Herrera-Martin}, {Kaczmarek}, {Lanman}, {Lazda}, {Leung}, {Meyers}, {Michilli}, {Pandhi}, {Pearlman}, {Pleunis}, {Ransom}, {Rahman}, {Sand}, {Shin}, {Smith}, {Stairs}, \& {Stenning}}]{2023ApJ...946...58C}
{Cook}, A.~M., {Bhardwaj}, M., {Gaensler}, B.~M., {et~al.} 2023, \bibinfo{title}{{An FRB Sent Me a DM: Constraining the Electron Column of the Milky Way Halo with Fast Radio Burst Dispersion Measures from CHIME/FRB},} \apj, 946, 58, \dodoi{10.3847/1538-4357/acbbd0}

\bibitem[{D.~O. {Cook} {et~al.}(2023){Cook}, {Mazzarella}, {Helou}, {Alcala}, {Chen}, {Ebert}, {Frayer}, {Kim}, {Lo}, {Madore}, {Ogle}, {Schmitz}, {Singer}, {Terek}, {Valladon}, \& {Wu}}]{Cook+2023_NED}
{Cook}, D.~O., {Mazzarella}, J.~M., {Helou}, G., {et~al.} 2023, \bibinfo{title}{{Completeness of the NASA/IPAC Extragalactic Database (NED) Local Volume Sample},} \apjs, 268, 14, \dodoi{10.3847/1538-4365/acdd06}

\bibitem[{J.~M. {Cordes} \& T.~J.~W. {Lazio}(2002){Cordes} \& {Lazio}}]{ne2001}
{Cordes}, J.~M., \& {Lazio}, T.~J.~W. 2002, \bibinfo{title}{{NE2001.I. A New Model for the Galactic Distribution of Free Electrons and its Fluctuations},} arXiv e-prints, astro, \dodoi{10.48550/arXiv.astro-ph/0207156}

\bibitem[{J.~M. {Cordes} \& M.~A. {McLaughlin}(2003){Cordes} \& {McLaughlin}}]{2003ApJ...596.1142C}
{Cordes}, J.~M., \& {McLaughlin}, M.~A. 2003, \bibinfo{title}{{Searches for Fast Radio Transients},} \apj, 596, 1142, \dodoi{10.1086/378231}

\bibitem[{J.~M. {Cordes} {et~al.}(2016){Cordes}, {Wharton}, {Spitler}, {Chatterjee}, \& {Wasserman}}]{cordes+2016}
{Cordes}, J.~M., {Wharton}, R.~S., {Spitler}, L.~G., {Chatterjee}, S., \& {Wasserman}, I. 2016, \bibinfo{title}{{Radio Wave Propagation and the Provenance of Fast Radio Bursts},} arXiv e-prints, arXiv:1605.05890, \dodoi{10.48550/arXiv.1605.05890}

\bibitem[{A.~P. {Curtin} {et~al.}(2024){Curtin}, {Sand}, {Pleunis}, {Jain}, {Kaspi}, {Michilli}, {Fonseca}, {Shin}, {Nimmo}, {Brar}, {Dong}, {Eadie}, {Gaensler}, {Herrera-Martin}, {Ibik}, {Joseph}, {Kaczmarek}, {Leung}, {Main}, {Masui}, {McKinven}, {Mena-Parra}, {Ng}, {Pandhi}, {Pearlman}, {Rafiei-Ravandi}, {Sammons}, {Scholz}, {Smith}, \& {Stairs}}]{2024arXiv241102870C}
{Curtin}, A.~P., {Sand}, K.~R., {Pleunis}, Z., {et~al.} 2024, \bibinfo{title}{{Morphology of 32 Repeating Fast Radio Burst Sources at Microsecond Time Scales with CHIME/FRB},} arXiv e-prints, arXiv:2411.02870, \dodoi{10.48550/arXiv.2411.02870}

\bibitem[{A.~P. {Curtin} {et~al.}(2025){Curtin}, {Andrew}, {Simha}, {Cai}, {Nimmo}, {Chatterjee}, {Cook}, {Dong}, {Dong}, {Eftekhari}, {Fong}, {Fonseca}, {Hessels}, {Joseph}, {Kaspi}, {Leung}, {Main}, {Masui}, {Mckinven}, {Michilli}, {Ng}, {Pandhi}, {Pearlman}, {Pleunis}, {Sammons}, {Sand}, {Scholz}, {Shah}, {Shin}, \& {Tohuvavohu}}]{curtinsubmitted}
{Curtin}, A.~P., {Andrew}, S., {Simha}, S., {et~al.} 2025, \bibinfo{title}{{Discovery and Localization of the Swift-Observed FRB 20241228A in a Star-forming Host Galaxy},} arXiv e-prints, arXiv:2506.10961, \dodoi{10.48550/arXiv.2506.10961}

\bibitem[{A.~T. Deller {et~al.}(2007)Deller, Tingay, Bailes, \& West}]{Deller_2007}
Deller, A.~T., Tingay, S.~J., Bailes, M., \& West, C. 2007, \bibinfo{title}{DiFX: A Software Correlator for Very Long Baseline Interferometry Using Multiprocessor Computing Environments,} Publications of the Astronomical Society of the Pacific, 119, 318, \dodoi{10.1086/513572}

\bibitem[{G. {Desvignes} {et~al.}(2011){Desvignes}, {Barott}, {Cognard}, {Lespagnol}, \& {Theureau}}]{2011AIPC.1357..349D}
{Desvignes}, G., {Barott}, W.~C., {Cognard}, I., {Lespagnol}, P., \& {Theureau}, G. 2011, in American Institute of Physics Conference Series, Vol. 1357, Radio Pulsars: An Astrophysical Key to Unlock the Secrets of the Universe, ed. M.~{Burgay}, N.~{D'Amico}, P.~{Esposito}, A.~{Pellizzoni}, \& A.~{Possenti} (AIP), 349--350, \dodoi{10.1063/1.3615154}

\bibitem[{A. {Dey} {et~al.}(2019){Dey}, {Schlegel}, {Lang}, {Blum}, {Burleigh}, {Fan}, {Findlay}, {Finkbeiner}, {Herrera}, {Juneau}, {Landriau}, {Levi}, {McGreer}, {Meisner}, {Myers}, {Moustakas}, {Nugent}, {Patej}, {Schlafly}, {Walker}, {Valdes}, {Weaver}, {Y{\`e}che}, {Zou}, {Zhou}, {Abareshi}, {Abbott}, {Abolfathi}, {Aguilera}, {Alam}, {Allen}, {Alvarez}, {Annis}, {Ansarinejad}, {Aubert}, {Beechert}, {Bell}, {BenZvi}, {Beutler}, {Bielby}, {Bolton}, {Brice{\~n}o}, {Buckley-Geer}, {Butler}, {Calamida}, {Carlberg}, {Carter}, {Casas}, {Castander}, {Choi}, {Comparat}, {Cukanovaite}, {Delubac}, {DeVries}, {Dey}, {Dhungana}, {Dickinson}, {Ding}, {Donaldson}, {Duan}, {Duckworth}, {Eftekharzadeh}, {Eisenstein}, {Etourneau}, {Fagrelius}, {Farihi}, {Fitzpatrick}, {Font-Ribera}, {Fulmer}, {G{\"a}nsicke}, {Gaztanaga}, {George}, {Gerdes}, {Gontcho}, {Gorgoni}, {Green}, {Guy}, {Harmer}, {Hernandez}, {Honscheid}, {Huang}, {James}, {Jannuzi}, {Jiang}, {Joyce}, {Karcher}, {Karkar}, {Kehoe}, {Kneib}, {Kueter-Young}, {Lan},
  {Lauer}, {Le Guillou}, {Le Van Suu}, {Lee}, {Lesser}, {Perreault Levasseur}, {Li}, {Mann}, {Marshall}, {Mart{\'\i}nez-V{\'a}zquez}, {Martini}, {du Mas des Bourboux}, {McManus}, {Meier}, {M{\'e}nard}, {Metcalfe}, {Mu{\~n}oz-Guti{\'e}rrez}, {Najita}, {Napier}, {Narayan}, {Newman}, {Nie}, {Nord}, {Norman}, {Olsen}, {Paat}, {Palanque-Delabrouille}, {Peng}, {Poppett}, {Poremba}, {Prakash}, {Rabinowitz}, {Raichoor}, {Rezaie}, {Robertson}, {Roe}, {Ross}, {Ross}, {Rudnick}, {Safonova}, {Saha}, {S{\'a}nchez}, {Savary}, {Schweiker}, {Scott}, {Seo}, {Shan}, {Silva}, {Slepian}, {Soto}, {Sprayberry}, {Staten}, {Stillman}, {Stupak}, {Summers}, {Sien Tie}, {Tirado}, {Vargas-Maga{\~n}a}, {Vivas}, {Wechsler}, {Williams}, {Yang}, {Yang}, {Yapici}, {Zaritsky}, {Zenteno}, {Zhang}, {Zhang}, {Zhou}, \& {Zhou}}]{DECALs_2019}
{Dey}, A., {Schlegel}, D.~J., {Lang}, D., {et~al.} 2019, \bibinfo{title}{{Overview of the DESI Legacy Imaging Surveys},} \aj, 157, 168, \dodoi{10.3847/1538-3881/ab089d}

\bibitem[{H. {Ding} {et~al.}(2020){Ding}, {Deller}, {Lower}, {Flynn}, {Chatterjee}, {Brisken}, {Hurley-Walker}, {Camilo}, {Sarkissian}, \& {Gupta}}]{2020MNRAS.498.3736D}
{Ding}, H., {Deller}, A.~T., {Lower}, M.~E., {et~al.} 2020, \bibinfo{title}{{A magnetar parallax},} \mnras, 498, 3736, \dodoi{10.1093/mnras/staa2531}

\bibitem[{P. {Disberg} \& I. {Mandel}(2025){Disberg} \& {Mandel}}]{2025arXiv250522102D}
{Disberg}, P., \& {Mandel}, I. 2025, \bibinfo{title}{{The Kick Velocity Distribution of Isolated Neutron Stars},} arXiv e-prints, arXiv:2505.22102, \dodoi{10.48550/arXiv.2505.22102}

\bibitem[{Y. {Dong} \&  {CHIME/FRB Collaboration}(2025){Dong} \& {CHIME/FRB Collaboration}}]{ATel17112}
{Dong}, Y., \& {CHIME/FRB Collaboration}. 2025, \bibinfo{title}{{Deep MMT upper limit on transient optical emission in the host galaxy of FRB 20250316A},} The Astronomer's Telegram, 17112, 1

\bibitem[{Y. {Dong} {et~al.}(2024){Dong}, {Eftekhari}, {Fong}, {Deller}, {Mannings}, {Simha}, {Sridhar}, {Rafelski}, {Gordon}, {Bhandari}, {Day}, {Heintz}, {Hessels}, {Leja}, {James}, {Kilpatrick}, {Mahony}, {Marcote}, {Margalit}, {Nimmo}, {Prochaska}, {Escorial}, {Ryder}, {Schroeder}, {Shannon}, \& {Tejos}}]{2024ApJ...961...44D}
{Dong}, Y., {Eftekhari}, T., {Fong}, W.-f., {et~al.} 2024, \bibinfo{title}{{Mapping Obscured Star Formation in the Host Galaxy of FRB 20201124A},} \apj, 961, 44, \dodoi{10.3847/1538-4357/ad0cbd}

\bibitem[{Y. {Dong} {et~al.}(2025){Dong}, {Kilpatrick}, {Fong}, {Curtin}, {Opoku}, {Andersen}, {Cook}, {Eftekhari}, {Fonseca}, {Gaensler}, {Joseph}, {Kaczmarek}, {Kahinga}, {Kaspi}, {Lanman}, {Lazda}, {Leung}, {Masui}, {Michilli}, {Nimmo}, {Pandhi}, {Pearlman}, {Sammons}, {Scholz}, {Shah}, {Shin}, \& {Smith}}]{Dong25}
{Dong}, Y., {Kilpatrick}, C.~D., {Fong}, W., {et~al.} 2025, \bibinfo{title}{{Searching for Historical Extragalactic Optical Transients Associated with Fast Radio Bursts},} arXiv e-prints, arXiv:2506.06420.
\newblock \doarXiv{2506.06420}

\bibitem[{G. Dumas {et~al.}(2011)Dumas, Schinnerer, Tabatabaei, Beck, Velusamy, \& Murphy}]{Dumas2011}
Dumas, G., Schinnerer, E., Tabatabaei, F.~S., {et~al.} 2011, \bibinfo{title}{THE LOCAL RADIO-IR RELATION IN M51,} The Astronomical Journal, 141, 41, \dodoi{10.1088/0004-6256/141/2/41}

\bibitem[{O.~J. Dunn(1961)Dunn}]{bonferroni}
Dunn, O.~J. 1961, \bibinfo{title}{Multiple Comparisons among Means,} Journal of the American Statistical Association, 56, 52, \dodoi{10.1080/01621459.1961.10482090}

\bibitem[{T. {Eftekhari} {et~al.}(2023){Eftekhari}, {Fong}, {Gordon}, {Sridhar}, {Kilpatrick}, {Bhandari}, {Deller}, {Dong}, {Rouco Escorial}, {Heintz}, {Leja}, {Margalit}, {Metzger}, {Pearlman}, {Prochaska}, {Ryder}, {Scholz}, {Shannon}, \& {Tejos}}]{2023ApJ...958...66E}
{Eftekhari}, T., {Fong}, W., {Gordon}, A.~C., {et~al.} 2023, \bibinfo{title}{{An X-Ray Census of Fast Radio Burst Host Galaxies: Constraints on Active Galactic Nuclei and X-Ray Counterparts},} \apj, 958, 66, \dodoi{10.3847/1538-4357/acf843}

\bibitem[{T. {Eftekhari} {et~al.}(2025){Eftekhari}, {Dong}, {Fong}, {Shah}, {Simha}, {Andersen}, {Andrew}, {Bhardwaj}, {Cassanelli}, {Chatterjee}, {Coulter}, {Fonseca}, {Gaensler}, {Gordon}, {Hessels}, {Ibik}, {Joseph}, {Kahinga}, {Kaspi}, {Kharel}, {Kilpatrick}, {Lanman}, {Lazda}, {Leung}, {Liu}, {Mas-Ribas}, {Masui}, {Mckinven}, {Mena-Parra}, {Miller}, {Nimmo}, {Pandhi}, {Patil}, {Pearlman}, {Pleunis}, {Prochaska}, {Rafiei-Ravandi}, {Sammons}, {Scholz}, {Shin}, {Smith}, \& {Stairs}}]{Eftekhari+2025}
{Eftekhari}, T., {Dong}, Y., {Fong}, W., {et~al.} 2025, \bibinfo{title}{{The Massive and Quiescent Elliptical Host Galaxy of the Repeating Fast Radio Burst FRB 20240209A},} \apjl, 979, L22, \dodoi{10.3847/2041-8213/ad9de2}

\bibitem[{A.~V. {Filippenko} {et~al.}(2001){Filippenko}, {Li}, {Treffers}, \& {Modjaz}}]{2001ASPC..246..121F}
{Filippenko}, A.~V., {Li}, W.~D., {Treffers}, R.~R., \& {Modjaz}, M. 2001, in Astronomical Society of the Pacific Conference Series, Vol. 246, IAU Colloq. 183: Small Telescope Astronomy on Global Scales, ed. B.~{Paczynski}, W.-P. {Chen}, \& C.~{Lemme}, 121

\bibitem[{H.~A. {Flewelling} {et~al.}(2020){Flewelling}, {Magnier}, {Chambers}, {Heasley}, {Holmberg}, {Huber}, {Sweeney}, {Waters}, {Calamida}, {Casertano}, {Chen}, {Farrow}, {Hasinger}, {Henderson}, {Long}, {Metcalfe}, {Narayan}, {Nieto-Santisteban}, {Norberg}, {Rest}, {Saglia}, {Szalay}, {Thakar}, {Tonry}, {Valenti}, {Werner}, {White}, {Denneau}, {Draper}, {Hodapp}, {Jedicke}, {Kaiser}, {Kudritzki}, {Price}, {Wainscoat}, {Chastel}, {McLean}, {Postman}, \& {Shiao}}]{Flewelling+2020}
{Flewelling}, H.~A., {Magnier}, E.~A., {Chambers}, K.~C., {et~al.} 2020, \bibinfo{title}{{The Pan-STARRS1 Database and Data Products},} \apjs, 251, 7, \dodoi{10.3847/1538-4365/abb82d}

\bibitem[{W. {Fong} {et~al.}(2015){Fong}, {Berger}, {Margutti}, \& {Zauderer}}]{2015ApJ...815..102F}
{Fong}, W., {Berger}, E., {Margutti}, R., \& {Zauderer}, B.~A. 2015, \bibinfo{title}{{A Decade of Short-duration Gamma-Ray Burst Broadband Afterglows: Energetics, Circumburst Densities, and Jet Opening Angles},} \apj, 815, 102, \dodoi{10.1088/0004-637X/815/2/102}

\bibitem[{E. {Fonseca} {et~al.}(2024){Fonseca}, {Pleunis}, {Breitman}, {Sand}, {Kharel}, {Boyle}, {Brar}, {Giri}, {Kaspi}, {Masui}, {Meyers}, {Patel}, {Scholz}, \& {Smith}}]{2024ApJS..271...49F}
{Fonseca}, E., {Pleunis}, Z., {Breitman}, D., {et~al.} 2024, \bibinfo{title}{{Modeling the Morphology of Fast Radio Bursts and Radio Pulsars with fitburst},} \apjs, 271, 49, \dodoi{10.3847/1538-4365/ad27d6}

\bibitem[{B.~M. {Gaensler} {et~al.}(2008){Gaensler}, {Madsen}, {Chatterjee}, \& {Mao}}]{2008PASA...25..184G}
{Gaensler}, B.~M., {Madsen}, G.~J., {Chatterjee}, S., \& {Mao}, S.~A. 2008, \bibinfo{title}{{The Vertical Structure of Warm Ionised Gas in the Milky Way},} \pasa, 25, 184, \dodoi{10.1071/AS08004}

\bibitem[{ {\!\!Gaia Collaboration} {et~al.}(2023){\!\!Gaia Collaboration}, {Vallenari}, {Brown}, {Prusti}, {de Bruijne}, {Arenou}, {Babusiaux}, {Biermann}, {Creevey}, {Ducourant}, {Evans}, {Eyer}, {Guerra}, {Hutton}, {Jordi}, {Klioner}, {Lammers}, {Lindegren}, {Luri}, {Mignard}, {Panem}, {Pourbaix}, {Randich}, {Sartoretti}, {Soubiran}, {Tanga}, {Walton}, {Bailer-Jones}, {Bastian}, {Drimmel}, {Jansen}, {Katz}, {Lattanzi}, {van Leeuwen}, {Bakker}, {Cacciari}, {Casta{\~n}eda}, {De Angeli}, {Fabricius}, {Fouesneau}, {Fr{\'e}mat}, {Galluccio}, {Guerrier}, {Heiter}, {Masana}, {Messineo}, {Mowlavi}, {Nicolas}, {Nienartowicz}, {Pailler}, {Panuzzo}, {Riclet}, {Roux}, {Seabroke}, {Sordo}, {Th{\'e}venin}, {Gracia-Abril}, {Portell}, {Teyssier}, {Altmann}, {Andrae}, {Audard}, {Bellas-Velidis}, {Benson}, {Berthier}, {Blomme}, {Burgess}, {Busonero}, {Busso}, {C{\'a}novas}, {Carry}, {Cellino}, {Cheek}, {Clementini}, {Damerdji}, {Davidson}, {de Teodoro}, {Nu{\~n}ez Campos}, {Delchambre}, {Dell'Oro}, {Esquej},
  {Fern{\'a}ndez-Hern{\'a}ndez}, {Fraile}, {Garabato}, {Garc{\'\i}a-Lario}, {Gosset}, {Haigron}, {Halbwachs}, {Hambly}, {Harrison}, {Hern{\'a}ndez}, {Hestroffer}, {Hodgkin}, {Holl}, {Jan{\ss}en}, {Jevardat de Fombelle}, {Jordan}, {Krone-Martins}, {Lanzafame}, {L{\"o}ffler}, {Marchal}, {Marrese}, {Moitinho}, {Muinonen}, {Osborne}, {Pancino}, {Pauwels}, {Recio-Blanco}, {Reyl{\'e}}, {Riello}, {Rimoldini}, {Roegiers}, {Rybizki}, {Sarro}, {Siopis}, {Smith}, {Sozzetti}, {Utrilla}, {van Leeuwen}, {Abbas}, {{\'A}brah{\'a}m}, {Abreu Aramburu}, {Aerts}, {Aguado}, {Ajaj}, {Aldea-Montero}, {Altavilla}, {{\'A}lvarez}, {Alves}, {Anders}, {Anderson}, {Anglada Varela}, {Antoja}, {Baines}, {Baker}, {Balaguer-N{\'u}{\~n}ez}, {Balbinot}, {Balog}, {Barache}, {Barbato}, {Barros}, {Barstow}, {Bartolom{\'e}}, {Bassilana}, {Bauchet}, {Becciani}, {Bellazzini}, {Berihuete}, {Bernet}, {Bertone}, {Bianchi}, {Binnenfeld}, {Blanco-Cuaresma}, {Blazere}, {Boch}, {Bombrun}, {Bossini}, {Bouquillon}, {Bragaglia}, {Bramante}, {Breedt},
  {Bressan}, {Brouillet}, {Brugaletta}, {Bucciarelli}, {Burlacu}, {Butkevich}, {Buzzi}, {Caffau}, {Cancelliere}, {Cantat-Gaudin}, {Carballo}, {Carlucci}, {Carnerero}, {Carrasco}, {Casamiquela}, {Castellani}, {Castro-Ginard}, {Chaoul}, {Charlot}, {Chemin}, {Chiaramida}, {Chiavassa}, {Chornay}, {Comoretto}, {Contursi}, {Cooper}, {Cornez}, {Cowell}, {Crifo}, {Cropper}, {Crosta}, {Crowley}, {Dafonte}, {Dapergolas}, {David}, {David}, {de Laverny}, {De Luise}, \& {De March}}]{GaiaDR3}
{\!\!Gaia Collaboration}, {Vallenari}, A., {Brown}, A.~G.~A., {et~al.} 2023, \bibinfo{title}{{Gaia Data Release 3. Summary of the content and survey properties},} \aap, 674, A1, \dodoi{10.1051/0004-6361/202243940}

\bibitem[{D.~W. {Gardenier} {et~al.}(2021){Gardenier}, {Connor}, {van Leeuwen}, {Oostrum}, \& {Petroff}}]{2021A&A...647A..30G}
{Gardenier}, D.~W., {Connor}, L., {van Leeuwen}, J., {Oostrum}, L.~C., \& {Petroff}, E. 2021, \bibinfo{title}{{Synthesising the repeating FRB population using frbpoppy},} \aap, 647, A30, \dodoi{10.1051/0004-6361/202039626}

\bibitem[{W.~W. {Golay} {et~al.}(2025){Golay}, {Berger}, {Kumar}, \& {Hiramatsu}}]{2025ATel17111....1G}
{Golay}, W.~W., {Berger}, E., {Kumar}, H., \& {Hiramatsu}, D. 2025, \bibinfo{title}{{Archival Very Large Array Non-Detection of a Persistent Radio Source associated with FRB 20250316A},} The Astronomer's Telegram, 17111, 1

\bibitem[{D.~C. {Good} {et~al.}(2023){Good}, {Chawla}, {Fonseca}, {Kaspi}, {Meyers}, {Pleunis}, {Sand}, {Scholz}, {Stairs}, \& {Tendulkar}}]{2023ApJ...944...70G}
{Good}, D.~C., {Chawla}, P., {Fonseca}, E., {et~al.} 2023, \bibinfo{title}{{Nondetection of CHIME/Fast Radio Burst Sources with the Arecibo Observatory},} \apj, 944, 70, \dodoi{10.3847/1538-4357/acb139}

\bibitem[{A.~C. {Gordon} {et~al.}(2023){Gordon}, {Fong}, {Kilpatrick}, {Eftekhari}, {Leja}, {Prochaska}, {Nugent}, {Bhandari}, {Blanchard}, {Caleb}, {Day}, {Deller}, {Dong}, {Glowacki}, {Gourdji}, {Mannings}, {Mahoney}, {Marnoch}, {Miller}, {Paterson}, {Rastinejad}, {Ryder}, {Sadler}, {Scott}, {Sears}, {Shannon}, {Simha}, {Stappers}, \& {Tejos}}]{Gordon+2023}
{Gordon}, A.~C., {Fong}, W.-f., {Kilpatrick}, C.~D., {et~al.} 2023, \bibinfo{title}{{The Demographics, Stellar Populations, and Star Formation Histories of Fast Radio Burst Host Galaxies: Implications for the Progenitors},} \apj, 954, 80, \dodoi{10.3847/1538-4357/ace5aa}

\bibitem[{A.~C. {Gordon} {et~al.}(2025){Gordon}, {Fong}, {Deller}, {Marnoch}, {Lim}, {Peng}, {Bannister}, {Bera}, {Bhat}, {Dial}, {Dong}, {Eftekhari}, {Glowacki}, {Gourdji}, {Gupta}, {Jahns-Schindler}, {Jaini}, {Kilpatrick}, {Liu}, {Prochaska}, {Ryder}, {Shannon}, {Simha}, {Tejos}, {Wang}, \& {Wang}}]{Gordon2025}
{Gordon}, A.~C., {Fong}, W.-f., {Deller}, A.~T., {et~al.} 2025, \bibinfo{title}{{Mapping the Spatial Distribution of Fast Radio Bursts within their Host Galaxies},} arXiv e-prints, arXiv:2506.06453, \dodoi{10.48550/arXiv.2506.06453}

\bibitem[{K. Gordon(2024)Gordon}]{gordon_2024}
Gordon, K. 2024, \bibinfo{title}{dust\_extinction,}, \url{https://doi.org/10.5281/zenodo.11235336}

\bibitem[{K.~D. {Gordon} {et~al.}(2023){Gordon}, {Clayton}, {Decleir}, {Fitzpatrick}, {Massa}, {Misselt}, \& {Tollerud}}]{G23}
{Gordon}, K.~D., {Clayton}, G.~C., {Decleir}, M., {et~al.} 2023, \bibinfo{title}{{One Relation for All Wavelengths: The Far-ultraviolet to Mid-infrared Milky Way Spectroscopic R(V)-dependent Dust Extinction Relationship},} \apj, 950, 86, \dodoi{10.3847/1538-4357/accb59}

\bibitem[{J. {Granot} \& R. {Sari}(2002){Granot} \& {Sari}}]{2002ApJ...568..820G}
{Granot}, J., \& {Sari}, R. 2002, \bibinfo{title}{{The Shape of Spectral Breaks in Gamma-Ray Burst Afterglows},} \apj, 568, 820, \dodoi{10.1086/338966}

\bibitem[{J. {Greiner} {et~al.}(2016){Greiner}, {Micha{\l}owski}, {Klose}, {Hunt}, {Gentile}, {Kamphuis}, {Herrero-Illana}, {Wieringa}, {Kr{\"u}hler}, {Schady}, {Elliott}, {Graham}, {Ibar}, {Knust}, {Nicuesa Guelbenzu}, {Palazzi}, {Rossi}, \& {Savaglio}}]{greiner2016}
{Greiner}, J., {Micha{\l}owski}, M.~J., {Klose}, S., {et~al.} 2016, \bibinfo{title}{{Probing dust-obscured star formation in the most massive gamma-ray burst host galaxies},} \aap, 593, A17, \dodoi{10.1051/0004-6361/201628861}

\bibitem[{E.~W. {Greisen}(2003){Greisen}}]{greisen_2003}
{Greisen}, E.~W. 2003, in Astrophysics and Space Science Library, Vol. 285, Information Handling in Astronomy - Historical Vistas, ed. A.~{Heck}, 109, \dodoi{10.1007/0-306-48080-8_7}

\bibitem[{G. {Hallinan} {et~al.}(2019){Hallinan}, {Ravi}, {Weinreb}, {Kocz}, {Huang}, {Woody}, {Lamb}, {D'Addario}, {Catha}, {Law}, {Kulkarni}, {Phinney}, {Eastwood}, {Bouman}, {McLaughlin}, {Ransom}, {Siemens}, {Cordes}, {Lynch}, {Kaplan}, {Brazier}, {Bhatnagar}, {Myers}, {Walter}, \& {Gaensler}}]{2019BAAS...51g.255H}
{Hallinan}, G., {Ravi}, V., {Weinreb}, S., {et~al.} 2019, in Bulletin of the American Astronomical Society, Vol.~51, 255, \dodoi{10.48550/arXiv.1907.07648}

\bibitem[{D.~M. {Hewitt} {et~al.}(2024{\natexlab{a}}){Hewitt}, {Bhardwaj}, {Gordon}, {Kirichenko}, {Nimmo}, {Bhandari}, {Cognard}, {Fong}, {Gil de Paz}, {Gopinath}, {Hessels}, {Kirsten}, {Marcote}, {Bezrukovs}, {Blaauw}, {Bray}, {Buttaccio}, {Cassanelli}, {Chawla}, {Corongiu}, {Deng}, {Didehbani}, {Dong}, {Gawro{\'n}ski}, {Giroletti}, {Guillemot}, {Huang}, {Ivanov}, {Joseph}, {Kaspi}, {Kharinov}, {Lazda}, {Lindqvist}, {Maccaferri}, {Mas-Ribas}, {Masui}, {Mckinven}, {Melnikov}, {Michilli}, {Mikhailov}, {Nugent}, {Ould-Boukattine}, {Paragi}, {Pearlman}, {Pen}, {Pleunis}, {Sand}, {Shah}, {Shin}, {Snelders}, {Venturi}, {Wang}, {Williams-Baldwin}, {Yang}, \& {Yuan}}]{2024ApJ...977L...4H}
{Hewitt}, D.~M., {Bhardwaj}, M., {Gordon}, A.~C., {et~al.} 2024{\natexlab{a}}, \bibinfo{title}{{A Repeating Fast Radio Burst Source in a Low-luminosity Dwarf Galaxy},} \apjl, 977, L4, \dodoi{10.3847/2041-8213/ad8ce1}

\bibitem[{D.~M. {Hewitt} {et~al.}(2024{\natexlab{b}}){Hewitt}, {Bhandari}, {Marcote}, {Hessels}, {Nimmo}, {Kirsten}, {Bach}, {Bezrukovs}, {Bhardwaj}, {Blaauw}, {Bray}, {Buttaccio}, {Corongiu}, {Gawro{\'n}ski}, {Giroletti}, {Keimpema}, {Maccaferri}, {Paragi}, {Trudu}, {Snelders}, {Venturi}, {Wang}, {Williams-Baldwin}, {Wrigley}, {Yang}, \& {Yuan}}]{2024MNRAS.529.1814H}
{Hewitt}, D.~M., {Bhandari}, S., {Marcote}, B., {et~al.} 2024{\natexlab{b}}, \bibinfo{title}{{Milliarcsecond localization of the hyperactive repeating FRB 20220912A},} \mnras, 529, 1814, \dodoi{10.1093/mnras/stae632}

\bibitem[{ {\!\!HI4PI Collaboration} {et~al.}(2016){\!\!HI4PI Collaboration}, {Ben Bekhti}, {Fl{\"o}er}, {Keller}, {Kerp}, {Lenz}, {Winkel}, {Bailin}, {Calabretta}, {Dedes}, {Ford}, {Gibson}, {Haud}, {Janowiecki}, {Kalberla}, {Lockman}, {McClure-Griffiths}, {Murphy}, {Nakanishi}, {Pisano}, \& {Staveley-Smith}}]{2016A&A...594A.116H}
{\!\!HI4PI Collaboration}, {Ben Bekhti}, N., {Fl{\"o}er}, L., {et~al.} 2016, \bibinfo{title}{{HI4PI: A full-sky H I survey based on EBHIS and GASS},} \aap, 594, A116, \dodoi{10.1051/0004-6361/201629178}

\bibitem[{A.~Y.~Q. {Ho} {et~al.}(2023){Ho}, {Perley}, {Gal-Yam}, {Lunnan}, {Sollerman}, {Schulze}, {Das}, {Dobie}, {Yao}, {Fremling}, {Adams}, {Anand}, {Andreoni}, {Bellm}, {Bruch}, {Burdge}, {Castro-Tirado}, {Dahiwale}, {De}, {Dekany}, {Drake}, {Duev}, {Graham}, {Helou}, {Kaplan}, {Karambelkar}, {Kasliwal}, {Kool}, {Kulkarni}, {Mahabal}, {Medford}, {Miller}, {Nordin}, {Ofek}, {Petitpas}, {Riddle}, {Sharma}, {Smith}, {Stewart}, {Taggart}, {Tartaglia}, {Tzanidakis}, \& {Winters}}]{Ho+2023}
{Ho}, A. Y.~Q., {Perley}, D.~A., {Gal-Yam}, A., {et~al.} 2023, \bibinfo{title}{{A Search for Extragalactic Fast Blue Optical Transients in ZTF and the Rate of AT2018cow-like Transients},} \apj, 949, 120, \dodoi{10.3847/1538-4357/acc533}

\bibitem[{G. {Hobbs} {et~al.}(2005){Hobbs}, {Lorimer}, {Lyne}, \& {Kramer}}]{2005MNRAS.360..974H}
{Hobbs}, G., {Lorimer}, D.~R., {Lyne}, A.~G., \& {Kramer}, M. 2005, \bibinfo{title}{{A statistical study of 233 pulsar proper motions},} \mnras, 360, 974, \dodoi{10.1111/j.1365-2966.2005.09087.x}

\bibitem[{A. {Horowicz} \& B. {Margalit}(2025){Horowicz} \& {Margalit}}]{2025arXiv250408038H}
{Horowicz}, A., \& {Margalit}, B. 2025, \bibinfo{title}{{The Host Galaxies of Fast Radio Bursts Track a Combination of Stellar Mass and Star Formation, Similar to Type Ia Supernovae},} arXiv e-prints, arXiv:2504.08038, \dodoi{10.48550/arXiv.2504.08038}

\bibitem[{Y. {Huang} {et~al.}(2025){Huang}, {Lee}, {Libeskind}, {Simha}, {Valade}, \& {Prochaska}}]{Huang+2025}
{Huang}, Y., {Lee}, K.-G., {Libeskind}, N.~I., {et~al.} 2025, \bibinfo{title}{{Modelling the cosmic dispersion measure in the D < 120 Mpc Local Universe},} \mnras, 538, 2785, \dodoi{10.1093/mnras/staf417}

\bibitem[{S. {Hutschenreuter} {et~al.}(2024){Hutschenreuter}, {Haverkorn}, {Frank}, {Raycheva}, \& {En{\ss}lin}}]{2024A&A...690A.314H}
{Hutschenreuter}, S., {Haverkorn}, M., {Frank}, P., {Raycheva}, N.~C., \& {En{\ss}lin}, T.~A. 2024, \bibinfo{title}{{Disentangling the Faraday rotation sky},} \aap, 690, A314, \dodoi{10.1051/0004-6361/202346740}

\bibitem[{W.~V. {Jacobson-Gal{\'a}n} {et~al.}(2025){Jacobson-Gal{\'a}n}, {Dessart}, {Davis}, {Bostroem}, {Kilpatrick}, {Margutti}, {Filippenko}, {Foley}, {Chornock}, {Terreran}, {Hiramatsu}, {Newsome}, {Padilla Gonzalez}, {Pellegrino}, {Howell}, {Anderson}, {Angus}, {Auchettl}, {Brink}, {Cartier}, {Coulter}, {de Boer}, {Drout}, {Earl}, {Ertini}, {Farah}, {Farias}, {Gall}, {Gao}, {Gerlach}, {Guo}, {Haynie}, {Hosseinzadeh}, {Ibik}, {Jha}, {Jones}, {Langeroodi}, {LeBaron}, {Magnier}, {Piro}, {Raimundo}, {Rest}, {Rest}, {Rich}, {Rojas-Bravo}, {Sears}, {Taggart}, {Villar}, {Wainscoat}, {Wang}, {Wasserman}, {Yan}, {Yang}, {Zhang}, \& {Zheng}}]{JacobsonGalan+2025}
{Jacobson-Gal{\'a}n}, W.~V., {Dessart}, L., {Davis}, K.~W., {et~al.} 2025, \bibinfo{title}{{Final Moments III: Explosion Properties and Progenitor Constraints of CSM-Interacting Type II Supernovae},} arXiv e-prints, arXiv:2505.04698, \dodoi{10.48550/arXiv.2505.04698}

\bibitem[{C.~W. {James}(2023){James}}]{2023PASA...40...57J}
{James}, C.~W. 2023, \bibinfo{title}{{Modelling repetition in zDM: A single population of repeating fast radio bursts can explain CHIME data},} \pasa, 40, e057, \dodoi{10.1017/pasa.2023.51}

\bibitem[{R. Kale \& C.~H. Ishwara-Chandra(2020)Kale \& Ishwara-Chandra}]{Kale_2020}
Kale, R., \& Ishwara-Chandra, C.~H. 2020, \bibinfo{title}{CAPTURE: a continuum imaging pipeline for the uGMRT,} Experimental Astronomy, 51, 95–108, \dodoi{10.1007/s10686-020-09677-6}

\bibitem[{V.~M. {Kaspi} \& A.~M. {Beloborodov}(2017){Kaspi} \& {Beloborodov}}]{2017ARA&A..55..261K}
{Kaspi}, V.~M., \& {Beloborodov}, A.~M. 2017, \bibinfo{title}{{Magnetars},} \araa, 55, 261, \dodoi{10.1146/annurev-astro-081915-023329}

\bibitem[{G. {Kauffmann} {et~al.}(2003){Kauffmann}, {Heckman}, {Tremonti}, {Brinchmann}, {Charlot}, {White}, {Ridgway}, {Brinkmann}, {Fukugita}, {Hall}, {Ivezi{\'c}}, {Richards}, \& {Schneider}}]{Kauffmann+2003}
{Kauffmann}, G., {Heckman}, T.~M., {Tremonti}, C., {et~al.} 2003, \bibinfo{title}{{The host galaxies of active galactic nuclei},} \mnras, 346, 1055, \dodoi{10.1111/j.1365-2966.2003.07154.x}

\bibitem[{A. {Keimpema} {et~al.}(2015){Keimpema}, {Kettenis}, {Pogrebenko}, {Campbell}, {Cim{\'o}}, {Duev}, {Eldering}, {Kruithof}, {van Langevelde}, {Marchal}, {Molera Calv{\'e}s}, {Ozdemir}, {Paragi}, {Pidopryhora}, {Szomoru}, \& {Yang}}]{2015ExA....39..259K}
{Keimpema}, A., {Kettenis}, M.~M., {Pogrebenko}, S.~V., {et~al.} 2015, \bibinfo{title}{{The SFXC software correlator for very long baseline interferometry: algorithms and implementation},} Experimental Astronomy, 39, 259, \dodoi{10.1007/s10686-015-9446-1}

\bibitem[{R.~C. {Kennicutt}(1998){Kennicutt}}]{Kennicutt1998}
{Kennicutt}, Jr., R.~C. 1998, \bibinfo{title}{{The Global Schmidt Law in Star-forming Galaxies},} \apj, 498, 541, \dodoi{10.1086/305588}

\bibitem[{L.~J. {Kewley} {et~al.}(2001){Kewley}, {Dopita}, {Sutherland}, {Heisler}, \& {Trevena}}]{Kewley+2001}
{Kewley}, L.~J., {Dopita}, M.~A., {Sutherland}, R.~S., {Heisler}, C.~A., \& {Trevena}, J. 2001, \bibinfo{title}{{Theoretical Modeling of Starburst Galaxies},} \apj, 556, 121, \dodoi{10.1086/321545}

\bibitem[{L.~J. {Kewley} {et~al.}(2005){Kewley}, {Jansen}, \& {Geller}}]{2005PASP..117..227K}
{Kewley}, L.~J., {Jansen}, R.~A., \& {Geller}, M.~J. 2005, \bibinfo{title}{{Aperture Effects on Star Formation Rate, Metallicity, and Reddening},} \pasp, 117, 227, \dodoi{10.1086/428303}

\bibitem[{F. {Kirsten} {et~al.}(2022){Kirsten}, {Marcote}, {Nimmo}, {Hessels}, {Bhardwaj}, {Tendulkar}, {Keimpema}, {Yang}, {Snelders}, {Scholz}, {Pearlman}, {Law}, {Peters}, {Giroletti}, {Paragi}, {Bassa}, {Hewitt}, {Bach}, {Bezrukovs}, {Burgay}, {Buttaccio}, {Conway}, {Corongiu}, {Feiler}, {Forss{\'e}n}, {Gawro{\'n}ski}, {Karuppusamy}, {Kharinov}, {Lindqvist}, {Maccaferri}, {Melnikov}, {Ould-Boukattine}, {Possenti}, {Surcis}, {Wang}, {Yuan}, {Aggarwal}, {Anna-Thomas}, {Bower}, {Blaauw}, {Burke-Spolaor}, {Cassanelli}, {Clarke}, {Fonseca}, {Gaensler}, {Gopinath}, {Kaspi}, {Kassim}, {Lazio}, {Leung}, {Li}, {Lin}, {Masui}, {Mckinven}, {Michilli}, {Mikhailov}, {Ng}, {Orbidans}, {Pen}, {Petroff}, {Rahman}, {Ransom}, {Shin}, {Smith}, {Stairs}, \& {Vlemmings}}]{Kirsten+2022}
{Kirsten}, F., {Marcote}, B., {Nimmo}, K., {et~al.} 2022, \bibinfo{title}{{A repeating fast radio burst source in a globular cluster},} \nat, 602, 585, \dodoi{10.1038/s41586-021-04354-w}

\bibitem[{F. {Kirsten} {et~al.}(2024){Kirsten}, {Ould-Boukattine}, {Herrmann}, {Gawro{\'n}ski}, {Hessels}, {Lu}, {Snelders}, {Chawla}, {Yang}, {Blaauw}, {Nimmo}, {Puchalska}, {Wolak}, \& {van Ruiten}}]{2024NatAs...8..337K}
{Kirsten}, F., {Ould-Boukattine}, O.~S., {Herrmann}, W., {et~al.} 2024, \bibinfo{title}{{A link between repeating and non-repeating fast radio bursts through their energy distributions},} Nature Astronomy, 8, 337, \dodoi{10.1038/s41550-023-02153-z}

\bibitem[{J. {Kocz} {et~al.}(2019){Kocz}, {Ravi}, {Catha}, {D'Addario}, {Hallinan}, {Hobbs}, {Kulkarni}, {Shi}, {Vedantham}, {Weinreb}, \& {Woody}}]{2019MNRAS.489..919K}
{Kocz}, J., {Ravi}, V., {Catha}, M., {et~al.} 2019, \bibinfo{title}{{DSA-10: a prototype array for localizing fast radio bursts},} \mnras, 489, 919, \dodoi{10.1093/mnras/stz2219}

\bibitem[{D.~C. {Konijn} {et~al.}(2024){Konijn}, {Hewitt}, {Hessels}, {Cognard}, {Huang}, {Ould-Boukattine}, {Chawla}, {Nimmo}, {Snelders}, {Gopinath}, \& {Manaswini}}]{2024MNRAS.534.3331K}
{Konijn}, D.~C., {Hewitt}, D.~M., {Hessels}, J. W.~T., {et~al.} 2024, \bibinfo{title}{{A Nan{\c{c}}ay Radio Telescope study of the hyperactive repeating FRB 20220912A},} \mnras, 534, 3331, \dodoi{10.1093/mnras/stae2296}

\bibitem[{R. {Kothes} {et~al.}(2018){Kothes}, {Sun}, {Gaensler}, \& {Reich}}]{2018ApJ...852...54K}
{Kothes}, R., {Sun}, X., {Gaensler}, B., \& {Reich}, W. 2018, \bibinfo{title}{{A Radio Continuum and Polarization Study of SNR G57.2+0.8 Associated with Magnetar SGR 1935+2154},} \apj, 852, 54, \dodoi{10.3847/1538-4357/aa9e89}

\bibitem[{A.~E. {Lanman} {et~al.}(2022){Lanman}, {Andersen}, {Chawla}, {Josephy}, {Noble}, {Kaspi}, {Bandura}, {Bhardwaj}, {Boyle}, {Brar}, {Breitman}, {Cassanelli}, {Dong}, {Fonseca}, {Gaensler}, {Good}, {Kaczmarek}, {Leung}, {Masui}, {Meyers}, {Ng}, {Patel}, {Pearlman}, {Petroff}, {Pleunis}, {Rafiei-Ravandi}, {Rahman}, {Sanghavi}, {Scholz}, {Shin}, {Stairs}, {Tendulkar}, \& {Zwaniga}}]{2022ApJ...927...59L}
{Lanman}, A.~E., {Andersen}, B.~C., {Chawla}, P., {et~al.} 2022, \bibinfo{title}{{A Sudden Period of High Activity from Repeating Fast Radio Burst 20201124A},} \apj, 927, 59, \dodoi{10.3847/1538-4357/ac4bc7}

\bibitem[{A.~E. {Lanman} {et~al.}(2024){Lanman}, {Andrew}, {Lazda}, {Shah}, {Amiri}, {Balasubramanian}, {Bandura}, {Boyle}, {Brar}, {Carlson}, {Cliche}, {Gusinskaia}, {Hendricksen}, {Kaczmarek}, {Landecker}, {Leung}, {Mckinven}, {Mena-Parra}, {Milutinovic}, {Nimmo}, {Pearlman}, {Renard}, {Rahman}, {Shaw}, {Siegel}, {Smegal}, {Cassanelli}, {Chatterjee}, {Curtin}, {Dobbs}, {Dong}, {Halpern}, {Hopkins}, {Kaspi}, {Khairy}, {Masui}, {Meyers}, {Michilli}, {Petroff}, {Pinsonneault-Marotte}, {Pleunis}, {Rafiei-Ravandi}, {Shin}, {Smith}, {Vanderlinde}, \& {Zegmott}}]{lanman_kko}
{Lanman}, A.~E., {Andrew}, S., {Lazda}, M., {et~al.} 2024, \bibinfo{title}{{CHIME/FRB Outriggers: KKO Station System and Commissioning Results},} arXiv e-prints, arXiv:2402.07898, \dodoi{10.48550/arXiv.2402.07898}

\bibitem[{C.~J. {Law} {et~al.}(2024){Law}, {Sharma}, {Ravi}, {Chen}, {Catha}, {Connor}, {Faber}, {Hallinan}, {Harnach}, {Hellbourg}, {Hobbs}, {Hodge}, {Hodges}, {Lamb}, {Rasmussen}, {Sherman}, {Shi}, {Simard}, {Squillace}, {Weinreb}, {Woody}, \& {Yurk}}]{2024ApJ...967...29L}
{Law}, C.~J., {Sharma}, K., {Ravi}, V., {et~al.} 2024, \bibinfo{title}{{Deep Synoptic Array Science: First FRB and Host Galaxy Catalog},} \apj, 967, 29, \dodoi{10.3847/1538-4357/ad3736}

\bibitem[{C. {Leung} \&  {CHIME/FRB Collaboration}(2025){Leung} \& {CHIME/FRB Collaboration}}]{ATel17086}
{Leung}, C., \& {CHIME/FRB Collaboration}. 2025, \bibinfo{title}{{FRB 20250316A is likely associated with NGC 4141, but is not spatially coincident with either SN 2008X or SN 2009E},} The Astronomer's Telegram, 17086, 1

\bibitem[{C. {Leung} {et~al.}(2024){Leung}, {Andrew}, {Masui}, {Brar}, {Cassanelli}, {Chatterjee}, {Kaspi}, {Khairy}, {Lanman}, {Lazda}, {Mena-Parra}, {Noble}, {Pearlman}, {Rahman}, {Sanghavi}, \& {Shah}}]{Leung_2024}
{Leung}, C., {Andrew}, S., {Masui}, K.~W., {et~al.} 2024, \bibinfo{title}{{A VLBI Software Correlator for Fast Radio Transients},} arXiv e-prints, arXiv:2403.05631, \dodoi{10.48550/arXiv.2403.05631}

\bibitem[{X.~H. {Li} \& J.~L. {Han}(2003){Li} \& {Han}}]{2003A&A...410..253L}
{Li}, X.~H., \& {Han}, J.~L. 2003, \bibinfo{title}{{The effect of scattering on pulsar polarization angle},} \aap, 410, 253, \dodoi{10.1051/0004-6361:20031190}

\bibitem[{H.-H. {Lin} {et~al.}(2024){Lin}, {Scholz}, {Ng}, {Pen}, {Bhardwaj}, {Chawla}, {Curtin}, {Li}, {Newburgh}, {Reda}, {Sand}, {Tendulkar}, {Andersen}, {Bandura}, {Brar}, {Cassanelli}, {Cook}, {Dobbs}, {Dong}, {Eadie}, {Fonseca}, {Gaensler}, {Giri}, {Herrera-Martin}, {Hill}, {Kaczmarek}, {Kania}, {Kaspi}, {Khairy}, {Lanman}, {Leung}, {Masui}, {Mena-Parra}, {Meyers}, {Michilli}, {Milutinovic}, {Ordog}, {Pearlman}, {Pleunis}, {Rafiei-Ravandi}, {Rahman}, {Ransom}, {Sanghavi}, {Shin}, {Smith}, {Stairs}, {Stenning}, {Vanderlinde}, \& {Wulf}}]{2024ApJ...975...75L}
{Lin}, H.-H., {Scholz}, P., {Ng}, C., {et~al.} 2024, \bibinfo{title}{{Do All Fast Radio Bursts Repeat? Constraints from CHIME/FRB Far Sidelobe FRBs},} \apj, 975, 75, \dodoi{10.3847/1538-4357/ad779d}

\bibitem[{M.~S. {Longair} \& P.~A.~G. {Scheuer}(1970){Longair} \& {Scheuer}}]{1970MNRAS.151...45L}
{Longair}, M.~S., \& {Scheuer}, P.~A.~G. 1970, \bibinfo{title}{{The Luminosity-Volume Test for Quasi-Stellar Objects},} \mnras, 151, 45, \dodoi{10.1093/mnras/151.1.45}

\bibitem[{D.~R. {Lorimer} {et~al.}(2007){Lorimer}, {Bailes}, {McLaughlin}, {Narkevic}, \& {Crawford}}]{lorimer}
{Lorimer}, D.~R., {Bailes}, M., {McLaughlin}, M.~A., {Narkevic}, D.~J., \& {Crawford}, F. 2007, \bibinfo{title}{{A Bright Millisecond Radio Burst of Extragalactic Origin},} Science, 318, 777, \dodoi{10.1126/science.1147532}

\bibitem[{N. {Loudas} {et~al.}(2025){Loudas}, {Li}, {Strauss}, \& {Leja}}]{2025arXiv250215566L}
{Loudas}, N., {Li}, D., {Strauss}, M.~A., \& {Leja}, J. 2025, \bibinfo{title}{{Unveiling the origin of fast radio bursts by modeling the stellar mass and star formation distributions of their host galaxies},} arXiv e-prints, arXiv:2502.15566, \dodoi{10.48550/arXiv.2502.15566}

\bibitem[{J.~P. {Macquart} \& R. {Ekers}(2018){Macquart} \& {Ekers}}]{2018MNRAS.480.4211M}
{Macquart}, J.~P., \& {Ekers}, R. 2018, \bibinfo{title}{{FRB event rate counts - II. Fluence, redshift, and dispersion measure distributions},} \mnras, 480, 4211, \dodoi{10.1093/mnras/sty2083}

\bibitem[{J.~P. {Macquart} {et~al.}(2020){Macquart}, {Prochaska}, {McQuinn}, {Bannister}, {Bhandari}, {Day}, {Deller}, {Ekers}, {James}, {Marnoch}, {Os{\l}owski}, {Phillips}, {Ryder}, {Scott}, {Shannon}, \& {Tejos}}]{2020Natur.581..391M}
{Macquart}, J.~P., {Prochaska}, J.~X., {McQuinn}, M., {et~al.} 2020, \bibinfo{title}{{A census of baryons in the Universe from localized fast radio bursts},} \nat, 581, 391, \dodoi{10.1038/s41586-020-2300-2}

\bibitem[{D. {Madison} {et~al.}(2008){Madison}, {Li}, \& {Filippenko}}]{2008CBET.1239....2M}
{Madison}, D., {Li}, W., \& {Filippenko}, A.~V. 2008, \bibinfo{title}{{Supernova 2008X in NGC 4141},} Central Bureau Electronic Telegrams, 1239, 2

\bibitem[{A.~G. {Mannings} {et~al.}(2021){Mannings}, {Fong}, {Simha}, {Prochaska}, {Rafelski}, {Kilpatrick}, {Tejos}, {Heintz}, {Bannister}, {Bhandari}, {Day}, {Deller}, {Ryder}, {Shannon}, \& {Tendulkar}}]{Mannings+2021}
{Mannings}, A.~G., {Fong}, W.-f., {Simha}, S., {et~al.} 2021, \bibinfo{title}{{A High-resolution View of Fast Radio Burst Host Environments},} \apj, 917, 75, \dodoi{10.3847/1538-4357/abff56}

\bibitem[{B. Marcote {et~al.}(2017)Marcote, Paragi, Hessels, Keimpema, Langevelde, Huang, Bassa, Bogdanov, Bower, Burke-Spolaor, Butler, Campbell, Chatterjee, Cordes, Demorest, Garrett, Ghosh, Kaspi, Law, Lazio, McLaughlin, Ransom, Salter, Scholz, Seymour, Siemion, Spitler, Tendulkar, \& Wharton}]{Marcote_2017}
Marcote, B., Paragi, Z., Hessels, J. W.~T., {et~al.} 2017, \bibinfo{title}{The Repeating Fast Radio Burst FRB 121102 as Seen on Milliarcsecond Angular Scales,} The Astrophysical Journal Letters, 834, L8, \dodoi{10.3847/2041-8213/834/2/l8}

\bibitem[{B. {Marcote} {et~al.}(2020{\natexlab{a}}){Marcote}, {Nimmo}, {Hessels}, {Tendulkar}, {Bassa}, {Paragi}, {Keimpema}, {Bhardwaj}, {Karuppusamy}, {Kaspi}, {Law}, {Michilli}, {Aggarwal}, {Andersen}, {Archibald}, {Bandura}, {Bower}, {Boyle}, {Brar}, {Burke-Spolaor}, {Butler}, {Cassanelli}, {Chawla}, {Demorest}, {Dobbs}, {Fonseca}, {Giri}, {Good}, {Gourdji}, {Josephy}, {Kirichenko}, {Kirsten}, {Landecker}, {Lang}, {Lazio}, {Li}, {Lin}, {Linford}, {Masui}, {Mena-Parra}, {Naidu}, {Ng}, {Patel}, {Pen}, {Pleunis}, {Rafiei-Ravandi}, {Rahman}, {Renard}, {Scholz}, {Siegel}, {Smith}, {Stairs}, {Vanderlinde}, \& {Zwaniga}}]{2020Natur.577..190M}
{Marcote}, B., {Nimmo}, K., {Hessels}, J.~W.~T., {et~al.} 2020{\natexlab{a}}, \bibinfo{title}{{A repeating fast radio burst source localized to a nearby spiral galaxy},} \nat, 577, 190, \dodoi{10.1038/s41586-019-1866-z}

\bibitem[{B. {Marcote} {et~al.}(2020{\natexlab{b}}){Marcote}, {Nimmo}, {Hessels}, {Tendulkar}, {Bassa}, {Paragi}, {Keimpema}, {Bhardwaj}, {Karuppusamy}, {Kaspi}, {Law}, {Michilli}, {Aggarwal}, {Andersen}, {Archibald}, {Bandura}, {Bower}, {Boyle}, {Brar}, {Burke-Spolaor}, {Butler}, {Cassanelli}, {Chawla}, {Demorest}, {Dobbs}, {Fonseca}, {Giri}, {Good}, {Gourdji}, {Josephy}, {Kirichenko}, {Kirsten}, {Landecker}, {Lang}, {Lazio}, {Li}, {Lin}, {Linford}, {Masui}, {Mena-Parra}, {Naidu}, {Ng}, {Patel}, {Pen}, {Pleunis}, {Rafiei-Ravandi}, {Rahman}, {Renard}, {Scholz}, {Siegel}, {Smith}, {Stairs}, {Vanderlinde}, \& {Zwaniga}}]{Marcote+2020}
{Marcote}, B., {Nimmo}, K., {Hessels}, J.~W.~T., {et~al.} 2020{\natexlab{b}}, \bibinfo{title}{{A Repeating Fast Radio Burst Source Localized to a Nearby Spiral Galaxy},} \nat, 577, 190, \dodoi{10.1038/s41586-019-1866-z}

\bibitem[{K. {McGregor} \& D.~R. {Lorimer}(2024){McGregor} \& {Lorimer}}]{2024ApJ...961...10M}
{McGregor}, K., \& {Lorimer}, D.~R. 2024, \bibinfo{title}{{Modeling Current and Future High-cadence Surveys of Repeating Fast Radio Burst Populations},} \apj, 961, 10, \dodoi{10.3847/1538-4357/ad1184}

\bibitem[{R. {Mckinven} {et~al.}(2021){Mckinven}, {Michilli}, {Masui}, {Cubranic}, {Gaensler}, {Ng}, {Bhardwaj}, {Leung}, {Boyle}, {Brar}, {Cassanelli}, {Li}, {Mena-Parra}, {Rahman}, \& {Stairs}}]{2021ApJ...920..138M}
{Mckinven}, R., {Michilli}, D., {Masui}, K., {et~al.} 2021, \bibinfo{title}{{Polarization Pipeline for Fast Radio Bursts Detected by CHIME/FRB},} \apj, 920, 138, \dodoi{10.3847/1538-4357/ac126a}

\bibitem[{R. {Mckinven} {et~al.}(2023){Mckinven}, {Gaensler}, {Michilli}, {Masui}, {Kaspi}, {Bhardwaj}, {Cassanelli}, {Chawla}, {Dong}, {Fonseca}, {Leung}, {Li}, {Ng}, {Patel}, {Petroff}, {Pearlman}, {Pleunis}, {Rafiei-Ravandi}, {Rahman}, {Sand}, {Shin}, {Scholz}, {Stairs}, {Smith}, {Su}, \& {Tendulkar}}]{2023ApJ...950...12M}
{Mckinven}, R., {Gaensler}, B.~M., {Michilli}, D., {et~al.} 2023, \bibinfo{title}{{A Large-scale Magneto-ionic Fluctuation in the Local Environment of Periodic Fast Radio Burst Source FRB 20180916B},} \apj, 950, 12, \dodoi{10.3847/1538-4357/acc65f}

\bibitem[{J.~P. {McMullin} {et~al.}(2007){McMullin}, {Waters}, {Schiebel}, {Young}, \& {Golap}}]{McMullin2007}
{McMullin}, J.~P., {Waters}, B., {Schiebel}, D., {Young}, W., \& {Golap}, K. 2007, in Astronomical Society of the Pacific Conference Series, Vol. 376, Astronomical Data Analysis Software and Systems XVI, ed. R.~A. {Shaw}, F.~{Hill}, \& D.~J. {Bell}, 127

\bibitem[{I. {Medlock} {et~al.}(2024){Medlock}, {Nagai}, {Singh}, {Oppenheimer}, {Angl{\'e}s-Alc{\'a}zar}, \& {Villaescusa-Navarro}}]{2024ApJ...967...32M}
{Medlock}, I., {Nagai}, D., {Singh}, P., {et~al.} 2024, \bibinfo{title}{{Probing the Circumgalactic Medium with Fast Radio Bursts: Insights from CAMELS},} \apj, 967, 32, \dodoi{10.3847/1538-4357/ad3070}

\bibitem[{B.~D. {Metzger} {et~al.}(2017){Metzger}, {Berger}, \& {Margalit}}]{2017ApJ...841...14M}
{Metzger}, B.~D., {Berger}, E., \& {Margalit}, B. 2017, \bibinfo{title}{{Millisecond Magnetar Birth Connects FRB 121102 to Superluminous Supernovae and Long-duration Gamma-Ray Bursts},} \apj, 841, 14, \dodoi{10.3847/1538-4357/aa633d}

\bibitem[{M.~J. {Micha{\l}owski} {et~al.}(2020){Micha{\l}owski}, {Th{\"o}ne}, {de Ugarte Postigo}, {Hjorth}, {Le{\'s}niewska}, {Gotkiewicz}, {Dimitrov}, {Koprowski}, \& {Kamphuis}}]{Michalowski2020}
{Micha{\l}owski}, M.~J., {Th{\"o}ne}, C., {de Ugarte Postigo}, A., {et~al.} 2020, \bibinfo{title}{{NGC 2770: High supernova rate due to interaction},} \aap, 642, A84, \dodoi{10.1051/0004-6361/202038719}

\bibitem[{D. {Michilli} {et~al.}(2018){Michilli}, {Seymour}, {Hessels}, {Spitler}, {Gajjar}, {Archibald}, {Bower}, {Chatterjee}, {Cordes}, {Gourdji}, {Heald}, {Kaspi}, {Law}, {Sobey}, {Adams}, {Bassa}, {Bogdanov}, {Brinkman}, {Demorest}, {Fernandez}, {Hellbourg}, {Lazio}, {Lynch}, {Maddox}, {Marcote}, {McLaughlin}, {Paragi}, {Ransom}, {Scholz}, {Siemion}, {Tendulkar}, {van Rooy}, {Wharton}, \& {Whitlow}}]{2018Natur.553..182M}
{Michilli}, D., {Seymour}, A., {Hessels}, J.~W.~T., {et~al.} 2018, \bibinfo{title}{{An extreme magneto-ionic environment associated with the fast radio burst source FRB 121102},} \nat, 553, 182, \dodoi{10.1038/nature25149}

\bibitem[{J.~S. Morgan \& R. Ekers(2021)Morgan \& Ekers}]{Morgan_Ekers_2021}
Morgan, J.~S., \& Ekers, R. 2021, \bibinfo{title}{A measurement of source noise at low frequency: Implications for modern interferometers,} Publications of the Astronomical Society of Australia, 38, e013, \dodoi{10.1017/pasa.2021.3}

\bibitem[{B.~P. {Moster} {et~al.}(2013){Moster}, {Naab}, \& {White}}]{Moster+2013}
{Moster}, B.~P., {Naab}, T., \& {White}, S. D.~M. 2013, \bibinfo{title}{{Galactic star formation and accretion histories from matching galaxies to dark matter haloes},} \mnras, 428, 3121, \dodoi{10.1093/mnras/sts261}

\bibitem[{T. {Nagao} {et~al.}(2006){Nagao}, {Maiolino}, \& {Marconi}}]{Nagao+2006}
{Nagao}, T., {Maiolino}, R., \& {Marconi}, A. 2006, \bibinfo{title}{{Gas metallicity diagnostics in star-forming galaxies},} \aap, 459, 85, \dodoi{10.1051/0004-6361:20065216}

\bibitem[{K. {Nakajima} {et~al.}(2022){Nakajima}, {Ouchi}, {Xu}, {Rauch}, {Harikane}, {Nishigaki}, {Isobe}, {Kusakabe}, {Nagao}, {Ono}, {Onodera}, {Sugahara}, {Kim}, {Komiyama}, {Lee}, \& {Zahedy}}]{Nakajima+2022}
{Nakajima}, K., {Ouchi}, M., {Xu}, Y., {et~al.} 2022, \bibinfo{title}{{EMPRESS. V. Metallicity Diagnostics of Galaxies over 12 + log(O/H) ≃ 6.9-8.9 Established by a Local Galaxy Census: Preparing for JWST Spectroscopy},} \apjs, 262, 3, \dodoi{10.3847/1538-4365/ac7710}

\bibitem[{M. {Ng} \&  {CHIME/FRB Collaboration}(2025){Ng} \& {CHIME/FRB Collaboration}}]{Ng+2025}
{Ng}, M., \& {CHIME/FRB Collaboration}. 2025, \bibinfo{title}{{Discovery of FRB 20250316A, a bright fast radio burst in the direction of the nearby galaxy NGC 4141},} The Astronomer's Telegram, 17081, 1

\bibitem[{K. {Nimmo} {et~al.}(2022){Nimmo}, {Hewitt}, {Hessels}, {Kirsten}, {Marcote}, {Bach}, {Blaauw}, {Burgay}, {Corongiu}, {Feiler}, {Gawro{\'n}ski}, {Giroletti}, {Karuppusamy}, {Keimpema}, {Kharinov}, {Lindqvist}, {Maccaferri}, {Melnikov}, {Mikhailov}, {Ould-Boukattine}, {Paragi}, {Pilia}, {Possenti}, {Snelders}, {Surcis}, {Trudu}, {Venturi}, {Vlemmings}, {Wang}, {Yang}, \& {Yuan}}]{2022ApJ...927L...3N}
{Nimmo}, K., {Hewitt}, D.~M., {Hessels}, J.~W.~T., {et~al.} 2022, \bibinfo{title}{{Milliarcsecond Localization of the Repeating FRB 20201124A},} \apjl, 927, L3, \dodoi{10.3847/2041-8213/ac540f}

\bibitem[{K. {Nimmo} {et~al.}(2025){Nimmo}, {Pleunis}, {Beniamini}, {Kumar}, {Lanman}, {Li}, {Main}, {Sammons}, {Andrew}, {Bhardwaj}, {Chatterjee}, {Curtin}, {Fonseca}, {Gaensler}, {Joseph}, {Kader}, {Kaspi}, {Lazda}, {Leung}, {Masui}, {Mckinven}, {Michilli}, {Pandhi}, {Pearlman}, {Rafiei-Ravandi}, {Sand}, {Shin}, {Smith}, \& {Stairs}}]{Nimmo+2025}
{Nimmo}, K., {Pleunis}, Z., {Beniamini}, P., {et~al.} 2025, \bibinfo{title}{{Magnetospheric origin of a fast radio burst constrained using scintillation},} \nat, 637, 48, \dodoi{10.1038/s41586-024-08297-w}

\bibitem[{C.~H. {Niu} {et~al.}(2022){Niu}, {Aggarwal}, {Li}, {Zhang}, {Chatterjee}, {Tsai}, {Yu}, {Law}, {Burke-Spolaor}, {Cordes}, {Zhang}, {Ocker}, {Yao}, {Wang}, {Feng}, {Niino}, {Bochenek}, {Cruces}, {Connor}, {Jiang}, {Dai}, {Luo}, {Li}, {Miao}, {Niu}, {Anna-Thomas}, {Sydnor}, {Stern}, {Wang}, {Yuan}, {Yue}, {Zhou}, {Yan}, {Zhu}, \& {Zhang}}]{Niu_2022}
{Niu}, C.~H., {Aggarwal}, K., {Li}, D., {et~al.} 2022, \bibinfo{title}{{A repeating fast radio burst associated with a persistent radio source},} \nat, 606, 873, \dodoi{10.1038/s41586-022-04755-5}

\bibitem[{S.~K. {Ocker} {et~al.}(2022){Ocker}, {Cordes}, {Chatterjee}, {Niu}, {Li}, {McKee}, {Law}, {Tsai}, {Anna-Thomas}, {Yao}, \& {Cruces}}]{KochOcker_2022}
{Ocker}, S.~K., {Cordes}, J.~M., {Chatterjee}, S., {et~al.} 2022, \bibinfo{title}{{The Large Dispersion and Scattering of FRB 20190520B Are Dominated by the Host Galaxy},} \apj, 931, 87, \dodoi{10.3847/1538-4357/ac6504}

\bibitem[{S.~A. {Olausen} \& V.~M. {Kaspi}(2014){Olausen} \& {Kaspi}}]{2014ApJS..212....6O}
{Olausen}, S.~A., \& {Kaspi}, V.~M. 2014, \bibinfo{title}{{The McGill Magnetar Catalog},} \apjs, 212, 6, \dodoi{10.1088/0067-0049/212/1/6}

\bibitem[{N. {Oppermann} {et~al.}(2018){Oppermann}, {Yu}, \& {Pen}}]{2018MNRAS.475.5109O}
{Oppermann}, N., {Yu}, H.-R., \& {Pen}, U.-L. 2018, \bibinfo{title}{{On the non-Poissonian repetition pattern of FRB121102},} \mnras, 475, 5109, \dodoi{10.1093/mnras/sty004}

\bibitem[{D.~E. {Osterbrock} \& G.~J. {Ferland}(2006){Osterbrock} \& {Ferland}}]{OsterbrockFerland2006}
{Osterbrock}, D.~E., \& {Ferland}, G.~J. 2006, {Astrophysics of gaseous nebulae and active galactic nuclei} (University Science Books)

\bibitem[{O.~S. {Ould-Boukattine} {et~al.}(2024){Ould-Boukattine}, {Chawla}, {Hessels}, {Cooper}, {Gawro{\'n}ski}, {Herrmann}, {Kirsten}, {Hewitt}, {Konijn}, {Nimmo}, {Pleunis}, {Puchalska}, \& {Snelders}}]{2024arXiv241017024O}
{Ould-Boukattine}, O.~S., {Chawla}, P., {Hessels}, J.~W.~T., {et~al.} 2024, \bibinfo{title}{{A probe of the maximum energetics of fast radio bursts through a prolific repeating source},} arXiv e-prints, arXiv:2410.17024, \dodoi{10.48550/arXiv.2410.17024}

\bibitem[{O.~S. {Ould-Boukattine} {et~al.}(2025){Ould-Boukattine}, {Blaauw}, {Buchsteiner}, {Beer}, {Dijkema}, {Fischer}, {Gawronski}, {Hessels}, {Hewitt}, {Herrmann}, {Kirsten}, {Kuiper}, {Loege}, {Marcote}, {Mulder}, {Ovinge}, {Puchalska}, {Pleunis}, {Ranguin}, {Moroianu}, {Snelders}, {Sluman}, {Wolf}, \& {Yang}}]{2025ATel17124....1O}
{Ould-Boukattine}, O.~S., {Blaauw}, R., {Buchsteiner}, T., {et~al.} 2025, \bibinfo{title}{{Non-detection of repeat radio bursts from FRB 20250316A in a high-cadence observing campaign},} The Astronomer's Telegram, 17124, 1

\bibitem[{A. {Pandhi} {et~al.}(2024){Pandhi}, {Pleunis}, {Mckinven}, {Gaensler}, {Su}, {Ng}, {Bhardwaj}, {Brar}, {Cassanelli}, {Cook}, {Curtin}, {Kaspi}, {Lazda}, {Leung}, {Li}, {Masui}, {Michilli}, {Nimmo}, {Pearlman}, {Petroff}, {Rafiei-Ravandi}, {Sand}, {Scholz}, {Shin}, {Smith}, \& {Stairs}}]{2024ApJ...968...50P}
{Pandhi}, A., {Pleunis}, Z., {Mckinven}, R., {et~al.} 2024, \bibinfo{title}{{Polarization Properties of 128 Nonrepeating Fast Radio Bursts from the First CHIME/FRB Baseband Catalog},} \apj, 968, 50, \dodoi{10.3847/1538-4357/ad40aa}

\bibitem[{A. {Pandhi} {et~al.}(2025){Pandhi}, {Gaensler}, {Pleunis}, {Hutschenreuter}, {Law}, {Mckinven}, {O'Sullivan}, {Petroff}, \& {Vernstrom}}]{2025ApJ...982..146P}
{Pandhi}, A., {Gaensler}, B.~M., {Pleunis}, Z., {et~al.} 2025, \bibinfo{title}{{Improved Constraints on the Faraday Rotation toward Eight Fast Radio Bursts Using Dense Grids of Polarized Radio Galaxies},} \apj, 982, 146, \dodoi{10.3847/1538-4357/adb8e3}

\bibitem[{A. {Pastorello} {et~al.}(2007){Pastorello}, {Smartt}, {Mattila}, {Eldridge}, {Young}, {Itagaki}, {Yamaoka}, {Navasardyan}, {Valenti}, {Patat}, {Agnoletto}, {Augusteijn}, {Benetti}, {Cappellaro}, {Boles}, {Bonnet-Bidaud}, {Botticella}, {Bufano}, {Cao}, {Deng}, {Dennefeld}, {Elias-Rosa}, {Harutyunyan}, {Keenan}, {Iijima}, {Lorenzi}, {Mazzali}, {Meng}, {Nakano}, {Nielsen}, {Smoker}, {Stanishev}, {Turatto}, {Xu}, \& {Zampieri}}]{pastorello2007giant}
{Pastorello}, A., {Smartt}, S.~J., {Mattila}, S., {et~al.} 2007, \bibinfo{title}{{A giant outburst two years before the core-collapse of a massive star},} \nat, 447, 829, \dodoi{10.1038/nature05825}

\bibitem[{A. {Pastorello} {et~al.}(2012){Pastorello}, {Pumo}, {Navasardyan}, {Zampieri}, {Turatto}, {Sollerman}, {Taddia}, {Kankare}, {Mattila}, {Nicolas}, {Prosperi}, {San Segundo Delgado}, {Taubenberger}, {Boles}, {Bachini}, {Benetti}, {Bufano}, {Cappellaro}, {Cason}, {Cetrulo}, {Ergon}, {Germany}, {Harutyunyan}, {Howerton}, {Hurst}, {Patat}, {Stritzinger}, {Strolger}, \& {Wells}}]{2012A&A...537A.141P}
{Pastorello}, A., {Pumo}, M.~L., {Navasardyan}, H., {et~al.} 2012, \bibinfo{title}{{SN 2009E: a faint clone of SN 1987A},} \aap, 537, A141, \dodoi{10.1051/0004-6361/201118112}

\bibitem[{A.~B. {Pearlman} {et~al.}(2025{\natexlab{a}}){Pearlman}, {CHIME/FRB Collaboration}, {Gendreau}, {Arzoumanian}, {Ferrara}, {Chakraborty}, {Enoto}, {Guver}, \& {Hu}}]{Pearlman+2025_ATel17117}
{Pearlman}, A.~B., {CHIME/FRB Collaboration}, {Gendreau}, K.~C., {et~al.} 2025{\natexlab{a}}, \bibinfo{title}{{FRB 20250316A: X-ray Flux Upper Limits with NICER and Simultaneous Radio Observations with CHIME/Pulsar},} The Astronomer's Telegram, 17117, 1

\bibitem[{A.~B. {Pearlman} {et~al.}(2025{\natexlab{b}}){Pearlman}, {Scholz}, {Bethapudi}, {Hessels}, {Kaspi}, {Kirsten}, {Nimmo}, {Spitler}, {Fonseca}, {Meyers}, {Stairs}, {Tan}, {Bhardwaj}, {Chatterjee}, {Cook}, {Curtin}, {Dong}, {Eftekhari}, {Gaensler}, {G{\"u}ver}, {Kaczmarek}, {Leung}, {Masui}, {Michilli}, {Prince}, {Sand}, {Shin}, {Smith}, \& {Tendulkar}}]{2025NatAs...9..111P}
{Pearlman}, A.~B., {Scholz}, P., {Bethapudi}, S., {et~al.} 2025{\natexlab{b}}, \bibinfo{title}{{Multiwavelength constraints on the origin of a nearby repeating fast radio burst source in a globular cluster},} Nature Astronomy, 9, 111, \dodoi{10.1038/s41550-024-02386-6}

\bibitem[{R.~A. Perley \& B.~J. Butler(2017)Perley \& Butler}]{Perley_2017}
Perley, R.~A., \& Butler, B.~J. 2017, \bibinfo{title}{An Accurate Flux Density Scale from 50 MHz to 50 GHz,} The Astrophysical Journal Supplement Series, 230, 7, \dodoi{10.3847/1538-4365/aa6df9}

\bibitem[{L.~Y. {Petrov} \& Y.~Y. {Kovalev}(2025){Petrov} \& {Kovalev}}]{Petrov_2025}
{Petrov}, L.~Y., \& {Kovalev}, Y.~Y. 2025, \bibinfo{title}{{The Radio Fundamental Catalog. I. Astrometry},} \apjs, 276, 38, \dodoi{10.3847/1538-4365/ad8c36}

\bibitem[{L. {Piro} {et~al.}(2021){Piro}, {Bruni}, {Troja}, {O'Connor}, {Panessa}, {Ricci}, {Zhang}, {Burgay}, {Dichiara}, {Lee}, {Lotti}, {Niu}, {Pilia}, {Possenti}, {Trudu}, {Xu}, {Zhu}, {Kutyrev}, \& {Veilleux}}]{2021A&A...656L..15P}
{Piro}, L., {Bruni}, G., {Troja}, E., {et~al.} 2021, \bibinfo{title}{{The fast radio burst FRB 20201124A in a star-forming region: Constraints to the progenitor and multiwavelength counterparts},} \aap, 656, L15, \dodoi{10.1051/0004-6361/202141903}

\bibitem[{ {\!\!Planck Collaboration} {et~al.}(2020){\!\!Planck Collaboration}, {Aghanim}, {Akrami}, {Ashdown}, {Aumont}, {Baccigalupi}, {Ballardini}, {Banday}, {Barreiro}, {Bartolo}, {Basak}, {Battye}, {Benabed}, {Bernard}, {Bersanelli}, {Bielewicz}, {Bock}, {Bond}, {Borrill}, {Bouchet}, {Boulanger}, {Bucher}, {Burigana}, {Butler}, {Calabrese}, {Cardoso}, {Carron}, {Challinor}, {Chiang}, {Chluba}, {Colombo}, {Combet}, {Contreras}, {Crill}, {Cuttaia}, {de Bernardis}, {de Zotti}, {Delabrouille}, {Delouis}, {Di Valentino}, {Diego}, {Dor{\'e}}, {Douspis}, {Ducout}, {Dupac}, {Dusini}, {Efstathiou}, {Elsner}, {En{\ss}lin}, {Eriksen}, {Fantaye}, {Farhang}, {Fergusson}, {Fernandez-Cobos}, {Finelli}, {Forastieri}, {Frailis}, {Fraisse}, {Franceschi}, {Frolov}, {Galeotta}, {Galli}, {Ganga}, {G{\'e}nova-Santos}, {Gerbino}, {Ghosh}, {Gonz{\'a}lez-Nuevo}, {G{\'o}rski}, {Gratton}, {Gruppuso}, {Gudmundsson}, {Hamann}, {Handley}, {Hansen}, {Herranz}, {Hildebrandt}, {Hivon}, {Huang}, {Jaffe}, {Jones}, {Karakci},
  {Keih{\"a}nen}, {Keskitalo}, {Kiiveri}, {Kim}, {Kisner}, {Knox}, {Krachmalnicoff}, {Kunz}, {Kurki-Suonio}, {Lagache}, {Lamarre}, {Lasenby}, {Lattanzi}, {Lawrence}, {Le Jeune}, {Lemos}, {Lesgourgues}, {Levrier}, {Lewis}, {Liguori}, {Lilje}, {Lilley}, {Lindholm}, {L{\'o}pez-Caniego}, {Lubin}, {Ma}, {Mac{\'\i}as-P{\'e}rez}, {Maggio}, {Maino}, {Mandolesi}, {Mangilli}, {Marcos-Caballero}, {Maris}, {Martin}, {Martinelli}, {Mart{\'\i}nez-Gonz{\'a}lez}, {Matarrese}, {Mauri}, {McEwen}, {Meinhold}, {Melchiorri}, {Mennella}, {Migliaccio}, {Millea}, {Mitra}, {Miville-Desch{\^e}nes}, {Molinari}, {Montier}, {Morgante}, {Moss}, {Natoli}, {N{\o}rgaard-Nielsen}, {Pagano}, {Paoletti}, {Partridge}, {Patanchon}, {Peiris}, {Perrotta}, {Pettorino}, {Piacentini}, {Polastri}, {Polenta}, {Puget}, {Rachen}, {Reinecke}, {Remazeilles}, {Renzi}, {Rocha}, {Rosset}, {Roudier}, {Rubi{\~n}o-Mart{\'\i}n}, {Ruiz-Granados}, {Salvati}, {Sandri}, {Savelainen}, {Scott}, {Shellard}, {Sirignano}, {Sirri}, {Spencer}, {Sunyaev}, {Suur-Uski},
  {Tauber}, {Tavagnacco}, {Tenti}, {Toffolatti}, {Tomasi}, {Trombetti}, {Valenziano}, {Valiviita}, {Van Tent}, {Vibert}, {Vielva}, {Villa}, {Vittorio}, {Wandelt}, {Wehus}, {White}, {White}, {Zacchei}, \& {Zonca}}]{2020A&A...641A...6P}
{\!\!Planck Collaboration}, {Aghanim}, N., {Akrami}, Y., {et~al.} 2020, \bibinfo{title}{{Planck 2018 results. VI. Cosmological parameters},} \aap, 641, A6, \dodoi{10.1051/0004-6361/201833910}

\bibitem[{Z. {Pleunis} {et~al.}(2021){Pleunis}, {Good}, {Kaspi}, {Mckinven}, {Ransom}, {Scholz}, {Bandura}, {Bhardwaj}, {Boyle}, {Brar}, {Cassanelli}, {Chawla}, {(Adam) Dong}, {Fonseca}, {Gaensler}, {Josephy}, {Kaczmarek}, {Leung}, {Lin}, {Masui}, {Mena-Parra}, {Michilli}, {Ng}, {Patel}, {Rafiei-Ravandi}, {Rahman}, {Sanghavi}, {Shin}, {Smith}, {Stairs}, \& {Tendulkar}}]{2021ApJ...923....1P}
{Pleunis}, Z., {Good}, D.~C., {Kaspi}, V.~M., {et~al.} 2021, \bibinfo{title}{{Fast Radio Burst Morphology in the First CHIME/FRB Catalog},} \apj, 923, 1, \dodoi{10.3847/1538-4357/ac33ac}

\bibitem[{S. {Pradeep E.~T.} {et~al.}(2025){Pradeep E.~T.}, {Sprenger}, {Wucknitz}, {Main}, \& {Spitler}}]{t2025scintillometryfastradiobursts}
{Pradeep E.~T.}, S., {Sprenger}, T., {Wucknitz}, O., {Main}, R.~A., \& {Spitler}, L.~G. 2025, \bibinfo{title}{{Scintillometry of Fast Radio Bursts: Resolution effects in two-screen models},} arXiv e-prints, arXiv:2505.04576, \dodoi{10.48550/arXiv.2505.04576}

\bibitem[{J.~X. {Prochaska} \& Y. {Zheng}(2019){Prochaska} \& {Zheng}}]{ProchaskaZheng2019}
{Prochaska}, J.~X., \& {Zheng}, Y. 2019, \bibinfo{title}{{Probing Galactic haloes with fast radio bursts},} \mnras, 485, 648, \dodoi{10.1093/mnras/stz261}

\bibitem[{J.~X. {Prochaska} {et~al.}(2019){Prochaska}, {Macquart}, {McQuinn}, {Simha}, {Shannon}, {Day}, {Marnoch}, {Ryder}, {Deller}, {Bannister}, {Bhandari}, {Bordoloi}, {Bunton}, {Cho}, {Flynn}, {Mahony}, {Phillips}, {Qiu}, \& {Tejos}}]{2019Sci...366..231P}
{Prochaska}, J.~X., {Macquart}, J.-P., {McQuinn}, M., {et~al.} 2019, \bibinfo{title}{{The low density and magnetization of a massive galaxy halo exposed by a fast radio burst},} Science, 366, 231, \dodoi{10.1126/science.aay0073}

\bibitem[{J.~X. Prochaska {et~al.}(2020)Prochaska, Hennawi, Westfall, Cooke, Wang, Hsyu, Davies, Farina, \& Pelliccia}]{pypeit}
Prochaska, J.~X., Hennawi, J.~F., Westfall, K.~B., {et~al.} 2020, \bibinfo{title}{PypeIt: The Python Spectroscopic Data Reduction Pipeline,} Journal of Open Source Software, 5, 2308, \dodoi{10.21105/joss.02308}

\bibitem[{H. {Qian}(2025){Qian}}]{ATel17126}
{Qian}, H. 2025, \bibinfo{title}{{Non-detection of repeat radio bursts from FRB 20250316A by the FAST telescope},} The Astronomer's Telegram, 17126, 1

\bibitem[{R.~J. {Reynolds}(1977){Reynolds}}]{Reynolds+1977}
{Reynolds}, R.~J. 1977, \bibinfo{title}{{Pulsar dispersion measures and Halpha emission measures: limits on the electron density and filling factor for the ionized interstellar gas.},} \apj, 216, 433, \dodoi{10.1086/155484}

\bibitem[{L. {Rhodes} {et~al.}(2020){Rhodes}, {van der Horst}, {Fender}, {Monageng}, {Anderson}, {Antoniadis}, {Bietenholz}, {B{\"o}ttcher}, {Bright}, {Green}, {Kouveliotou}, {Kramer}, {Motta}, {Wijers}, {Williams}, \& {Woudt}}]{2020MNRAS.496.3326R}
{Rhodes}, L., {van der Horst}, A.~J., {Fender}, R., {et~al.} 2020, \bibinfo{title}{{Radio afterglows of very high-energy gamma-ray bursts 190829A and 180720B},} \mnras, 496, 3326, \dodoi{10.1093/mnras/staa1715}

\bibitem[{S.~D. {Ryder} {et~al.}(2023){Ryder}, {Bannister}, {Bhandari}, {Deller}, {Ekers}, {Glowacki}, {Gordon}, {Gourdji}, {James}, {Kilpatrick}, {Lu}, {Marnoch}, {Moss}, {Prochaska}, {Qiu}, {Sadler}, {Simha}, {Sammons}, {Scott}, {Tejos}, \& {Shannon}}]{2023Sci...382..294R}
{Ryder}, S.~D., {Bannister}, K.~W., {Bhandari}, S., {et~al.} 2023, \bibinfo{title}{{A luminous fast radio burst that probes the Universe at redshift 1},} Science, 382, 294, \dodoi{10.1126/science.adf2678}

\bibitem[{M.~W. Sammons {et~al.}(2023)Sammons, Deller, Glowacki, Gourdji, James, Prochaska, Qiu, Scott, Shannon, \& Trott}]{sammons_two_screen_2023}
Sammons, M.~W., Deller, A.~T., Glowacki, M., {et~al.} 2023, \bibinfo{title}{Two-screen scattering in CRAFT FRBs,} Monthly Notices of the Royal Astronomical Society, 525, 5653, \dodoi{10.1093/mnras/stad2631}

\bibitem[{E.~F. {Schlafly} \& D.~P. {Finkbeiner}(2011){Schlafly} \& {Finkbeiner}}]{SchlaflyFinkbeiner2011}
{Schlafly}, E.~F., \& {Finkbeiner}, D.~P. 2011, \bibinfo{title}{{Measuring Reddening with Sloan Digital Sky Survey Stellar Spectra and Recalibrating SFD},} \apj, 737, 103, \dodoi{10.1088/0004-637X/737/2/103}

\bibitem[{G. {Schroeder} {et~al.}(2024){Schroeder}, {Rhodes}, {Laskar}, {Nugent}, {Rouco Escorial}, {Rastinejad}, {Fong}, {van der Horst}, {Veres}, {Alexander}, {Andersson}, {Berger}, {Blanchard}, {Chastain}, {Christensen}, {Fender}, {Green}, {Groot}, {Heywood}, {Horesh}, {Izzo}, {Kilpatrick}, {K{\"o}rding}, {Lien}, {Malesani}, {McBride}, {Mooley}, {Rowlinson}, {Sears}, {Stappers}, {Tanvir}, {Vergani}, {Wijers}, {Williams-Baldwin}, \& {Woudt}}]{2024ApJ...970..139S}
{Schroeder}, G., {Rhodes}, L., {Laskar}, T., {et~al.} 2024, \bibinfo{title}{{A Radio Flare in the Long-lived Afterglow of the Distant Short GRB 210726A: Energy Injection or a Reverse Shock from Shell Collisions?},} \apj, 970, 139, \dodoi{10.3847/1538-4357/ad49ab}

\bibitem[{A. Seymour {et~al.}(2019)Seymour, Michilli, \& Pleunis}]{2019ascl.soft10004S}
Seymour, A., Michilli, D., \& Pleunis, Z. 2019, \bibinfo{title}{{DM\_phase: Algorithm for correcting dispersion of radio signals},} \url{https://ui.adsabs.harvard.edu/abs/2019ascl.soft10004S}

\bibitem[{V. {Shah} {et~al.}(2025){Shah}, {Shin}, {Leung}, {Fong}, {Eftekhari}, {Amiri}, {Andersen}, {Andrew}, {Bhardwaj}, {Brar}, {Cassanelli}, {Chatterjee}, {Curtin}, {Dobbs}, {Dong}, {Dong}, {Fonseca}, {Gaensler}, {Halpern}, {Hessels}, {Ibik}, {Jain}, {Joseph}, {Kaczmarek}, {Kahinga}, {Kaspi}, {Kharel}, {Landecker}, {Lanman}, {Lazda}, {Main}, {Mas-Ribas}, {Masui}, {Mckinven}, {Mena-Parra}, {Meyers}, {Michilli}, {Nimmo}, {Pandhi}, {Patil}, {Pearlman}, {Pleunis}, {Prochaska}, {Rafiei-Ravandi}, {Sammons}, {Sand}, {Scholz}, {Smith}, \& {Stairs}}]{Shah+2025}
{Shah}, V., {Shin}, K., {Leung}, C., {et~al.} 2025, \bibinfo{title}{{A Repeating Fast Radio Burst Source in the Outskirts of a Quiescent Galaxy},} \apjl, 979, L21, \dodoi{10.3847/2041-8213/ad9ddc}

\bibitem[{K. {Sharma} {et~al.}(2024){Sharma}, {Ravi}, {Connor}, {Law}, {Ocker}, {Sherman}, {Kosogorov}, {Faber}, {Hallinan}, {Harnach}, {Hellbourg}, {Hobbs}, {Hodge}, {Hodges}, {Lamb}, {Rasmussen}, {Somalwar}, {Weinreb}, {Woody}, {Leja}, {Anand}, {Das}, {Qin}, {Rose}, {Dong}, {Miller}, \& {Yao}}]{Sharma+2024}
{Sharma}, K., {Ravi}, V., {Connor}, L., {et~al.} 2024, \bibinfo{title}{{Preferential occurrence of fast radio bursts in massive star-forming galaxies},} \nat, 635, 61, \dodoi{10.1038/s41586-024-08074-9}

\bibitem[{M.~C. {Shepherd} {et~al.}(1994){Shepherd}, {Pearson}, \& {Taylor}}]{1994BAAS...26..987S}
{Shepherd}, M.~C., {Pearson}, T.~J., \& {Taylor}, G.~B. 1994, in Bulletin of the American Astronomical Society, Vol.~26, 987--989

\bibitem[{K. {Shin} {et~al.}(2023){Shin}, {Masui}, {Bhardwaj}, {Cassanelli}, {Chawla}, {Dobbs}, {Dong}, {Fonseca}, {Gaensler}, {Herrera-Mart{\'\i}n}, {Kaczmarek}, {Kaspi}, {Leung}, {Merryfield}, {Michilli}, {M{\"u}nchmeyer}, {Pearlman}, {Rafiei-Ravandi}, {Smith}, {Stairs}, \& {Tendulkar}}]{2023ApJ...944..105S}
{Shin}, K., {Masui}, K.~W., {Bhardwaj}, M., {et~al.} 2023, \bibinfo{title}{{Inferring the Energy and Distance Distributions of Fast Radio Bursts Using the First CHIME/FRB Catalog},} \apj, 944, 105, \dodoi{10.3847/1538-4357/acaf06}

\bibitem[{S. {Simha} {et~al.}(2025){Simha}, {Eftekhari}, \& {\!\! CHIME/FRB Collaboration}}]{ATel17116}
{Simha}, S., {Eftekhari}, T., \& {\!\! CHIME/FRB Collaboration}. 2025, \bibinfo{title}{{Gemini Imaging and Spectroscopy of the Sub-arcsecond Localization Region of FRB 20250316A},} The Astronomer's Telegram, 17116, 1

\bibitem[{J.~G. {Sorce} {et~al.}(2014){Sorce}, {Tully}, {Courtois}, {Jarrett}, {Neill}, \& {Shaya}}]{2014MNRAS.444..527S}
{Sorce}, J.~G., {Tully}, R.~B., {Courtois}, H.~M., {et~al.} 2014, \bibinfo{title}{{From Spitzer Galaxy photometry to Tully-Fisher distances},} \mnras, 444, 527, \dodoi{10.1093/mnras/stu1450}

\bibitem[{J.~S. {Speagle}(2020){Speagle}}]{2020MNRAS.493.3132S}
{Speagle}, J.~S. 2020, \bibinfo{title}{{DYNESTY: a dynamic nested sampling package for estimating Bayesian posteriors and evidences},} \mnras, 493, 3132, \dodoi{10.1093/mnras/staa278}

\bibitem[{L.~G. {Spitler} {et~al.}(2016){Spitler}, {Scholz}, {Hessels}, {Bogdanov}, {Brazier}, {Camilo}, {Chatterjee}, {Cordes}, {Crawford}, {Deneva}, {Ferdman}, {Freire}, {Kaspi}, {Lazarus}, {Lynch}, {Madsen}, {McLaughlin}, {Patel}, {Ransom}, {Seymour}, {Stairs}, {Stappers}, {van Leeuwen}, \& {Zhu}}]{2016Natur.531..202S}
{Spitler}, L.~G., {Scholz}, P., {Hessels}, J.~W.~T., {et~al.} 2016, \bibinfo{title}{{A repeating fast radio burst},} \nat, 531, 202, \dodoi{10.1038/nature17168}

\bibitem[{L.-G. {Strolger} {et~al.}(2015){Strolger}, {Dahlen}, {Rodney}, {Graur}, {Riess}, {McCully}, {Ravindranath}, {Mobasher}, \& {Shahady}}]{Strolger2015}
{Strolger}, L.-G., {Dahlen}, T., {Rodney}, S.~A., {et~al.} 2015, \bibinfo{title}{{The Rate of Core Collapse Supernovae to Redshift 2.5 from the CANDELS and CLASH Supernova Surveys},} \apj, 813, 93, \dodoi{10.1088/0004-637X/813/2/93}

\bibitem[{H. {Sun} {et~al.}(2025{\natexlab{a}}){Sun}, {Cheng}, {Yi}, {Liu}, {Jin}, \& {Ling}}]{ATel17100}
{Sun}, H., {Cheng}, H.~Q., {Yi}, D.~Y., {et~al.} 2025{\natexlab{a}}, \bibinfo{title}{{FRB 20250316A: detection of a candidate associated X-ray source EP J120944.2+585060 by Einstein Probe},} The Astronomer's Telegram, 17100, 1

\bibitem[{H. {Sun} {et~al.}(2025{\natexlab{b}}){Sun}, {Li}, {Jin}, {Liu}, {Zhang}, {Li}, {Geng}, {Wu}, {Li}, {Tao}, {Feng}, {Rau}, {Rea}, \& {W. Yuan}}]{SunChandra}
{Sun}, H., {Li}, D.~Y., {Jin}, C.~C., {et~al.} 2025{\natexlab{b}}, \bibinfo{title}{{Chandra localization of EP J120944.2+585060: likely not associated with FRB 20250316A},} The Astronomer's Telegram, 17119, 1

\bibitem[{F. {Taddia} {et~al.}(2013){Taddia}, {Sollerman}, {Razza}, {Gafton}, {Pastorello}, {Fransson}, {Stritzinger}, {Leloudas}, \& {Ergon}}]{2013A&A...558A.143T}
{Taddia}, F., {Sollerman}, J., {Razza}, A., {et~al.} 2013, \bibinfo{title}{{A metallicity study of 1987A-like supernova host galaxies},} \aap, 558, A143, \dodoi{10.1051/0004-6361/201322276}

\bibitem[{S.~P. {Tendulkar} {et~al.}(2017){Tendulkar}, {Bassa}, {Cordes}, {Bower}, {Law}, {Chatterjee}, {Adams}, {Bogdanov}, {Burke-Spolaor}, {Butler}, {Demorest}, {Hessels}, {Kaspi}, {Lazio}, {Maddox}, {Marcote}, {McLaughlin}, {Paragi}, {Ransom}, {Scholz}, {Seymour}, {Spitler}, {van Langevelde}, \& {Wharton}}]{Tendulkar+2017}
{Tendulkar}, S.~P., {Bassa}, C.~G., {Cordes}, J.~M., {et~al.} 2017, \bibinfo{title}{{The Host Galaxy and Redshift of the Repeating Fast Radio Burst FRB 121102},} \apjl, 834, L7, \dodoi{10.3847/2041-8213/834/2/L7}

\bibitem[{S.~P. {Tendulkar} {et~al.}(2021){Tendulkar}, {Gil de Paz}, {Kirichenko}, {Hessels}, {Bhardwaj}, {{\'A}vila}, {Bassa}, {Chawla}, {Fonseca}, {Kaspi}, {Keimpema}, {Kirsten}, {Lazio}, {Marcote}, {Masui}, {Nimmo}, {Paragi}, {Rahman}, {Pay{\'a}}, {Scholz}, \& {Stairs}}]{Tendulkar+2021}
{Tendulkar}, S.~P., {Gil de Paz}, A., {Kirichenko}, A.~Y., {et~al.} 2021, \bibinfo{title}{{The 60 pc Environment of FRB 20180916B},} \apjl, 908, L12, \dodoi{10.3847/2041-8213/abdb38}

\bibitem[{D. {Thornton} {et~al.}(2013){Thornton}, {Stappers}, {Bailes}, {Barsdell}, {Bates}, {Bhat}, {Burgay}, {Burke-Spolaor}, {Champion}, {Coster}, {D'Amico}, {Jameson}, {Johnston}, {Keith}, {Kramer}, {Levin}, {Milia}, {Ng}, {Possenti}, \& {van Straten}}]{2013Sci...341...53T}
{Thornton}, D., {Stappers}, B., {Bailes}, M., {et~al.} 2013, \bibinfo{title}{{A Population of Fast Radio Bursts at Cosmological Distances},} Science, 341, 53, \dodoi{10.1126/science.1236789}

\bibitem[{J. {Tian} {et~al.}(2025){Tian}, {Pastor-Marazuela}, {Rajwade}, {Stappers}, {Shaji}, {Hanmer}, {Caleb}, {Bezuidenhout}, {Jankowski}, {Breton}, {Barr}, {Kramer}, {Groot}, {Bloemen}, {Vreeswijk}, {Pieterse}, {Woudt}, {Fender}, {Wijnands}, \& {Buckley}}]{2025MNRAS.tmp..751T}
{Tian}, J., {Pastor-Marazuela}, I., {Rajwade}, K.~M., {et~al.} 2025, \bibinfo{title}{{MeerKAT discovery of a hyperactive repeating fast radio burst source},} \mnras, \dodoi{10.1093/mnras/staf793}

\bibitem[{A. {Tohuvavohu} {et~al.}(2024){Tohuvavohu}, {Kennea}, {Roberts}, {DeLaunay}, {Ronchini}, {Cenko}, {Ewing}, {Magee}, {Messick}, {Sachdev}, \& {Singer}}]{2024ApJ...975L..19T}
{Tohuvavohu}, A., {Kennea}, J.~A., {Roberts}, C.~J., {et~al.} 2024, \bibinfo{title}{{Swiftly Chasing Gravitational Waves across the Sky in Real Time},} \apjl, 975, L19, \dodoi{10.3847/2041-8213/ad87ce}

\bibitem[{V. {Trimble}(1968){Trimble}}]{1968AJ.....73..535T}
{Trimble}, V. 1968, \bibinfo{title}{{Motions and Structure of the Filamentary Envelope of the Crab Nebula},} \aj, 73, 535, \dodoi{10.1086/110658}

\bibitem[{E. {Troja} {et~al.}(2025){Troja}, {Yang}, {Dichiara}, {Kabir}, \& {Sun}}]{2025ATel17109....1T}
{Troja}, E., {Yang}, Y.~H., {Dichiara}, S., {Kabir}, M.~E., \& {Sun}, H. 2025, \bibinfo{title}{{Swift follow-up observations of FRB250316A and EP J120944.2+585060},} The Astronomer's Telegram, 17109, 1

\bibitem[{R.~B. {Tully} {et~al.}(2016){Tully}, {Courtois}, \& {Sorce}}]{2016AJ....152...50T}
{Tully}, R.~B., {Courtois}, H.~M., \& {Sorce}, J.~G. 2016, \bibinfo{title}{{Cosmicflows-3},} \aj, 152, 50, \dodoi{10.3847/0004-6256/152/2/50}

\bibitem[{K. {Vanderlinde} {et~al.}(2019){Vanderlinde}, {Liu}, {Gaensler}, {Bond}, {Hinshaw}, {Ng}, {Chiang}, {Stairs}, {Brown}, {Sievers}, {Mena}, {Smith}, {Bandura}, {Masui}, {Spekkens}, {Belostotski}, {Dobbs}, {Turok}, {Boyle}, {Rupen}, {Landecker}, {Pen}, \& {Kaspi}}]{2019clrp.2020...28V}
{Vanderlinde}, K., {Liu}, A., {Gaensler}, B., {et~al.} 2019, in Canadian Long Range Plan for Astronomy and Astrophysics White Papers, Vol. 2020, 28, \dodoi{10.5281/zenodo.3765414}

\bibitem[{V.~A. {Villar} {et~al.}(2017){Villar}, {Berger}, {Metzger}, \& {Guillochon}}]{Villar+2017}
{Villar}, V.~A., {Berger}, E., {Metzger}, B.~D., \& {Guillochon}, J. 2017, \bibinfo{title}{{Theoretical Models of Optical Transients. I. A Broad Exploration of the Duration-Luminosity Phase Space},} \apj, 849, 70, \dodoi{10.3847/1538-4357/aa8fcb}

\bibitem[{P.~F. {Wang} {et~al.}(2023){Wang}, {Han}, {Xu}, {Wang}, {Yan}, {Jing}, {Su}, {Zhou}, \& {Wang}}]{2023RAA....23j4002W}
{Wang}, P.~F., {Han}, J.~L., {Xu}, J., {et~al.} 2023, \bibinfo{title}{{FAST Pulsar Database. I. Polarization Profiles of 682 Pulsars},} Research in Astronomy and Astrophysics, 23, 104002, \dodoi{10.1088/1674-4527/acea1f}

\bibitem[{K.~E. Whitaker {et~al.}(2012)Whitaker, van Dokkum, Brammer, \& Franx}]{Whitaker2012}
Whitaker, K.~E., van Dokkum, P.~G., Brammer, G., \& Franx, M. 2012, \bibinfo{title}{THE STAR FORMATION MASS SEQUENCE OUT TO z = 2.5,} The Astrophysical Journal, 754, L29, \dodoi{10.1088/2041-8205/754/2/l29}

\bibitem[{S. {Yamasaki} {et~al.}(2024){Yamasaki}, {Goto}, {Ling}, \& {Hashimoto}}]{2024MNRAS.52711158Y}
{Yamasaki}, S., {Goto}, T., {Ling}, C.-T., \& {Hashimoto}, T. 2024, \bibinfo{title}{{The true fraction of repeating fast radio bursts revealed through CHIME source count evolution},} \mnras, 527, 11158, \dodoi{10.1093/mnras/stad3844}

\bibitem[{Y.~J. {Yang} {et~al.}(2025){Yang}, {Aryan}, {Chen}, {Kong}, {Hashimoto}, \& {Jin}}]{2025ATel17101....1Y}
{Yang}, Y.~J., {Aryan}, A., {Chen}, T.~W., {et~al.} 2025, \bibinfo{title}{{Swift XRT and UVOT Upper Limits of FRB 20250316A},} The Astronomer's Telegram, 17101, 1

\bibitem[{M. {Zevin} {et~al.}(2022){Zevin}, {Nugent}, {Adhikari}, {Fong}, {Holz}, \& {Kelley}}]{2022ApJ...940L..18Z}
{Zevin}, M., {Nugent}, A.~E., {Adhikari}, S., {et~al.} 2022, \bibinfo{title}{{Observational Inference on the Delay Time Distribution of Short Gamma-Ray Bursts},} \apjl, 940, L18, \dodoi{10.3847/2041-8213/ac91cd}

\bibitem[{B. {Zhang}(2014){Zhang}}]{2014ApJ...780L..21Z}
{Zhang}, B. 2014, \bibinfo{title}{{A Possible Connection between Fast Radio Bursts and Gamma-Ray Bursts},} \apjl, 780, L21, \dodoi{10.1088/2041-8205/780/2/L21}

\bibitem[{Y.-K. {Zhang} {et~al.}(2023){Zhang}, {Li}, {Zhang}, {Cao}, {Feng}, {Wang}, {Qu}, {Niu}, {Zhu}, {Han}, {Jiang}, {Lee}, {Li}, {Luo}, {Niu}, {Tsai}, {Wang}, {Wang}, {Wu}, {Xu}, {Yang}, {Zhang}, {Zhou}, \& {Zhu}}]{2023ApJ...955..142Z}
{Zhang}, Y.-K., {Li}, D., {Zhang}, B., {et~al.} 2023, \bibinfo{title}{{FAST Observations of FRB 20220912A: Burst Properties and Polarization Characteristics},} \apj, 955, 142, \dodoi{10.3847/1538-4357/aced0b}

\end{thebibliography}
\bibliographystyle{aasjournalv7}

\end{CJK*}
\end{document}